\documentstyle[emulateapj,psfig]{article}

\makeatletter

\newenvironment{inlinefigure}{%
\def\@captype{figure}%
\noindent\begin{minipage}{0.999\linewidth}\begin{center}}
{\end{center}\end{minipage}\smallskip}
\makeatother

\lefthead{Barger et al.}

\begin{document}
\title{The Cosmic Evolution of Hard X-ray Selected
Active Galactic Nuclei{1}}
\author{
A.~J.~Barger,$\!$\altaffilmark{2,3,4}
L.~L.~Cowie,$\!$\altaffilmark{4}
R.~F.~Mushotzky,$\!$\altaffilmark{5}
Y.~Yang,$\!$\altaffilmark{5,6}
W.-H.~Wang,$\!$\altaffilmark{4}
A.~T.~Steffen,$\!$\altaffilmark{2}
P.~Capak$\!$\altaffilmark{4}
}

\altaffiltext{1}{Based in part on data obtained at the W. M. Keck
Observatory, which is operated as a scientific partnership among the
the California Institute of Technology, the University of
California, and NASA and was made possible by the generous financial
support of the W. M. Keck Foundation.}
\altaffiltext{2}{Department of Astronomy, University of
Wisconsin-Madison, 475 North Charter Street, Madison, WI 53706}
\altaffiltext{3}{Department of Physics and Astronomy,
University of Hawaii, 2505 Correa Road, Honolulu, HI 96822}
\altaffiltext{4}{Institute for Astronomy, University of Hawaii,
2680 Woodlawn Drive, Honolulu, HI 96822}
\altaffiltext{5}{NASA Goddard Space Flight Center, Code 662,
Greenbelt, MD 20771}
\altaffiltext{6}{Astronomy Department, University of Maryland,
College Park, MD 20742}

\slugcomment{Accepted by The Astronomical Journal (scheduled for 
Feb 2005)}

\begin{abstract}
We use highly spectroscopically complete deep and wide-area 
{\em Chandra\/} surveys to determine the cosmic evolution 
of hard X-ray--selected AGNs.
For the deep fields, we supplement the spectroscopic redshifts
with photometric redshifts to assess where the
unidentified sources are likely to lie. We find that the median 
redshifts are fairly constant with X-ray flux at $z\sim 1$.
We classify the optical spectra and measure the FWHM line widths. 
Most of the broad-line AGNs show essentially no visible 
absorption in X-rays, while the sources without broad
lines (${\rm FWHM} < 2000$~km~s$^{-1}$;
``optically-narrow'' AGNs) show a wide 
range of absorbing column densities. We determine hard X-ray 
luminosity functions for all spectral types with
$L_X\ge 10^{42}$~ergs~s$^{-1}$ and for broad-line AGNs 
alone. At $z<1.2$, both are well described by 
pure luminosity evolution, with $L_\ast$ evolving as $(1+z)^{3.2\pm 0.8}$
for all spectral types and as $(1+z)^{3.0\pm 1.0}$ for broad-line AGNs
alone. Thus, all AGNs drop in luminosity by almost an order of magnitude
over this redshift range. We show that this observed drop 
is due to AGN downsizing rather than to an evolution
in the accretion rates onto the supermassive black holes.

We directly compare our broad-line AGN hard 
X-ray luminosity functions with the optical QSO luminosity functions 
and find that at the bright end they agree extremely 
well at all redshifts. However, the optical QSO luminosity functions 
do not probe faint enough to see the downturn in the broad-line AGN 
hard X-ray luminosity functions and even appear to be missing some 
sources at the lowest luminosities they probe.

We find that broad-line AGNs dominate the number densities at the 
higher X-ray luminosities, while optically-narrow AGNs dominate at the 
lower X-ray luminosities. We rule out galaxy dilution as a partial
explanation for this effect by measuring the nuclear UV/optical
properties of the {\em Chandra\/} sources using the {\em HST\/} ACS 
GOODS-North data. The UV/optical nuclei of the optically-narrow
AGNs are much weaker than expected if the optically-narrow AGNs 
were similar to the broad-line AGNs. We therefore postulate the 
need for a luminosity dependent unified model.
An alternative possibility is that the broad-line AGNs and the 
optically-narrow AGNs are intrinsically different source populations. 
We cover both interpretations by constructing composite spectral 
energy distributions---including long-wavelength data from the 
mid-infrared to the submillimeter---by spectral type 
and by X-ray luminosity. We use these spectral energy 
distributions to infer the bolometric corrections (from hard X-ray 
luminosities to bolometric luminosities) needed to map the accretion 
history. 

We determine the accreted supermassive black hole 
mass density for all spectral types and for broad-line AGNs alone
using the observed evolution of the hard X-ray energy density 
production rate and our inferred bolometric corrections.
We find that only about one-half to one-quarter of the supermassive 
black hole mass density was fabricated in broad-line AGNs. Using 
either recent optical QSO luminosity function 
determinations or our broad-line AGN hard X-ray luminosity function 
determinations, we measure an accreted supermassive black hole mass 
density that is a factor of almost two lower 
than that measured by previous work, assuming $\epsilon=0.1$. 
This leaves room for the obscured accretion when compared with the
local supermassive black hole mass density.
In fact, we find reasonable agreement between the accreted 
supermassive black hole mass density from all spectral types and 
the local supermassive black hole mass density, assuming
$\epsilon\approx 0.1-0.2$. However, there is very little room for
further obscured sources or for any low efficiency accretion periods.
\end{abstract}


\section{Introduction}
\label{secintro}

The determination of the time-history of accretion 
is crucial to our understanding of how supermassive black holes 
form and evolve. However, much of the accretion power in 
the universe is absorbed (e.g., Almaini, Lawrence, \& Boyle 1999), 
making it difficult to measure at optical wavelengths.
The {\em Chandra\/} (Weisskopf et al.\ 2002)
and {\em XMM-Newton\/} (Jansen et al.\ 2001)
{\em X-ray Observatories\/} have revolutionized distant 
active galactic nucleus (AGN) studies
by making it possible to map the history of the AGN population
using hard ($2-8$~keV) X-ray surveys. Hard X-rays can directly 
probe AGN activity, are uncontaminated by star 
formation processes at the X-ray luminosities of interest, 
and detect all but the most absorbed sources. Thus,
hard X-ray surveys provide as complete and unbiased a sample of
AGNs as is presently possible (e.g., Mushotzky 2004), though
they will still miss Compton-thick sources.

Not surprisingly, hard X-ray luminosity functions constructed 
from {\em Chandra\/} and {\em XMM-Newton\/} samples have revealed 
that previous optically-selected and soft X-ray--selected AGN 
samples substantially undercount the AGN population,
at least at low and intermediate X-ray luminosities
($L_{2-8~{\rm keV}}<10^{44}$~ergs~s$^{-1}$; 
Cowie et al.\ 2003; Hasinger 2003; Steffen et al.\ 2003;
Fiore et al.\ 2003; Ueda et al.\ 2003). In fact,
optically-selected broad-line AGNs comprise 
only about a third (e.g., Barger et al.\ 2003b) 
of the X-ray background (Giacconi et al.\ 1962),
and many X-ray sources show no signs of AGN activity at all in 
their optical spectra (e.g., Barger et al.\ 2001b;
Tozzi et al.\ 2001; Hornschemeier et al.\ 2001).
Theoretical models of supermassive black hole formation 
(e.g., Haehnelt, Natarajan, \& Rees 1998;
Kauffmann \& Haehnelt 2000, 2002)
have historically relied on comparisons with the optical quasar 
luminosity function, so these models now need to be reworked.

In addition, the hard X-ray luminosity functions have revealed 
that broad-line AGNs 
dominate the number densities at the higher X-ray luminosities,
while non--broad-line AGNs dominate at the lower X-ray luminosities
(Steffen et al.\ 2003; we hereafter refer
to this as the ``Steffen effect''). Although we do not yet have a 
physical explanation for the Steffen effect, there are two
simple possibilities to consider. One possibility is that
the simple unified model for AGNs, where the differences between 
broad-line AGNs and non--broad-line AGNs are only due to orientation effects, 
needs to be modified to include an X-ray luminosity dependent 
covering factor. The second, more speculative possibility is that 
the Steffen effect is a consequence of the broad-line AGNs being
intrinsically different than the non--broad-line AGNs. 

We are now in a position to do a very 
thorough study of the nature and evolution of hard X-ray--selected 
AGNs. The primary aims of this paper are to explore the origin of
the Steffen effect and to determine the bolometric luminosities
of all AGNs in order to understand the energy release history
of supermassive black hole accretion. The recent advances that make this comprehensive 
study possible are the 
high-resolution, multicolor observations that have been 
obtained with the ACS camera on the {\em Hubble Space Telescope (HST)\/} 
of fields with deep {\em Chandra\/} X-ray data, and the extensive, 
high-quality optical spectroscopic follow-up observations that have 
been made of X-ray sources detected in both deep and wide-area 
{\em Chandra\/} surveys.

The structure of the paper is as follows. 
In \S\ref{secsamp}, we describe the X-ray samples used in our analysis 
and the completeness of the spectroscopic and photometric redshift 
identifications. We then determine the X-ray luminosities and median 
redshifts versus X-ray flux.
In \S\ref{secclass}, we spectrally classify the optical counterparts 
to the X-ray sources and investigate the dependence of the optical 
spectral types on X-ray obscuration. In \S\ref{secnuc}, we use multicolor 
{\em HST\/} ACS Great Observatories Origins Deep Survey (GOODS-North;
Giavalisco et al.\ 2004) observations of the CDF-N to separate 
the nuclear component of each source from the host galaxy
light, and we then compare the (nuclear -- galaxy) colors with 
the optical spectral types.

In \S\ref{sechxlf}--\S\ref{sechizhxlf}, we construct up-to-date 
low and high-redshift hard X-ray luminosity functions 
(all spectral types and broad-line AGNs alone) for our highly 
spectroscopically 
complete samples, do maximum likelihood fits over the redshift 
range $z=0-1.2$, and examine the evolution of the hard X-ray 
luminosity functions. In \S\ref{seclf}, we directly compare the 
broad-line AGN hard X-ray luminosity functions with the optical 
QSO luminosity functions.
In \S\ref{secenergy}, we determine the evolution of the rest-frame 
hard X-ray comoving energy density production rate
for all spectral types together as well as separated by spectral type.
In \S\ref{secsmbh}, we estimate black hole masses for a small sample 
of broad-line AGNs using the measured MgII 2800~\AA\ line widths and nuclear 
optical luminosities. We then explore ``mass starvation'' versus 
``AGN downsizing'' as explanations for the observed rapid decline 
in the energy density production rates between $z=1$ and $z=0$.

We show in \S\ref{secfailure} that the simple unified model cannot
explain the Steffen effect. In \S\ref{seclum}, we postulate that the 
simplest interpretation of the Steffen effect is a luminosity dependent 
unified model, although another interpretation might be intrinsic
differences in the source populations. In \S\ref{sechow}, we 
discuss how the bolometric corrections to go from X-ray 
luminosities to bolometric luminosities can be determined. 
To infer what these corrections are, we need long-wavelength data. 
In \S\ref{seclongwave}, we use mid-infrared 
(MIR) and far-infrared (FIR)/submillimeter
data obtained with the ISOPHOT and ISOCAM instruments on the
{\em Infrared Space Observatory (ISO)\/} and the SCUBA
bolometer array on the James Clerk Maxwell Telescope (JCMT)
to observe directly any enhancements at these wavelengths
due to absorption and reradiation by gas and dust. We note 
that the observational situation in the MIR/FIR may be expected 
to improve with the {\em Spitzer Space Telescope\/}.
Since we do not know for sure what the origin of the Steffen effect 
is, in \S\ref{secspectype} and \S\ref{sechxlum}, we cover both 
possibilities (intrinsic differences in the source populations or 
a luminosity dependent unified model) 
by constructing composite spectral energy distributions of the 
sources first by optical spectral type and then by X-ray luminosity. 
In \S\ref{secbolcorrtype} and 
\S\ref{secbolcorrlum}, we infer the 
bolometric corrections by spectral type and by X-ray luminosity. 
In \S\ref{secacc}, we use our bolometric 
corrections and the rest-frame hard X-ray comoving energy density 
production rate to determine the accretion history of the universe. 
We summarize our results in \S\ref{secsummary}.

We assume $\Omega_M=1/3$, $\Omega_\Lambda=2/3$, and
$H_o=65$~km~s$^{-1}$~Mpc$^{-1}$. All magnitudes are in the AB 
magnitude system.

\section{Hard and Soft X-ray Samples}
\label{secsamp}

A detailed study of the cosmic evolution of the hard X-ray source 
population requires the complementarity of deep and wide-field 
X-ray surveys. Fortunately, both types of X-ray surveys are now 
becoming available. Since the {\em Chandra\/} angular resolution is
critical to the identification of the optical counterparts to the
X-ray sources, in this paper we only consider {\em Chandra\/} 
surveys (except at the very brightest X-ray fluxes, where we
supplement our data with an {\em ASCA\/} sample).
To make sure our analysis is as robust as possible,
we include only the three most spectroscopically complete 
{\em Chandra\/} surveys available.

The two deep X-ray surveys that we use
are the {\em Chandra\/} Deep Field-North (CDF-N) and 
the {\em Chandra\/} Deep Field-South (CDF-S),
the deepest X-ray images ever taken.
The $\approx2$~Ms CDF-N survey samples a large, distant 
cosmological volume down to very faint X-ray flux limits of 
$f_{2-8~{\rm keV}} \approx1.4\times
10^{-16}$~ergs~cm$^{-2}$~s$^{-1}$ and 
$f_{0.5-2~{\rm keV}} \approx 1.5 \times
10^{-17}$~ergs~cm$^{-2}$~s$^{-1}$ 
(Alexander et al.\ 2003b), while
the $\approx1$~Ms CDF-S survey samples only a factor of two 
shallower (Giacconi et al.\ 2002). 

We supplement these deep surveys with a wide-area, intermediate
depth survey in order to sample a large, low-redshift cosmological 
volume and to detect the rare, high-luminosity population. 
The low-redshift volume enables us 
to probe robustly the evolution of AGNs between $z\sim 0$ 
and $z\sim 1$. The only wide-area, intermediate depth survey 
published to date that has a high level of spectroscopic completeness 
is the {\em Chandra\/} Large-Area Synoptic X-ray Survey, or CLASXS.
This survey covers an $\sim0.4$~deg$^2$ region in the 
Lockman Hole-Northwest, imaged to X-ray flux limits of 
$f_{2-8~{\rm keV}} \approx3\times 10^{-15}$~ergs~cm$^{-2}$~s$^{-1}$
and $f_{0.5-2~{\rm keV}}\approx5\times 10^{-16}$~ergs~cm$^{-2}$~s$^{-1}$
(Yang et al.\ 2004).

The high spatial resolution of the {\em Chandra\/} X-ray images
allows a generally unambiguous optical counterpart to be found
for the great majority of X-ray sources having counterparts $R<26$.
(Note that about 20\% of the optical counterpart identifications in 
the $R=24-26$ range will be spurious; see \S\ref{secphotz}.)
Only a few have multiple counterparts.
Most of the $R<24$ sources, and many of the $R<25$ sources, can be
spectroscopically identified (Barger et al.\ 2003b;
Szokoly et al.\ 2004; Steffen et al.\ 2004). In Table~\ref{tab1},
we show for the three {\em Chandra\/} fields in our sample
(the most intensively spectroscopically
observed {\em Chandra\/} fields to date) the total
number of sources in the X-ray catalogs, the number of such
sources observed spectroscopically, and the number of such
sources identified spectroscopically.

If we want to explore the highest X-ray luminosities, we also need
much wider-field, higher-flux samples than these {\em Chandra\/} 
data can provide. {\em ASCA\/}, {\em BeppoSAX\/},
and {\em RXTE\/} provide such samples, but with their
low spatial resolutions, cross-identifications to the optical
counterparts are more difficult and ambiguous. Ultimately, many of
the sources in these surveys will be pinned down with 
higher resolution {\em Chandra\/} or {\em XMM-Newton\/} observations
and their counterparts robustly identified. Fortunately, however,
since these high X-ray flux sources generally do have brighter optical 
counterparts than the sources in lower X-ray flux surveys,
in many cases it is already possible to identify the counterparts,
even with the positional uncertainties. In particular, the
{\em ASCA\/} sample of Akiyama et al.\ (2003)
have nearly complete identifications (see Table~\ref{tab1}), albeit
with some ambiguous cases, while roughly 70\% of the bright
{\em RXTE\/} sample have identifications
(Sazonov \& Revnivtsev 2004).

Hereafter, the CDF-N, CDF-S, CLASXS, and 
Akiyama et al.\ (2003) {\em ASCA\/} 
hard X-ray--selected samples will constitute this paper's ``total hard 
X-ray sample'', and the CDF-N, CDF-S, and CLASXS soft X-ray--selected 
samples will constitute this paper's ``total soft X-ray sample''.

%
%
\begin{deluxetable}{lrrrr}
\tablecaption{Breakdown of the X-ray Samples by Field and Spectral
Class}
\tablehead{Category & CDF-N & CDF-S & CLASXS & ASCA}
\startdata
total  &    503  &  346  &  525 & 32 \cr
observed  &    451  &  247  &  467 & 32  \cr
identified  &    306  &  137  &  272 & 31  \cr
\cr
broad-line  &    43  &  32  & 106 & 30  \cr
high-excitation &    39  &  23  &  45 & 0    \cr
star formers  &   148  &  55  &  73 & 0  \cr
absorbers  &    58  &  20  &  28 & 0  \cr
stars  &    14  & 7  &  20 & 1 \cr
\enddata
\label{tab1}
\end{deluxetable}

\subsection{Spectroscopic Completeness}
\label{seccomplete}

\begin{figure*}
\centerline{\psfig{figure=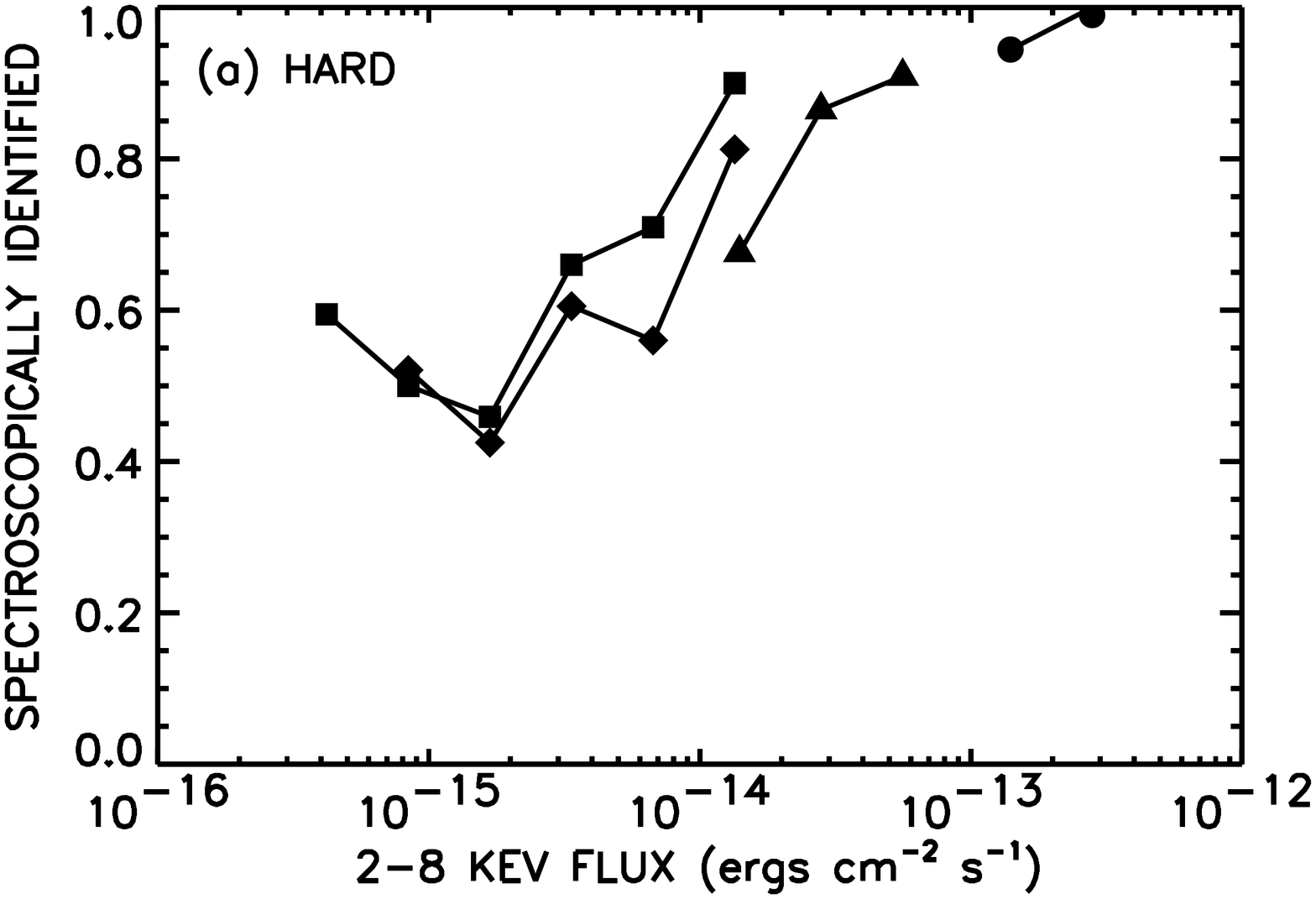,width=3.5in,angle=90}
\psfig{figure=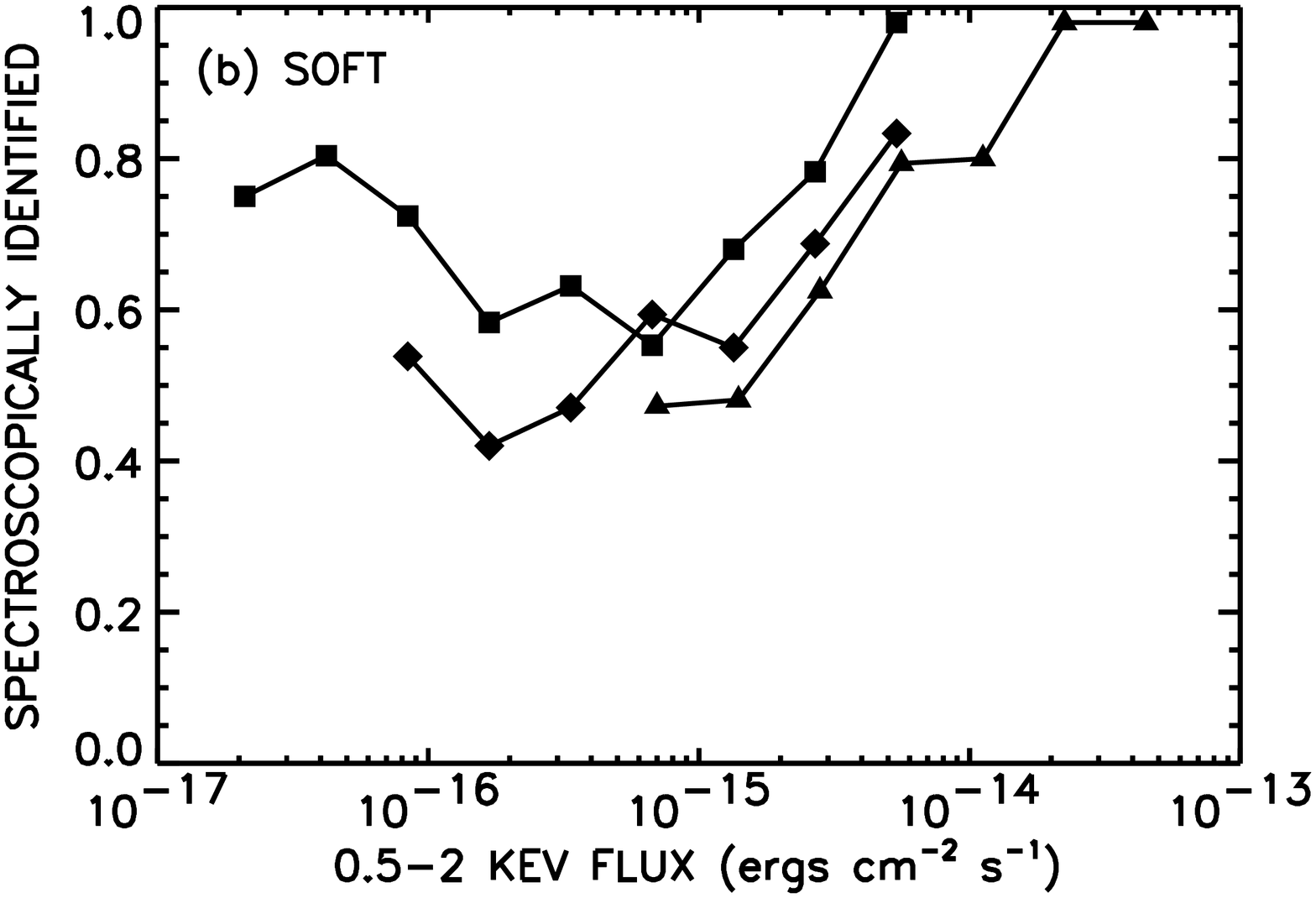,width=3.5in,angle=90}}
\figurenum{1}
\caption{
(a) Spectroscopic completeness (i.e., the
fraction of observed sources that are spectroscopically identified)
of the hard X-ray surveys that make up the total hard X-ray sample
({\em squares\/}---CDF-N, Barger et al.\ 2003b;
{\em diamonds\/}---CDF-S, Szokoly et al.\ 2004;
{\em triangles\/}---CLASXS, Steffen et al.\ 2004;
{\em circles\/}---{\em ASCA\/}, Akiyama et al.\ 2003).
Sources are grouped into flux bins that increase by a multiplicative
factor of 2. Only flux bins containing more than 10 sources are
plotted to illustrate the useful flux ranges of the various samples.
(b) Same, but now of the soft X-ray surveys that make up the total
soft X-ray sample.
\label{figsamp}
}
\end{figure*}

Figure~\ref{figsamp} shows the useful flux ranges of the three 
{\em Chandra\/} surveys and the {\em ASCA\/} survey (which is only 
in the hard X-ray sample) that make up this paper's total hard and 
soft X-ray samples. The surveys provide good coverage over the flux 
ranges 
$f_{2-8~{\rm keV}}\approx 10^{-16}-10^{-13}$~ergs~cm$^{-2}$~s$^{-1}$
and
$f_{0.5-2~{\rm keV}}\approx 10^{-17}-10^{-13}$~ergs~cm$^{-2}$~s$^{-1}$.
Note that for all of the {\em Chandra\/} surveys, the observed X-ray 
counts were converted to $2-8$~keV fluxes in the original papers 
using individual 
power-law indices determined from the ratios of the hard-to-soft 
X-ray counts (i.e., the hardness ratios). For the {\em ASCA\/} survey,
Akiyama et al.\ (2003) assumed an intrinsic
$\Gamma=1.7$ photon index to compute $2-10$~keV fluxes, so we
have converted their $2-10$~keV fluxes to $2-8$~keV assuming 
their $\Gamma=1.7$. 

\begin{figure*}
\centerline{\psfig{figure=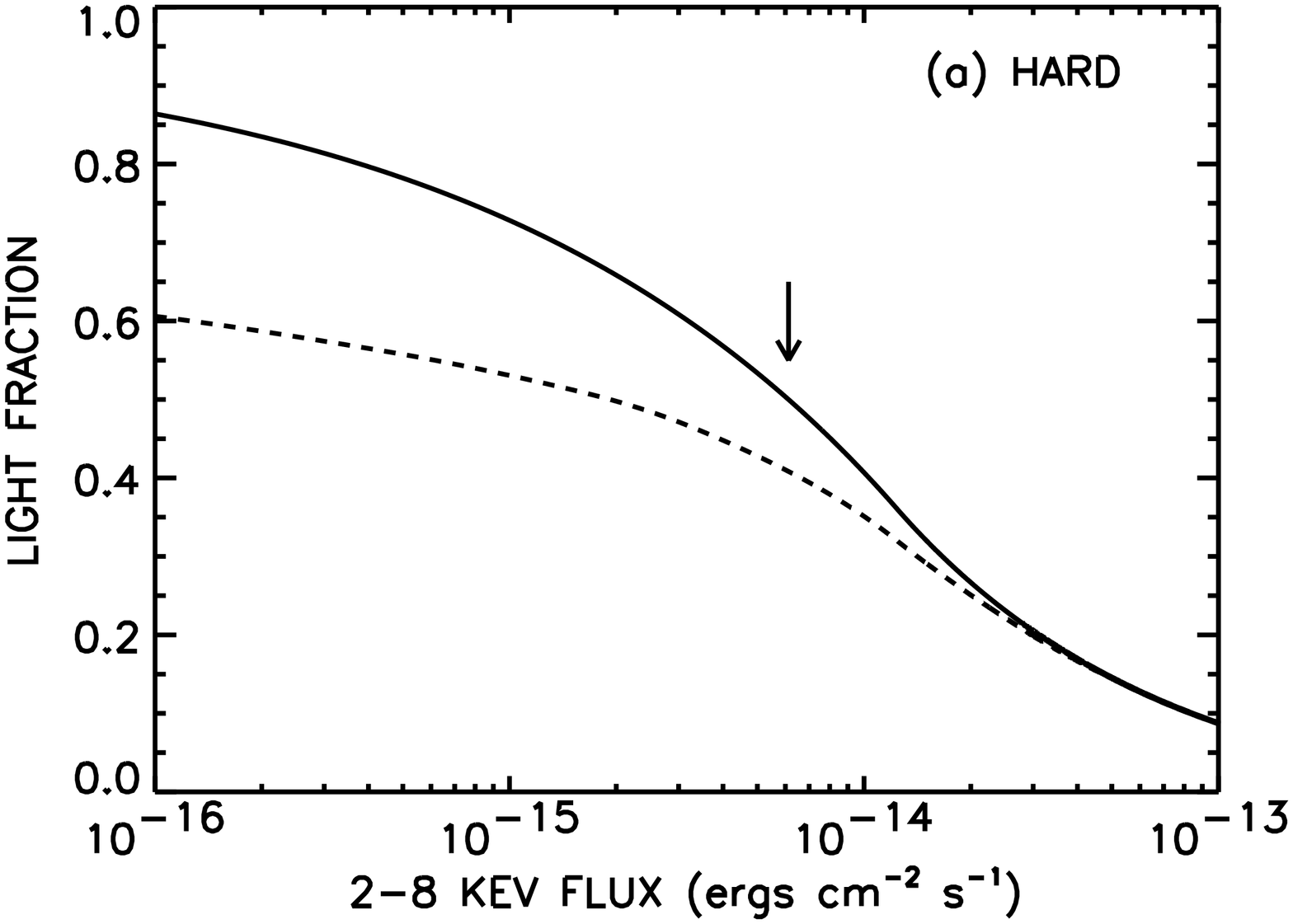,width=3.5in,angle=90}
\psfig{figure=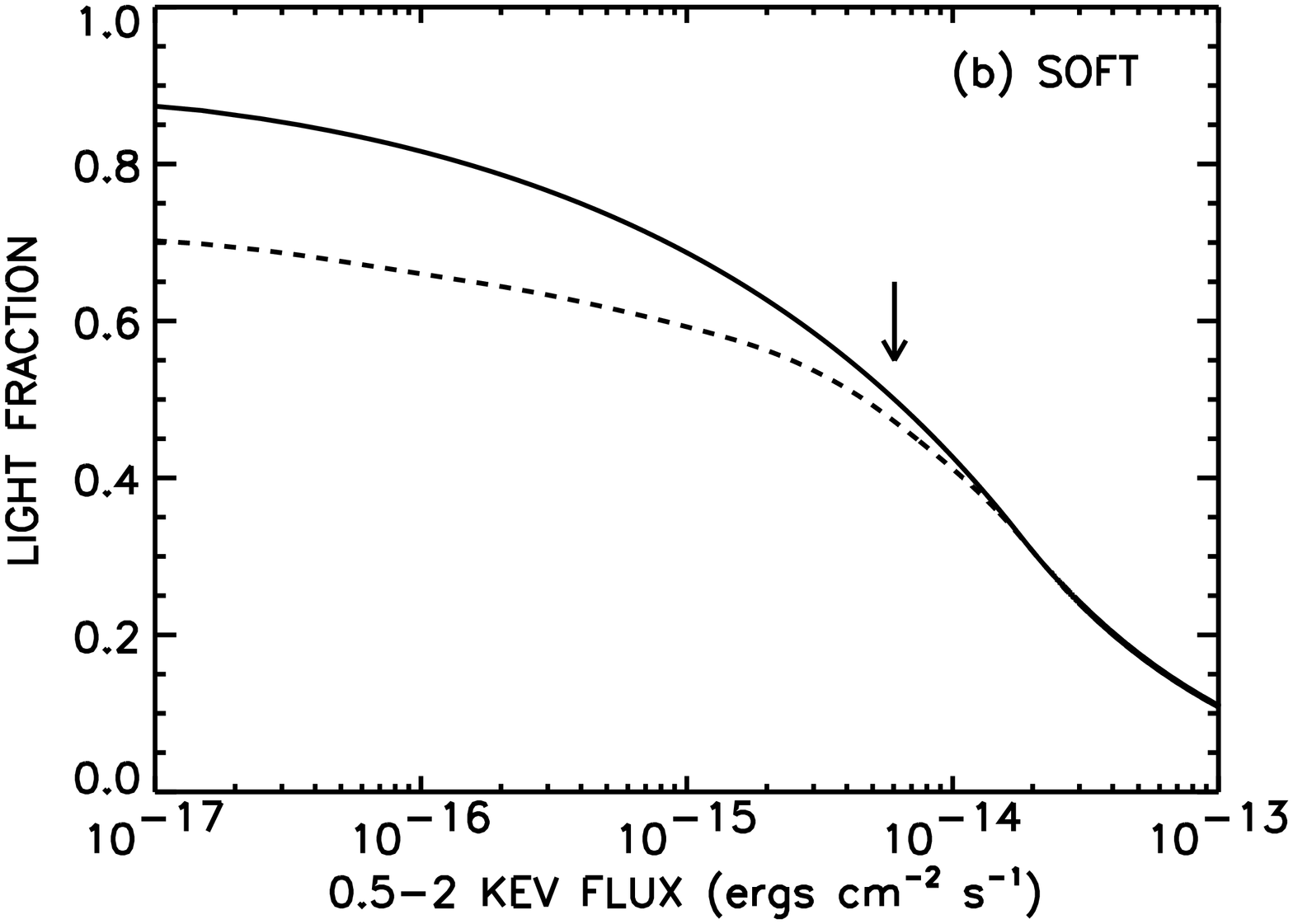,width=3.5in,angle=90}}
\figurenum{2}
\caption{
(a) Fraction of the $2-8$~keV light from the X-ray number
counts that lies above a given flux for the total hard X-ray sample.
Dashed line shows the fraction of the light that is spectroscopically
identified. Arrow marks the half-light flux.
(b) Same, but now fraction of the $0.5-2$~keV
light for the total soft X-ray sample.
\label{figfract}
}
\end{figure*}

Figure~\ref{figfract} shows how the contributions to
the $2-8$~keV total resolved X-ray background, obtained by extrapolating 
the number counts to faint and bright X-ray fluxes, are
strongly peaked around 
$f_{2-8~{\rm keV}}\approx 10^{-14}$~ergs~cm$^{-2}$~s$^{-1}$.
In fact, nearly 75\% arises in the
$f_{2-8~{\rm keV}}\approx 10^{-16}-10^{-13}$~ergs~cm$^{-2}$~s$^{-1}$
range (e.g., Campana et al.\ 2001;
Cowie et al.\ 2002;
Rosati et al.\ 2002;
Alexander et al.\ 2003b).
Approximately 70\% of the $2-8$~keV light in the flux range
$f_{2-8~{\rm keV}}\approx 10^{-16}-10^{-13}$~ergs~cm$^{-2}$~s$^{-1}$
is spectroscopically identified,
while about 80\% of the $0.5-2$~keV light in the flux range
$f_{0.5-2~{\rm keV}}\approx 10^{-17}-10^{-13}$~ergs~cm$^{-2}$~s$^{-1}$
is spectroscopically identified.

Figure~\ref{figsamp}a also shows that
the fraction of observed sources that can be spectroscopically
identified is a strong function of the hard X-ray flux.
Above $f_{2-8~{\rm keV}}\approx 10^{-14}$~ergs~cm$^{-2}$~s$^{-1}$,
$80-90$\% of the sources have spectroscopic redshifts, while below
this, the fraction drops to about 60\%. While this is partly
a consequence of the fainter X-ray sources being optically
fainter, it is also a consequence of the fraction of broad-line AGNs 
being much higher at the brighter X-ray luminosities.

\subsection{X-ray Luminosities}
\label{secxlum}

The hard X-ray flux limit of
$f_{2-8~{\rm keV}}\approx 10^{-14}$~ergs~cm$^{-2}$~s$^{-1}$,
above which the redshift identifications are very complete 
($80-90$\%), corresponds to a rest-frame $2-8$~keV luminosity 
of $L_X\approx 5\times10^{43}$~ergs~s$^{-1}$ at $z=1$, while 
the soft X-ray flux limit of
$f_{0.5-2~{\rm keV}}\approx 2\times10^{-15}$~ergs~cm$^{-2}$~s$^{-1}$,
above which the surveys are about 80\% spectroscopically complete,
corresponds to a rest-frame $2-8$~keV luminosity of
$L_X\approx 1.7\times10^{44}$~ergs~s$^{-1}$ at $z=3$. Thus, nearly 
all of the high X-ray luminosity sources have been identified, and
incompleteness has very little effect on the bright end of the
hard X-ray luminosity function determinations (see \S\ref{sechxlf}).

\begin{figure*}
\centerline{\psfig{figure=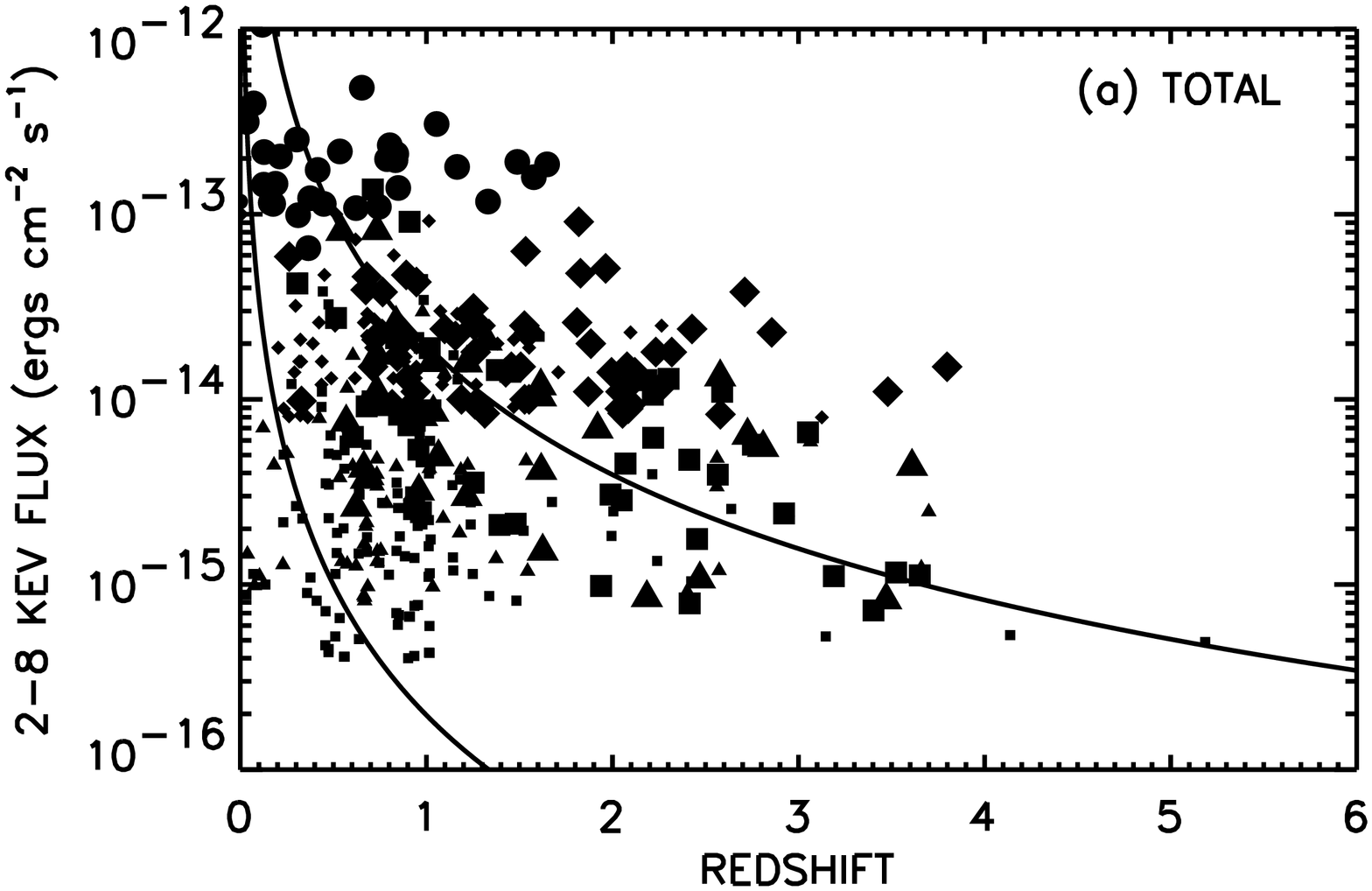,width=3.5in,angle=90}
\psfig{figure=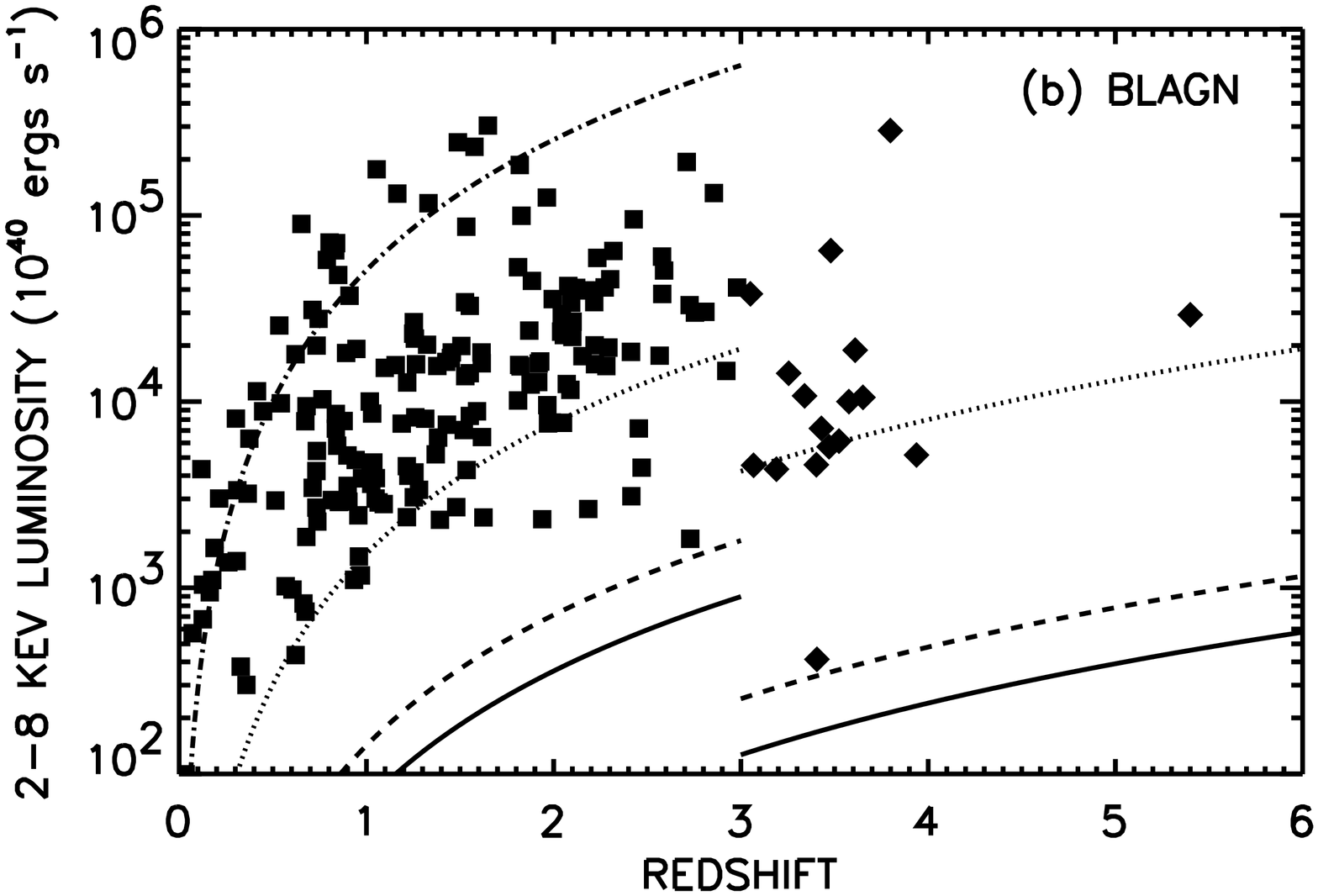,width=3.5in,angle=90}}
\figurenum{3}
\caption{
(a) $2-8$~keV flux vs. redshift for the hard
X-ray surveys that make up the total hard X-ray sample
({\em squares\/}---CDF-N, Barger et al.\ 2003b;
{\em triangles\/}---CDF-S, Szokoly et al.\ 2004;
{\em diamonds\/}---CLASXS, Steffen et al.\ 2004;
{\em circles\/}---{\em ASCA\/}, Akiyama et al.\ 2003).
Spectroscopically unidentified sources are not shown.
Curves correspond to loci of constant rest-frame $2-8$~keV
luminosity with $K$-corrections calculated using
a $\Gamma=1.8$ spectrum. Lower (higher) curve corresponds to
$10^{42}$~ergs~s$^{-1}$ ($10^{44}$~ergs~s$^{-1}$).
Broad-line AGNs (an optical spectroscopic classification; see \S3)
are denoted by large symbols.
(b) $2-8$~keV luminosity vs. redshift for the broad-line AGNs in
the total hard X-ray sample. At $z<3$, the luminosities were
calculated from the observed-frame $2-8$~keV fluxes {\em (squares)},
and at $z>3$, the luminosities were calculated from the observed-frame
$0.5-2$~keV fluxes {\em (diamonds)}. The $K$-corrections
were calculated using a $\Gamma=1.8$ spectrum. Curves were
determined using the on-axis flux limits of the different surveys
and the $K$-corrections
({\em solid\/}---CDF-N, Alexander et al.\ 2003b;
{\em dashed\/}---CDF-S, Giacconi et al.\ 2002;
{\em dotted\/}---CLASXS, Yang et al.\ 2004;
{\em dot-dashed\/}---{\em ASCA\/}, Akiyama et al.\ 2003).
Curves are disjoint at $z=3$, because, as for the individual sources,
these luminosities were calculated from the observed-frame $2-8$~keV
fluxes for $z<3$ and from the observed-frame $0.5-2$~keV fluxes
for $z>3$, and the $0.5-2$~keV {\em Chandra\/} images are deeper
than the $2-8$~keV images.
\label{figbroad}
}
\end{figure*}

In Figure~\ref{figbroad}a, we show $2-8$~keV flux versus redshift
for the spectroscopically observed sources in the total hard 
X-ray sample. The CLASXS sources nicely fill in the almost an order
of magnitude flux gap between the CDF-N/CDF-S and  {\em ASCA\/} samples.
The two solid curves correspond to loci of constant
rest-frame $2-8$~keV luminosity, $L_X$. Any source more luminous than
$L_X=10^{42}$~ergs~s$^{-1}$ {\em (lower curve)\/} is very likely to be
an AGN on energetic grounds (Zezas, Georgantopoulos, \& Ward 1998;
Moran, Lehnert, \& Helfand\ 1999), though many of
the intermediate luminosity sources do not show obvious AGN signatures
in their optical spectra. Sources with $L_X>10^{44}$~ergs~s$^{-1}$ 
{\em (upper curve)\/} are often called quasars,
since this X-ray luminosity roughly corresponds to the absolute
optical magnitude of $M_B=-23$ that is the traditional dividing
line between quasars and Seyfert~1 galaxies.
However, the distinction between Seyfert~1 galaxies and quasars seems
to have little physical significance. In this paper, we
simply refer to all objects which have lines with full width half
maximum (FWHM) greater than 2000~km~s$^{-1}$ in their optical spectra
as broad-line AGNs (see \S\ref{secclass}). The broad-line AGNs are denoted by large 
symbols in Figure~\ref{figbroad}a. 
From the figure, we can see that the great majority of the most luminous 
sources, where we are very spectroscopically complete, are broad-line AGNs. 

In Figure~\ref{figbroad}b, we show rest-frame
$2-8$~keV luminosity versus redshift for the broad-line AGNs alone.
At $z<3$, the luminosities were calculated from the observed-frame 
$2-8$~keV fluxes {\em (squares)}, and at $z>3$, the luminosities 
were calculated from the observed-frame $0.5-2$~keV fluxes 
{\em (diamonds)}. One advantage of using the observed-frame soft
X-ray fluxes at high redshifts is the increased sensitivity, since
the $0.5-2$~keV {\em Chandra\/} images are deeper than the $2-8$~keV 
images. In addition, at $z=3$, observed-frame $0.5-2$~keV corresponds
to rest-frame $2-8$~keV, providing the best possible
match to the lower redshift data. In calculating the rest-frame luminosities,
we have assumed an intrinsic $\Gamma=1.8$, for which there is only
a small differential $K$-correction to rest-frame $2-8$~keV.
Note that using the individual photon indices (rather than the
universal power-law index of $\Gamma=1.8$ adopted here) to calculate
the $K$-corrections would result in only a small difference in the
rest-frame luminosities (Barger et al.\ 2002).
The curves in Figure~\ref{figbroad}b show the luminosity limits 
(using the on-axis flux limits and the $K$-corrections) of the 
various X-ray surveys used in our analysis.
The break in the curves at $z=3$ results from switching from
observed-frame $2-8$~keV flux to observed-frame $0.5-2$~keV 
flux in calculating the rest-frame luminosities (due to
the increased sensitivity at $0.5-2$~keV). 

We can see the Steffen effect from Figure~\ref{figbroad}b---the 
population of broad-line AGNs drops off dramatically at the lower X-ray 
luminosities. Since broad-line AGNs are straightforward to identify 
spectroscopically, even in the so-called redshift ``desert'' at 
$z\sim 1.5-2$, and since the bulk of the sources in the total X-ray 
samples have now been spectroscopically observed (see Table~\ref{tab1}), 
we do not need to worry that broad-line AGNs are making up a substantial fraction 
of the unidentified population. In fact, at $16<R<24$, deep 
{\em Chandra\/} observations have picked up all of the color-selected 
quasars identified by the COMBO-17 survey in their fields of view 
(Wolf et al.\ 2004). Moreover, deep {\em Chandra\/} 
observations have picked up all of the spectroscopically identified 
broad-line AGNs in the highly spectroscopically complete (to $R=24.5$; 
Cowie et al.\ 2004b; Wirth et al.\ 2004) ACS GOODS-North region of 
the CDF-N (see \S\ref{secclass}). 

At $21<R<25.5$,
Steidel et al.\ (2002) proposed that there
might be a subsample of AGNs in their Lyman break galaxy
survey that are relatively X-ray faint and hence would not be 
detected in even the deepest X-ray pointings. They based this
on a comparison of the AGNs detected in their Lyman break galaxy survey 
with the X-ray sources detected in a restricted region of the 1~Ms 
exposure of the CDF-N. Four of their 148 Lyman break galaxy AGN candidates 
were detected in X-rays, and two of these were spectroscopically 
identified. One was found to be an optically-faint broad-line AGN at $z=3.406$. 
Using their estimated spectroscopic completeness for AGNs and their
detection of this one source, they concluded that there could be 
about 11 such optically-faint broad-line AGNs in a redshift interval 
of $\Delta z\simeq0.6$ near $z=3$ in a full {\em Chandra\/} ACIS-I 
field. As can be seen from Figure~\ref{figbroad}b, we also easily 
spectroscopically identified the $z=3.406$ broad-line AGN in the CDF-N, 
and it does indeed have a low X-ray luminosity. However, other than 
this one source, the low X-ray luminosity regime at these high 
redshifts is devoid of broad-line AGNs, despite the X-ray sensitivity to 
such sources and the ease of spectroscopic identifications of
broad-line AGNs. We therefore conclude that there are not 
very many low X-ray luminosity broad-line AGNs at $z\sim 3$. We will come 
back to the Steffen effect again when we 
construct the hard X-ray luminosity functions in \S\ref{sechxlf}.

\begin{figure*}
\centerline{\psfig{figure=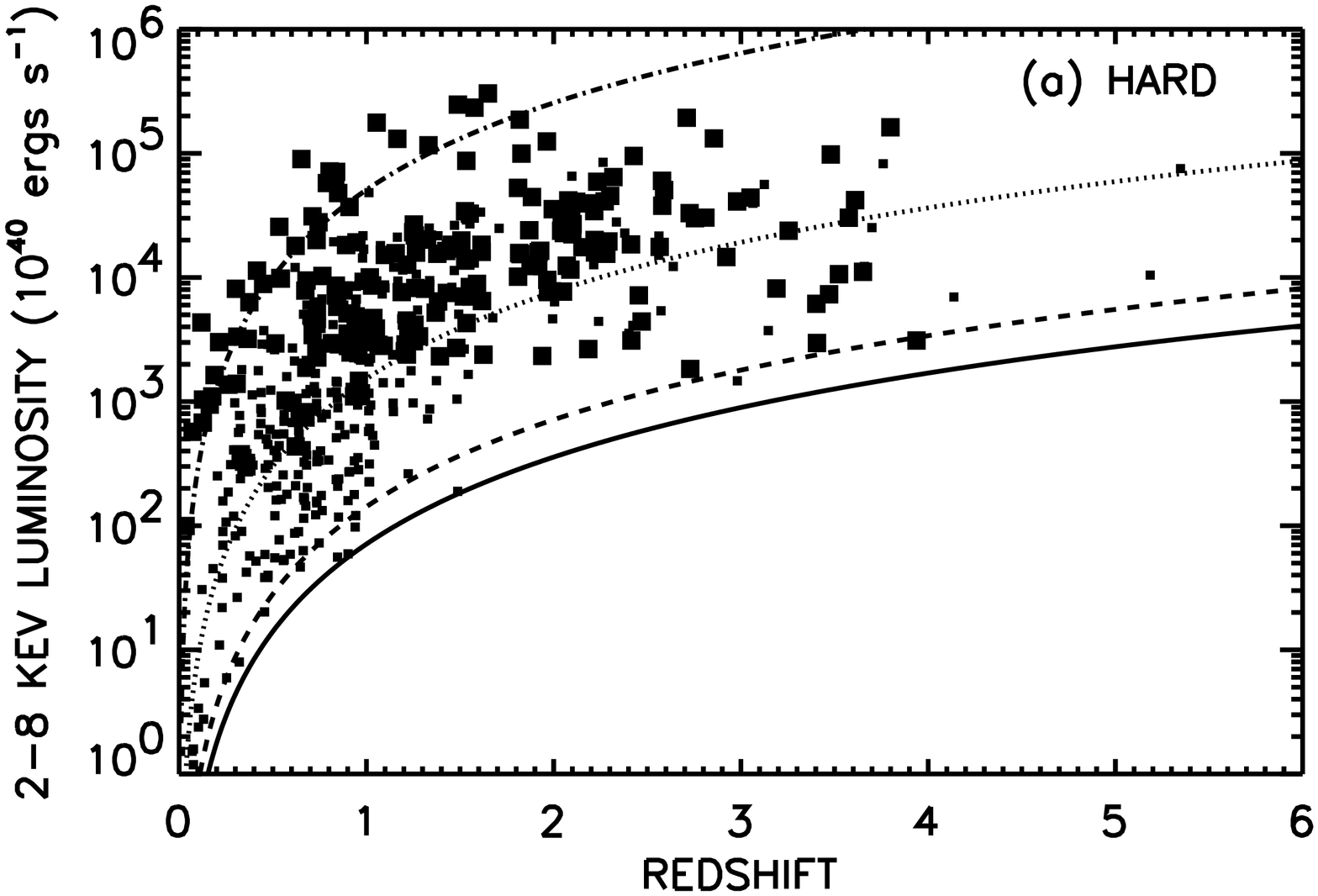,width=3.5in,angle=90}
\psfig{figure=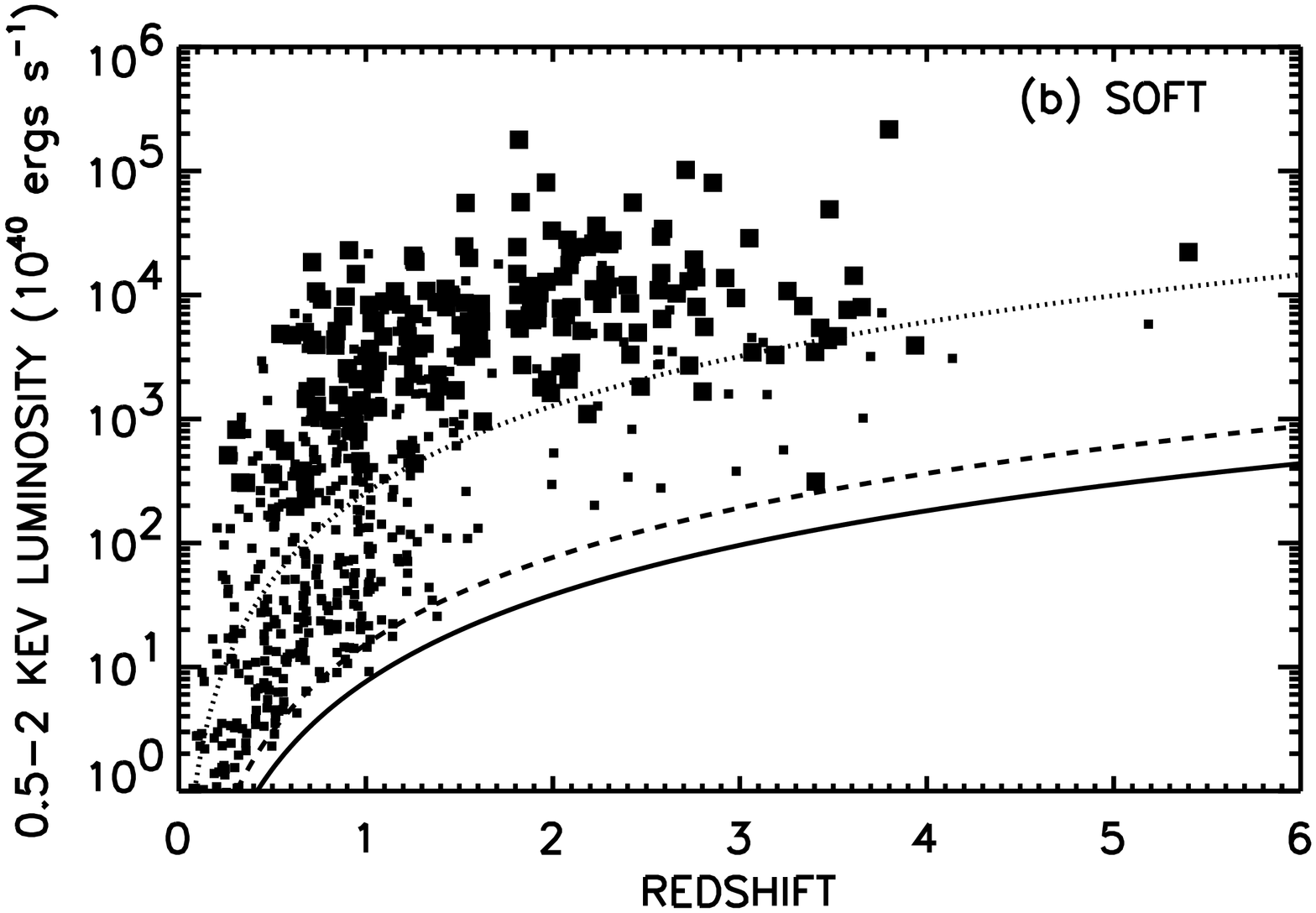,width=3.5in,angle=90}}
\figurenum{4}
\caption{
Rest-frame (a) $2-8$~keV and (b) $0.5-2$~keV luminosity
vs. redshift for the total hard and
soft X-ray samples, respectively. The luminosities were determined
using the observed-frame (a) $2-8$~keV and (b) $0.5-2$~keV X-ray fluxes
and $K$-corrections calculated using a $\Gamma = 1.8$ spectrum.
Spectroscopically unidentified sources and stars are not shown.
Broad-line AGNs are denoted by large symbols. Curves were determined using
the on-axis (a) $2-8$~keV and (b) $0.5-2$~keV flux limits of the
different surveys and the $K$-corrections
({\em solid\/}---CDF-N, Alexander et al.\ 2003b;
{\em dashed\/}---CDF-S, Giacconi et al.\ 2002;
{\em dotted\/}---CLASXS, Yang et al.\ 2004;
{\em dot-dashed\/}---{\em ASCA\/}, Akiyama et al.\ 2003, only
in (a)).
\label{figlum}
}
\end{figure*}

To illustrate the range of
luminosities covered by the hard and soft X-ray samples, 
in Figure~\ref{figlum} we show redshift versus rest-frame
(a) $2-8$~keV and (b) $0.5-2$~keV luminosity for the total 
hard and soft X-ray samples, respectively.
Here the rest-frame luminosities of the X-ray sources 
{\em (squares)\/} were determined from the observed-frame 
(a) hard and (b) soft X-ray fluxes and the $K$-corrections calculated 
using a $\Gamma=1.8$ spectrum. The rest-frame luminosity limits of 
the X-ray surveys {\em (curves)\/} were determined from the on-axis
(a) $2-8$~keV and (b) $0.5-2$~keV flux limits of the surveys and the 
$K$-corrections.

%
%

%
%

%
%

%
%

\subsection{Photometric Redshifts}
\label{secphotz}

It is possible to extend the redshift information to
fainter magnitudes using photometric redshifts rather
than spectroscopic redshifts, and this has been done
for both the CDF-N (Barger et al.\ 2002, 2003b; 
P. Capak et al., in preparation) and the CDF-S (Wolf et al.\ 2004;
Zheng et al.\ 2004).
These redshifts are robust and suprisingly accurate 
(often to better than 8\%) for non--broad-line AGNs. 

The CDF-N photometric redshifts were computed using broadband
galaxy colors and the Bayesian code of Ben\'{\i}tez (2000), 
and only sources with probabilities for the photometric redshift
of greater than 90\% were included. Zheng et al.\ (2004)
computed their photometric redshifts based on a variety of codes
and data sets and included sources with very large offsets
from the X-ray positions. However, the positional offsets should 
not be large, given the accuracy of the {\em Chandra\/} X-ray 
positions. Moreover, as we move to optically fainter sources, 
the cross-identifications to the X-ray sources become
progressively more insecure. Even with a $2''$
match radius, about 20\% of optical identifications
in the $R=24-26$ range will be spurious, and with larger
match radii, many of the identifications will be incorrect
(Barger et al.\ 2003b). 
Therefore, in our analysis, we only include a CDF-S photometric 
redshift from Zheng et al.\ (2004)
where the optical counterpart to the X-ray source 
lies within $2''$ of the X-ray position. Despite the different 
methodologies and the slightly fainter flux limits of the CDF-N, 
both samples have very similar photometric redshift success rates, 
with just under 85\% of the entire X-ray sample being identified in 
each of the fields. Treister et al.\ (2004) claimed that by using 
the photometric redshifts of Mobasher et al.\ (2004)
for the ACS GOODS-South region of the CDF-S, 
they could achieve 100\% spectroscopic 
plus photometric redshift completeness in that field.
However, Mobasher et al.\ do not include any photometric redshift
reliability measures (like the $>90$\% probabilities used in the CDF-N), 
and hence Treister et al.\ are likely including photometric redshifts 
for sources that are really too optically faint for reliable 
determinations.

%
%
\begin{inlinefigure}
\psfig{figure=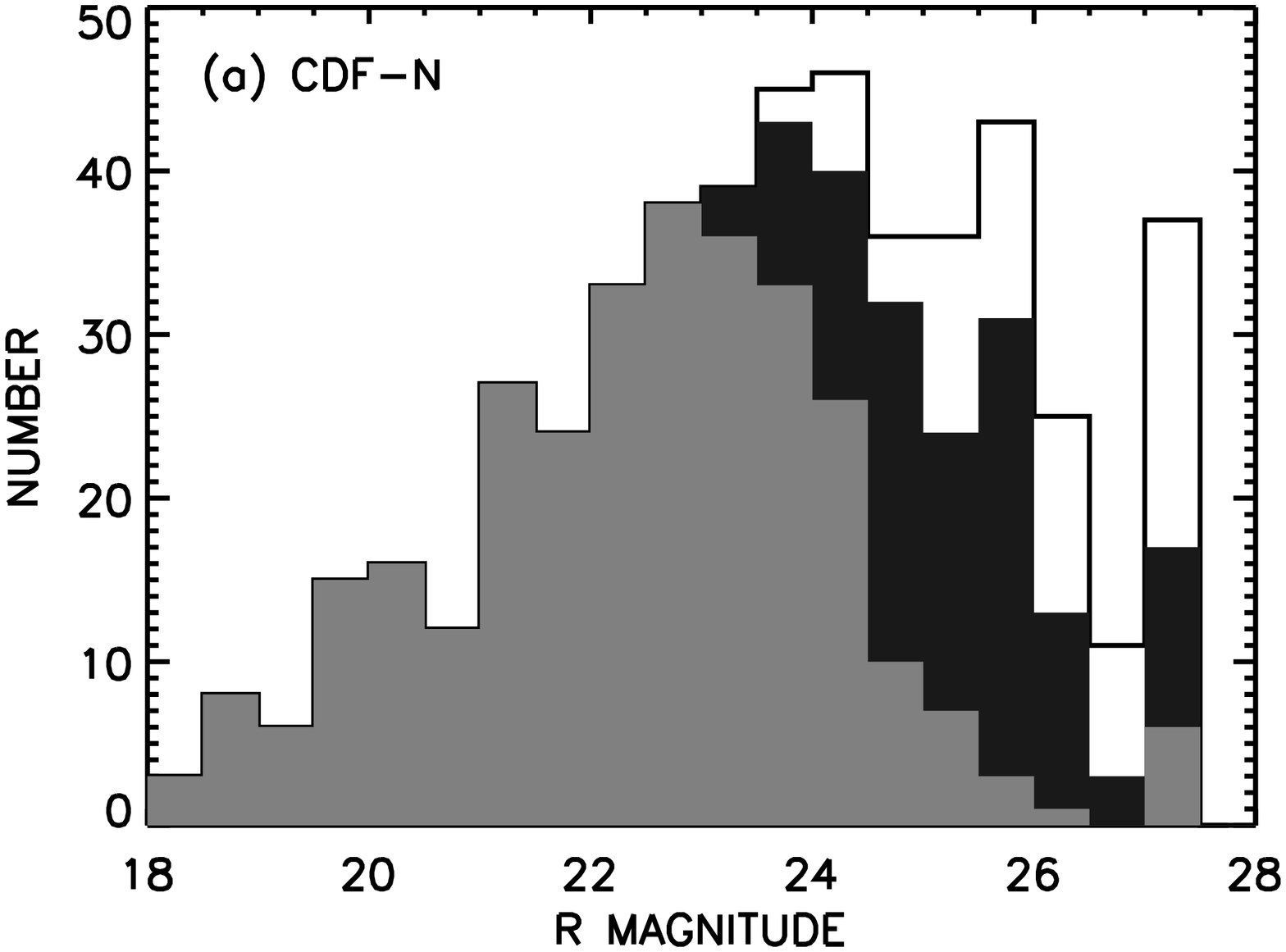,width=8.5cm,angle=90}
\psfig{figure=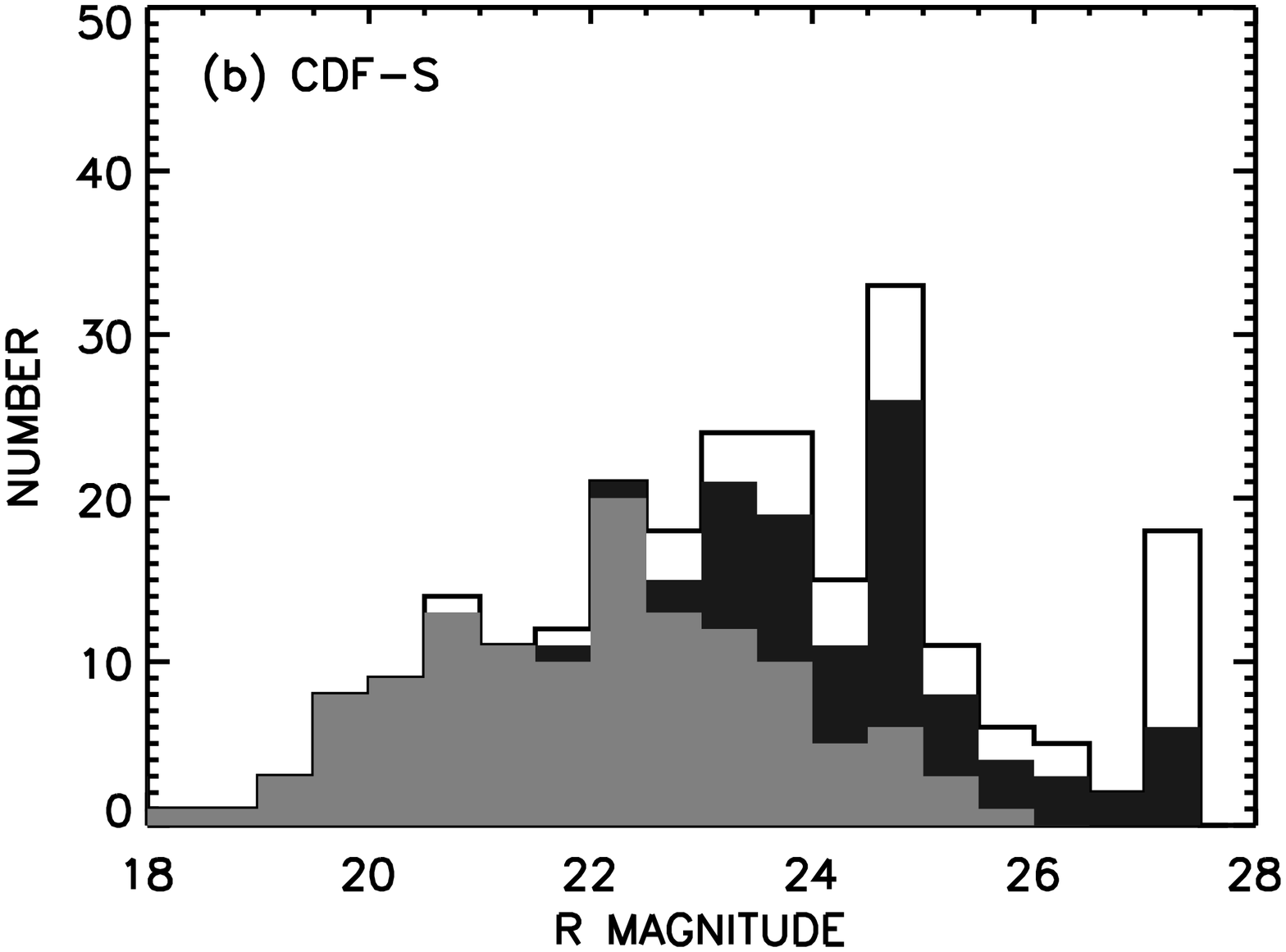,width=8.5cm,angle=90}
\figurenum{5}
\caption{
Fraction of the total (a) CDF-N and (b) CDF-S samples that
are spectroscopically {\em (light shading)\/}
and photometrically {\em (dark shading)\/}
identified vs. $R$ magnitude. All $R>27$ sources are placed in the
final magnitude bin. For both samples, the use of photometric
redshifts increases the identified fraction to about 85\%,
with most of the additional identifications lying in the
$R=24-26$ magnitude range.
\label{photz_rmags}
}
\addtolength{\baselineskip}{10pt}
\end{inlinefigure}
 
In Figure~\ref{photz_rmags}, we show the spectroscopic 
(Barger et al.\ 2003b; Szokoly et al.\ 2004) and photometric
(Barger et al.\ 2003b; Zheng et al.\ 2004) redshift
identifications versus $R$ magnitude for the CDF-N and 
CDF-S. As can be seen
from the figure, most sources brighter than $R=24$ can be 
spectroscopically identified, and photometric redshifts can extend 
this by almost 2 magnitudes. 

%
%
\begin{inlinefigure}
\psfig{figure=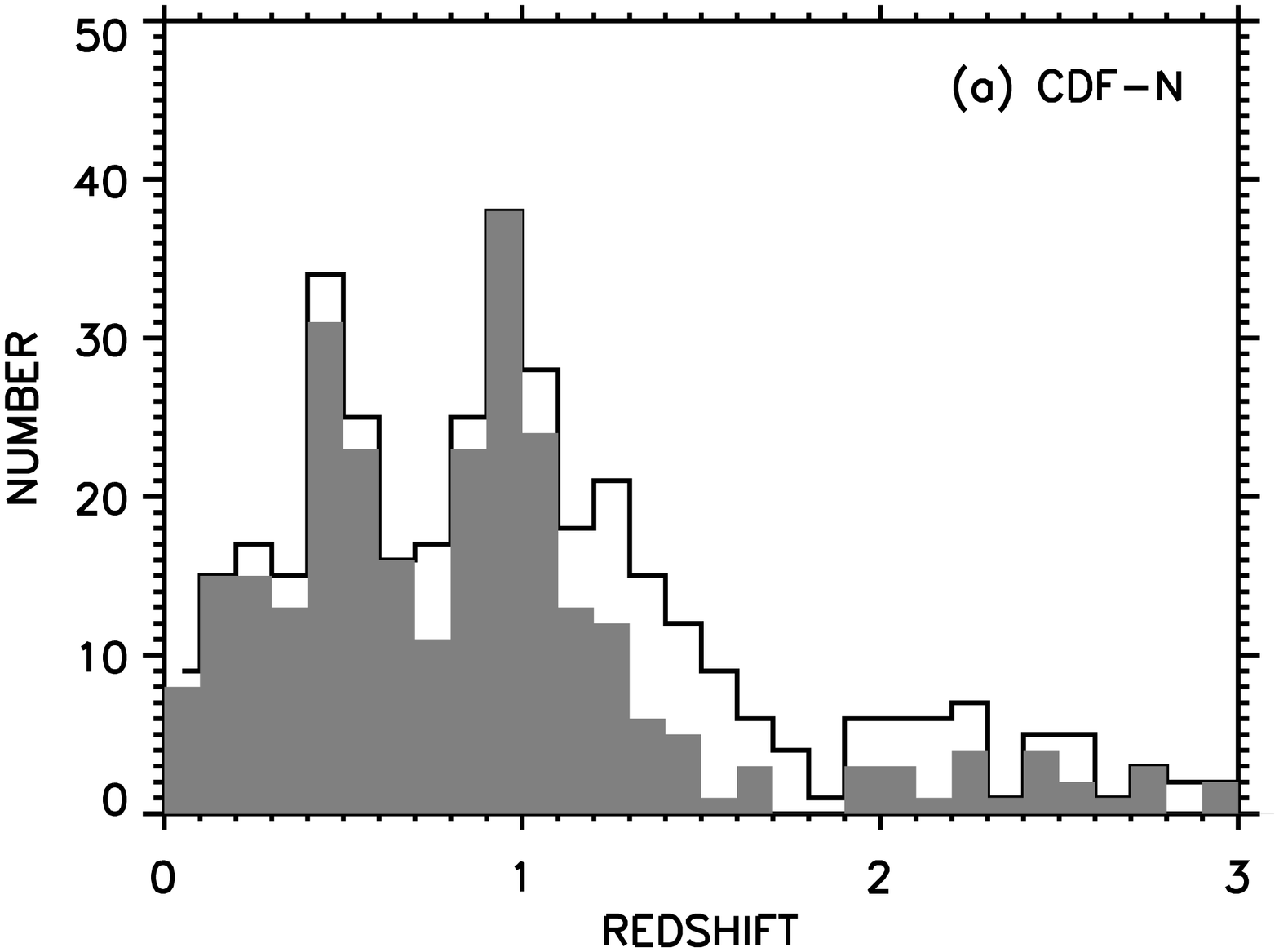,width=8.5cm,angle=90}
\psfig{figure=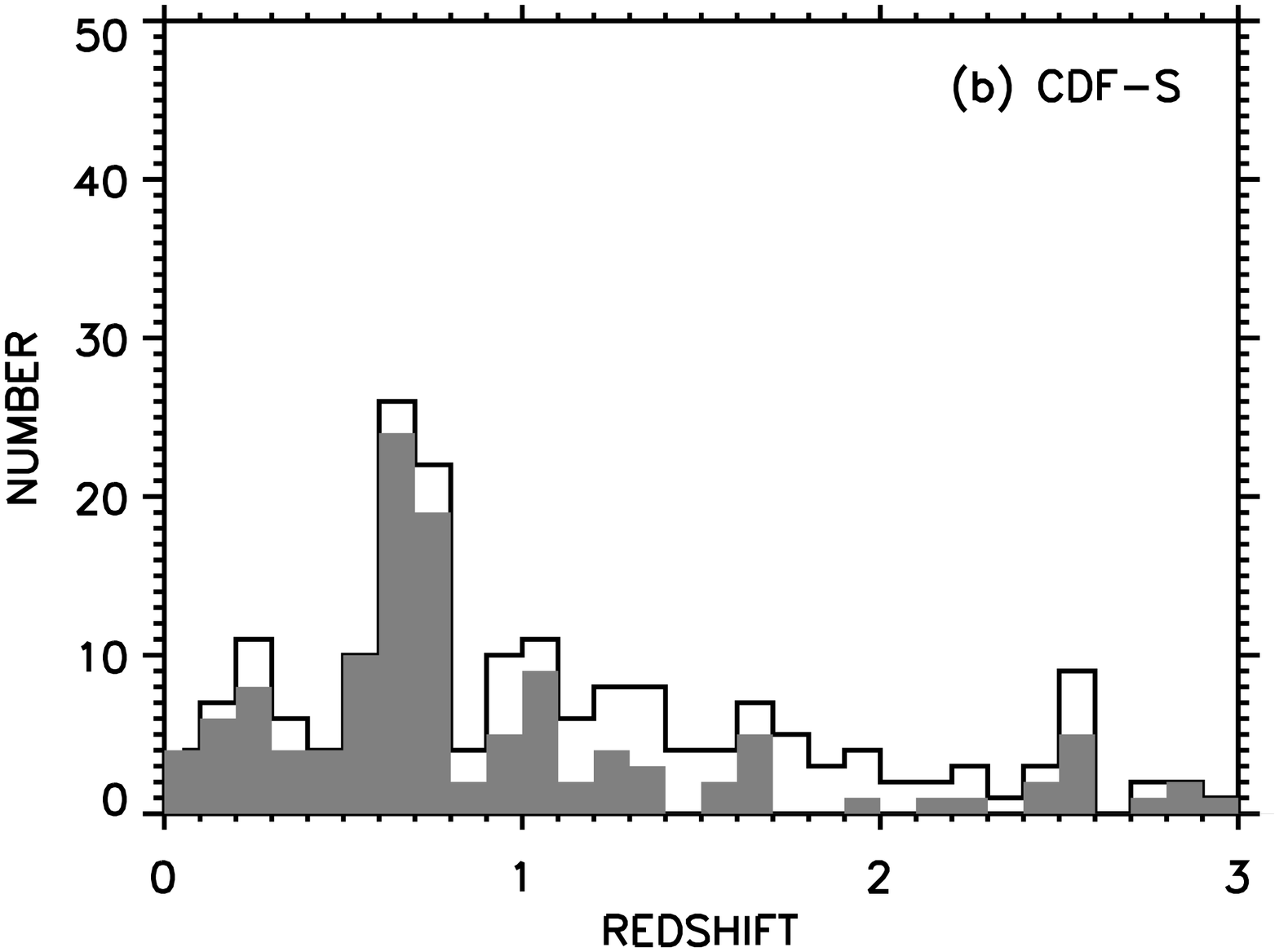,width=8.5cm,angle=90}
\figurenum{6}
\caption{
Fraction of spectroscopically {\em (shading)\/}
and photometrically {\em (open histogram)\/} identified sources
vs. redshift for the (a) CDF-N and (b) CDF-S samples.
As might be expected, the spectroscopic samples are highly complete
at $z<1.2$. The spectroscopic incompleteness at $z>1.2$ is due
to the absence of strong spectral features and the faintness of
the sources at these redshifts, making them harder to identify.
\label{photz_hist}
}
\addtolength{\baselineskip}{10pt}
\end{inlinefigure}

Given the absence, as yet, of a similar set of photometric 
redshifts for the CLASXS region, we have chosen to analyze our 
samples primarily by assuming that the redshifts of the optically 
fainter, spectroscopically unidentified sources are essentially 
unknown. Nevertheless, we may use the photometric redshifts to 
assess where the unidentified sources are likely to lie, and 
hence where incompleteness is likely to be a serious concern in
computing the luminosity functions (see \S\ref{sechxlf}). 
We show a comparison
of the photometrically identified sources with the
spectroscopically identified sources in Figure~\ref{photz_hist}.
As might be expected, nearly all of the $z<1.2$ sources with photometric
redshifts are also spectroscopically identified, and incompleteness
only becomes a serious concern at $z>1.2$, where the sources become 
optically faint and much harder to identify spectroscopically.

Figure~\ref{figlumz} shows the luminosity versus redshift
plot of Figure~\ref{figlum}a for just the CDF-N, with the 
spectroscopic and photometric redshifts denoted by solid and 
open squares, respectively. Many of the photometrically identified
(but spectroscopically unidentified) sources correspond to lower 
X-ray luminosity sources in the $z=1.2-2.5$ range. We shall 
assume hereafter that the spectroscopically unidentified sources 
lie primarily at $z>1.2$. 

%
%
\begin{inlinefigure}
\psfig{figure=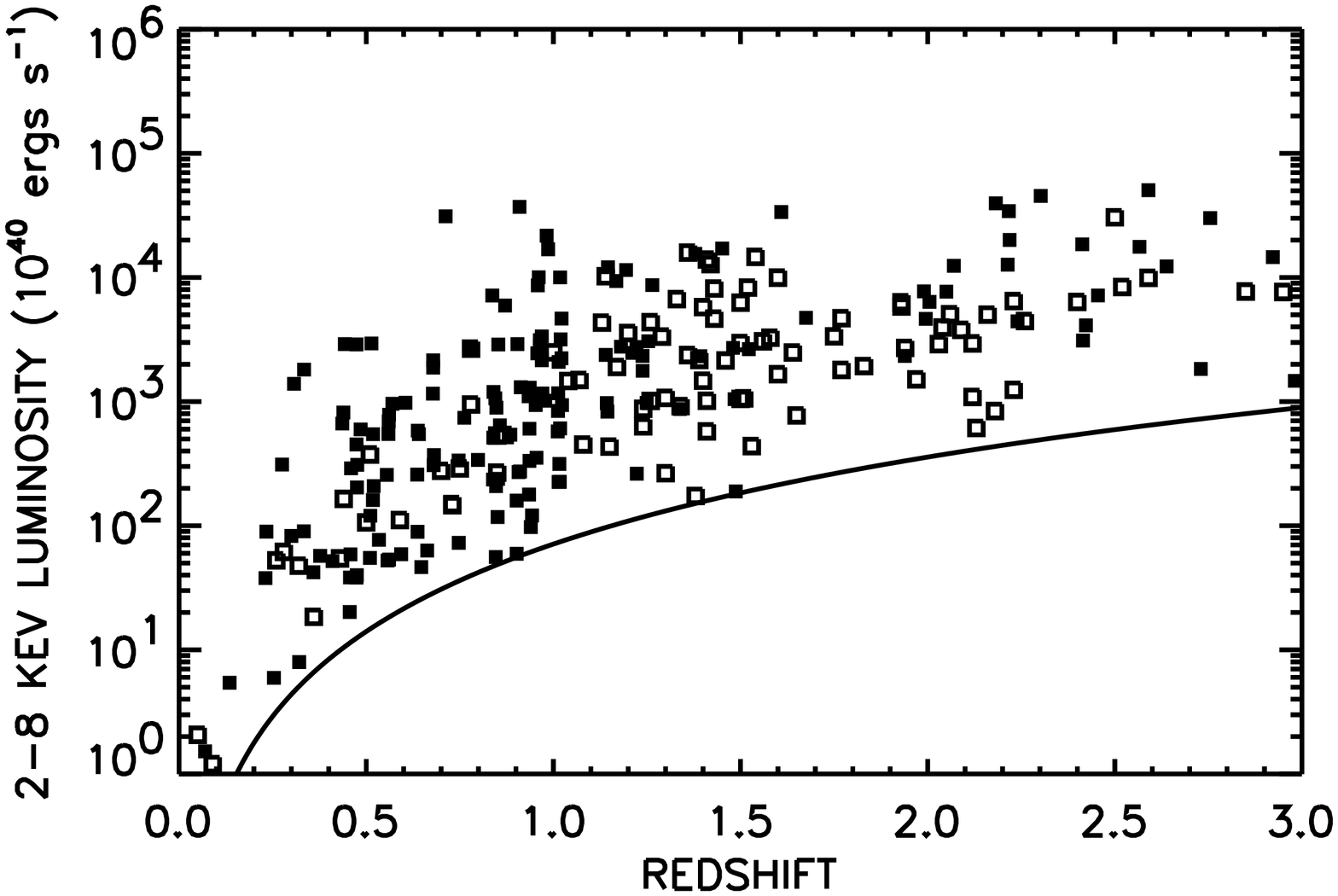,width=8.5cm,angle=90}
\figurenum{7}
\caption{
Rest-frame $2-8$~keV luminosity vs. redshift for the total hard
X-ray sample in the CDF-N. The luminosities were determined
using the observed-frame $2-8$~keV fluxes and $K$-corrections
calculated using a $\Gamma = 1.8$ spectrum. Stars are not shown.
Spectroscopically (photometrically) identified sources are denoted
by solid (open) squares. Solid curve was determined using the
on-axis $2-8$~keV flux limit of the CDF-N survey
(Alexander et al.\ 2003b) and the $K$-corrections.
\label{figlumz}
}
\addtolength{\baselineskip}{10pt}
\end{inlinefigure}

\subsection{Alternate Redshift Estimators}

Fiore et al.\ (2003) have pointed out 
a correlation between the X-ray--to--optical flux ratios 
and the hard X-ray luminosities of non--broad-line AGNs, such that 
higher X-ray luminosity sources tend to have higher 
X-ray--to--optical flux ratios. This correlation arises
due to the obscuration of the nuclear UV/optical light---but not
the hard X-ray light---in these systems, such that the host 
galaxy light dominates. The redshift determination then 
relies on the fact that the host galaxies have a narrow 
range in their rest-frame optical magnitudes. 

%
%
\begin{inlinefigure}
\psfig{figure=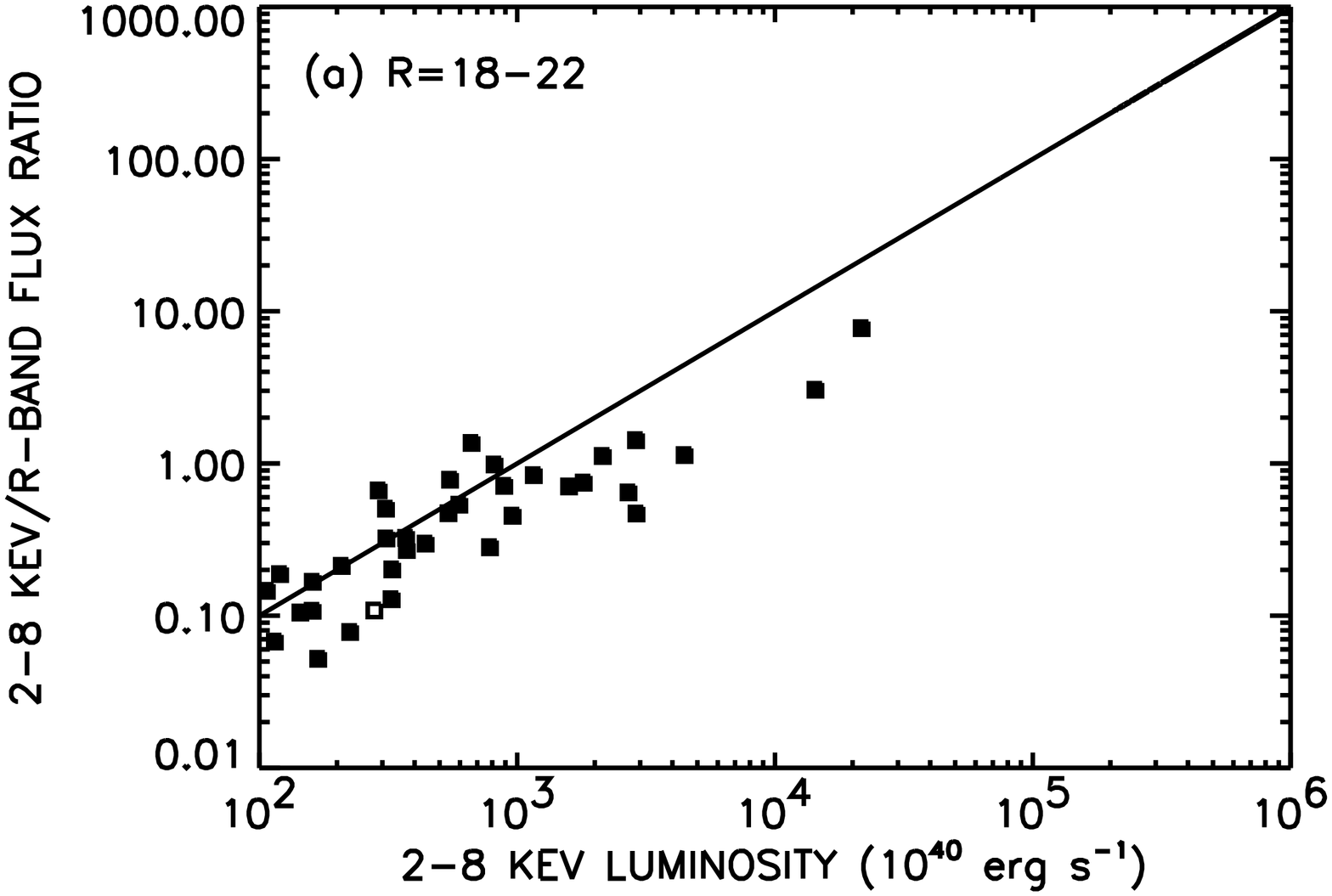,width=8.5cm,angle=90}
\psfig{figure=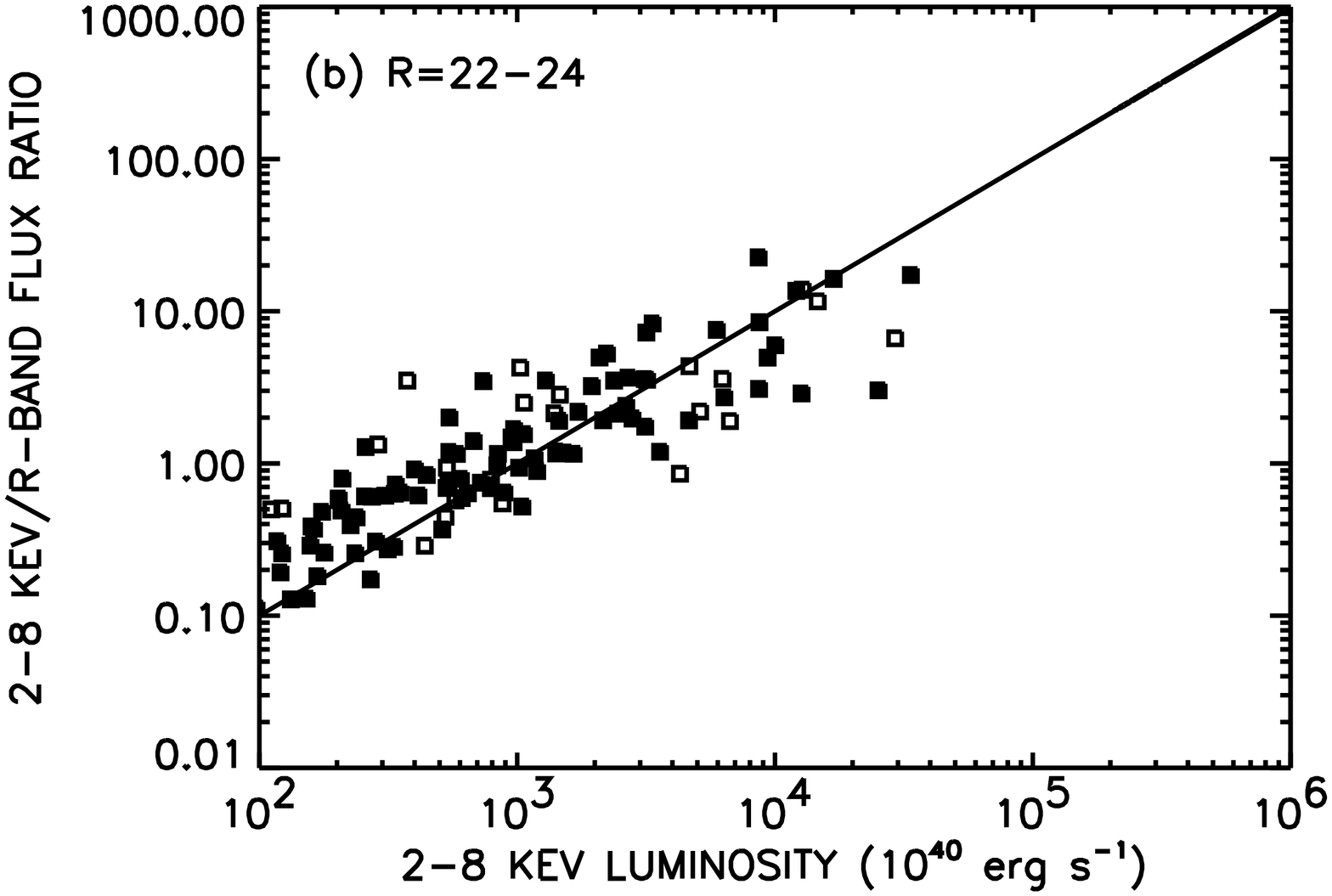,width=8.5cm,angle=90}
\psfig{figure=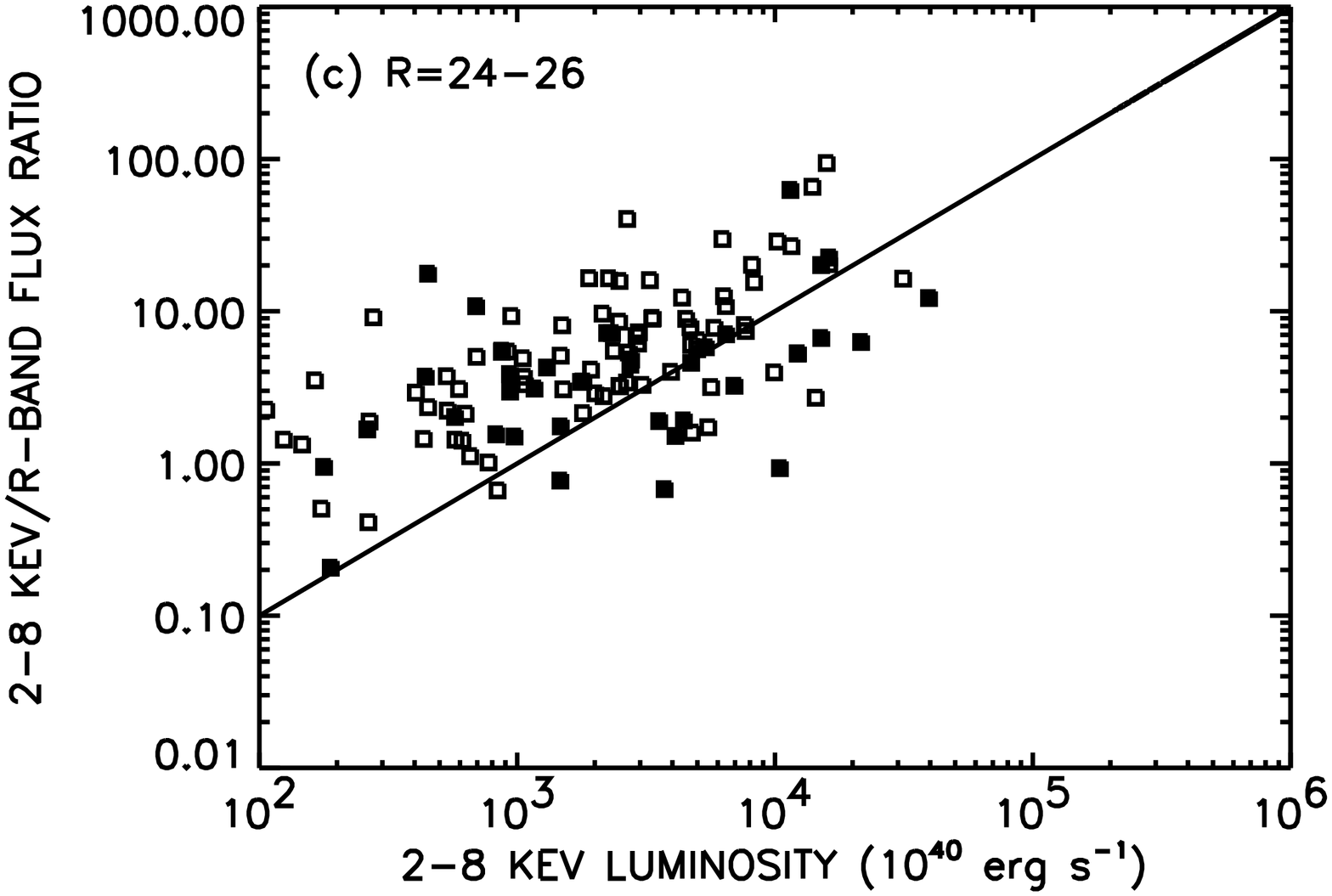,width=8.5cm,angle=90}
\figurenum{8}
\caption[]{
$f_X/f_R$ vs. $L_X$ for the non--broad-line AGN sample in the CDF-N
and CDF-S for $R$-magnitude ranges (a) $R=18-22$, (b) $R=22-24$,
and (c) $R=24-26$. Spectroscopic (photometric) redshifts are shown
as solid (open) squares. Solid line shows the linear relation
given by Fiore et al.\ (2004).
\label{fiore_test}
}
\addtolength{\baselineskip}{10pt}
\end{inlinefigure}

We can use the spectroscopic plus photometric redshift 
samples in the CDF-N and CDF-S 
to test the Fiore et al.\ (2003) 
relation. In Figure~\ref{fiore_test}, we show the X-ray--to--optical 
flux ratios versus $L_X$ for the non--broad-line AGNs, separated into three 
$R$ magnitude intervals. 
Unfortunately, we find that the relationship depends on both 
variables, with the fainter $R$ magnitude
sources having a higher normalization of the X-ray--to--optical 
flux ratio versus $L_X$. This means that applying the
Fiore et al.\ (2003) relation to the sources 
that do not have redshift determinations (generally because they 
are faint) using a calibration that is based only on 
sources with redshifts (generally because they are bright) will 
result in an overestimate of the redshifts of these sources. 
Thus, estimates of the number of obscured sources with quasar
X-ray luminosities based on this type of analysis will be too high
(see, e.g., Padovani et al.\ 2004).

It might be possible to make a better estimate of the
redshifts using the dependence of the rest-frame absolute 
magnitude on $L_X$, together with typical $K$-corrections.
However, given the small fraction of unidentified sources
in our sample, and the relatively large uncertainties in 
this type of determination, we do not pursue this in the 
present paper.

\subsection{Median Redshifts}

\begin{figure*}
\centerline{\psfig{figure=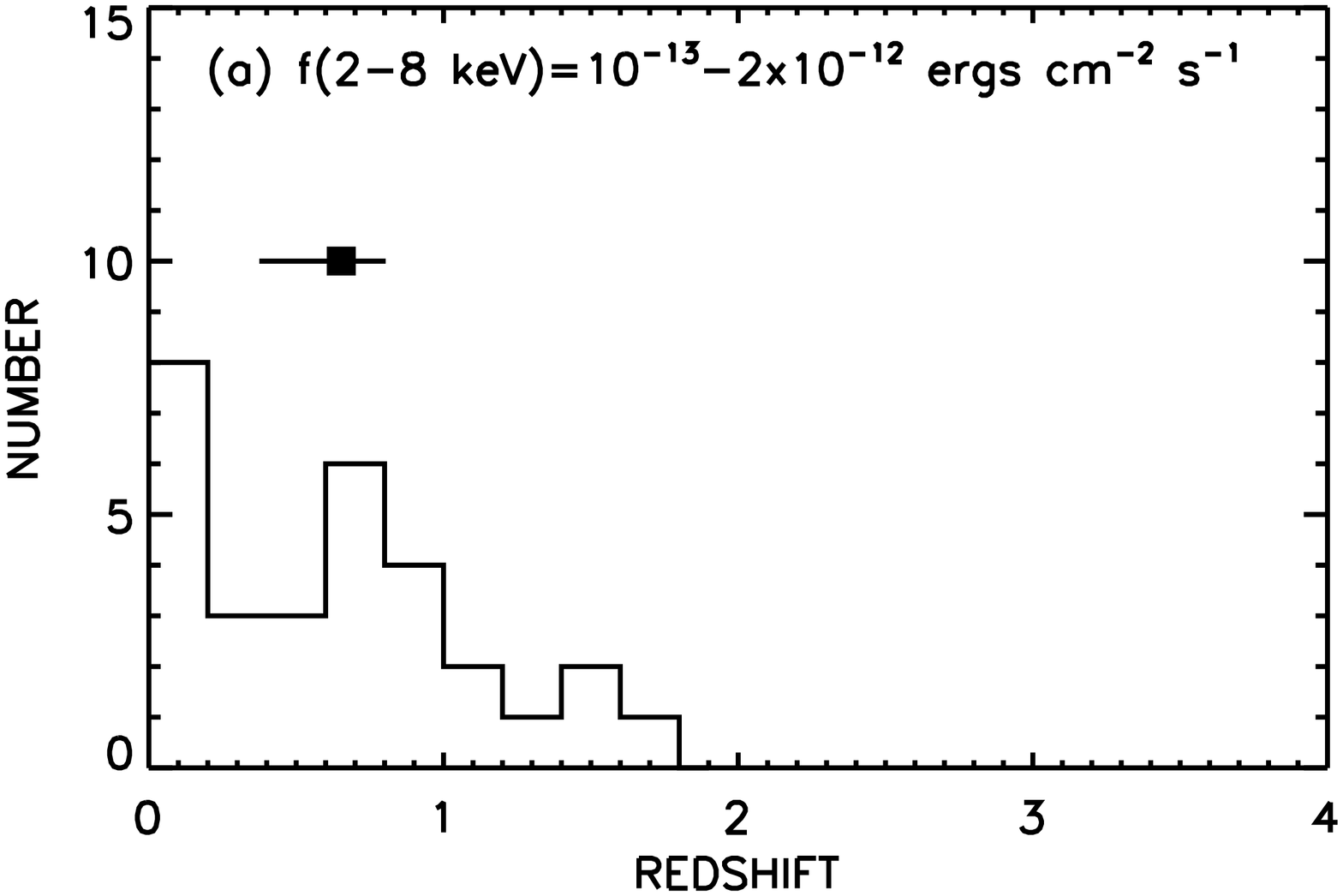,width=3.5in,angle=90}
\psfig{figure=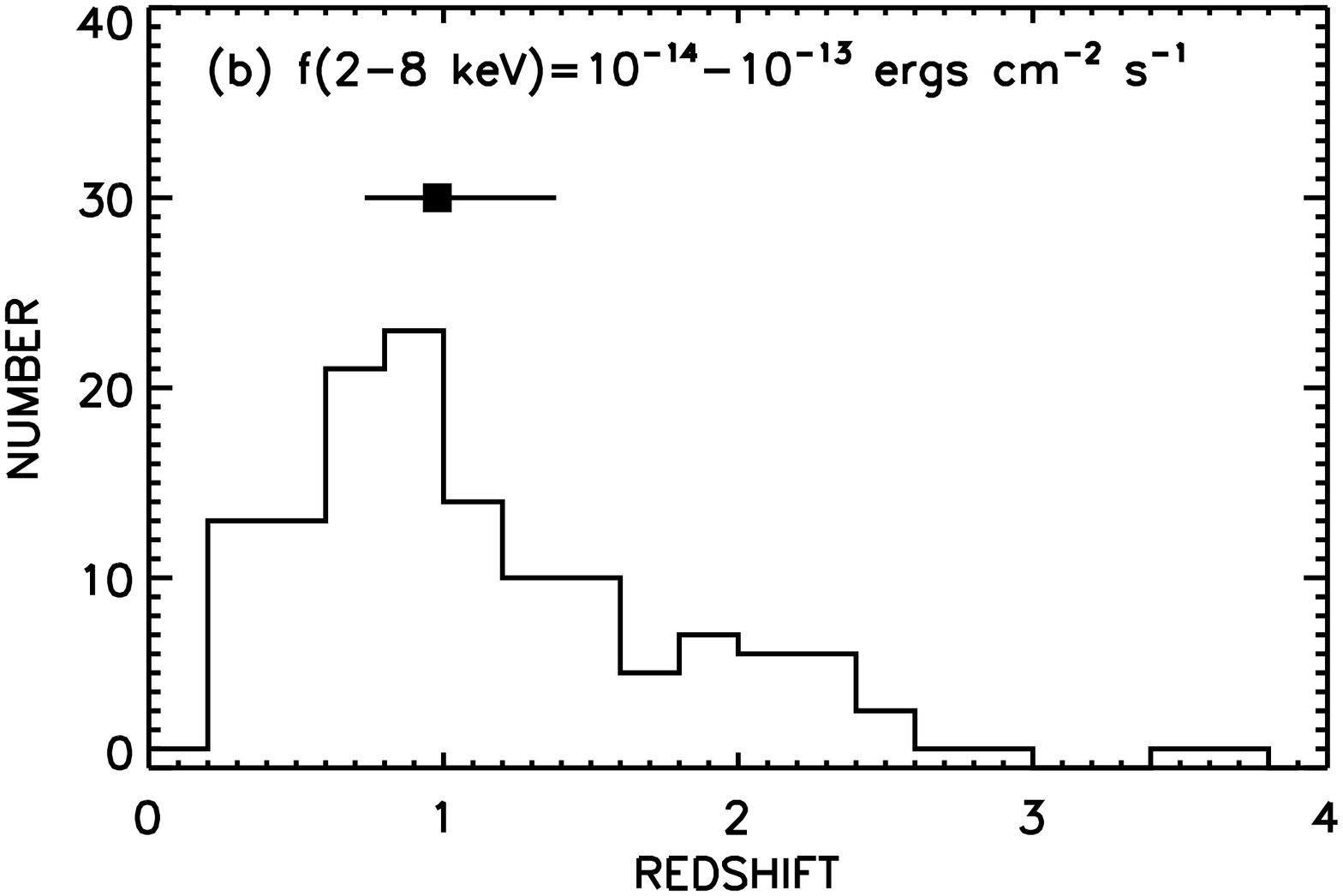,width=3.5in,angle=90}}
\centerline{\psfig{figure=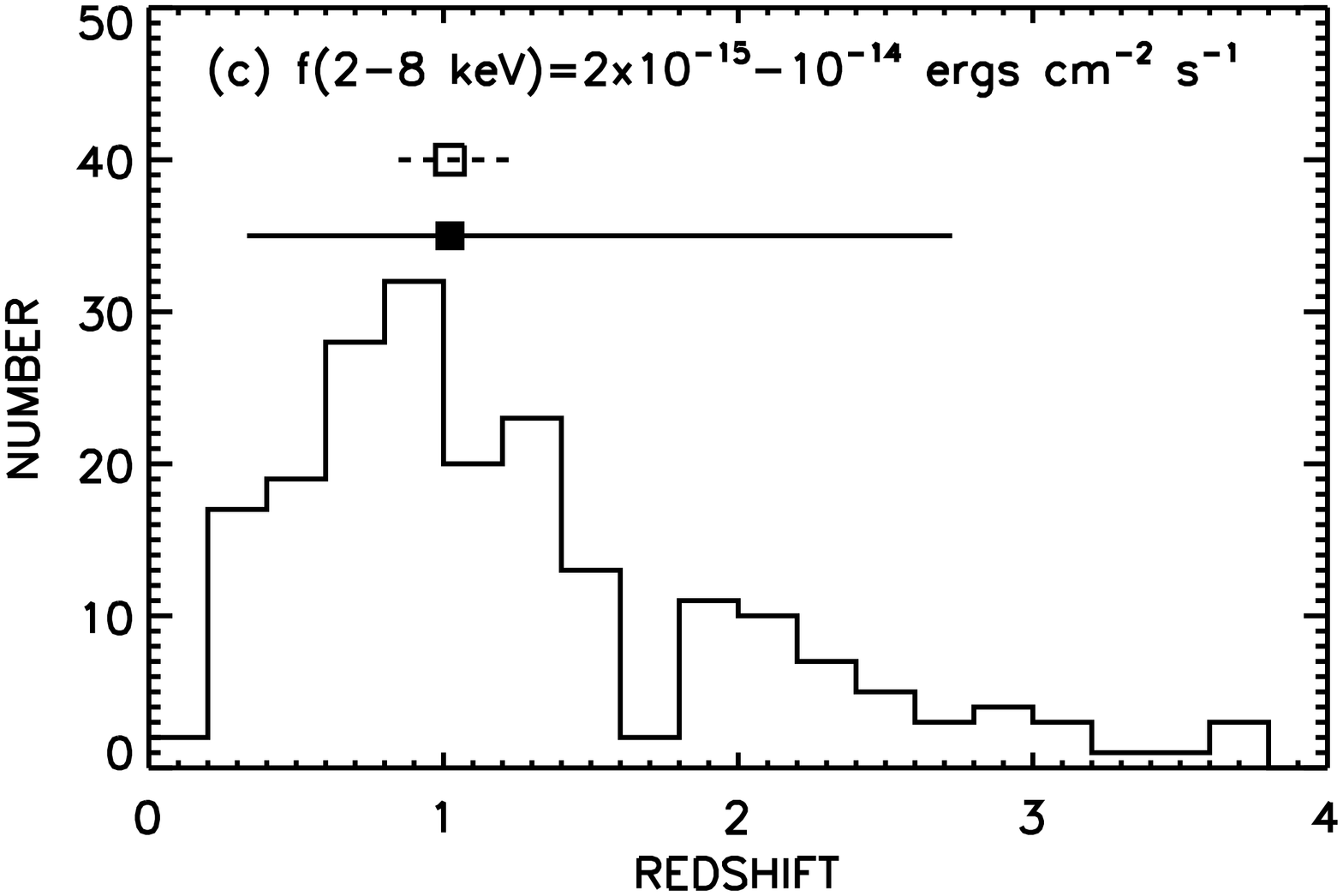,width=3.5in,angle=90}}
\figurenum{9}
\caption{
Spectroscopic redshift distributions for three
flux intervals using the total hard X-ray sample.
Median redshifts {\em (solid squares)\/}
with $1\sigma$ median redshift ranges {\em (solid bars)\/}
are also shown. The $1\sigma$ upper and lower limits on the median
redshifts
were determined by placing all of the unidentified sources at
arbitrarily high and low redshifts, respectively.
For the lowest flux bin, we also show the median redshift
{\em (open square)\/} and $1\sigma$ range {\em (dashed bar)\/}
when we use only the CDF-N and CDF-S data and
include the photometric redshifts in
the computation, placing sources with neither
spectroscopic nor photometric redshifts
at arbitrarily high and low redshifts, respectively.
\label{figmedz}
}
\end{figure*}

We are now in a position to be able to compare
the median redshifts of the total hard and soft X-ray samples
at a range of fluxes. In Figure~\ref{figmedz}, we show 
spectroscopic redshift distributions for three flux
intervals in the total hard X-ray sample. We also show
the median redshifts {\em (solid squares)\/} and $1\sigma$ 
median redshift ranges {\em (solid bars)}. The $1\sigma$ 
upper and lower limits on the median redshifts were determined 
by placing all of the spectroscopically unidentified sources at 
arbitrarily high and low redshifts, 
respectively. At the brighter X-ray
fluxes, where most of the sources have been spectroscopically
identified, the median redshifts are fairly well determined,
while at the fainter X-ray fluxes, where the spectroscopic
incompleteness is more substantial, the median redshifts are
less well determined. In Figure~\ref{figmedz}c, we also show
the median redshift {\em (open square)\/} and $1\sigma$ median 
redshift range {\em (dotted bars)\/} when we use only
the CDF-N and CDF-S data and include the photometric redshifts in 
the computation, allowing for sources with neither spectroscopic
nor photometric redshifts 
in the same way as above. This substantially improves the accuracy 
of the median redshift determination in the lowest X-ray flux bin.

Steffen et al.\ (2004) determined median
redshifts for the surveys used in the present paper combined 
with other {\em ROSAT\/} (Lehmann et al.\ 2001) 
and {\em XMM-Newton\/} (Mainieri et al.\ 2002;
Fiore et al.\ 2003) surveys. Their
redshift versus $0.5-2$~keV and $2-8$~keV 
flux diagrams (their Figures 10a and 10b) showed that the 
median redshifts of the surveys remained about constant at $z\sim 1$
and were not strongly correlated with either the soft or hard
X-ray fluxes. They cautioned, however, that since
the magnitudes of high-redshift sources may be faint,
the spectroscopic surveys may be biased against finding
high-redshift sources, particularly at the fainter X-ray fluxes
where the surveys are more incomplete.

\begin{figure*}
\centerline{\psfig{figure=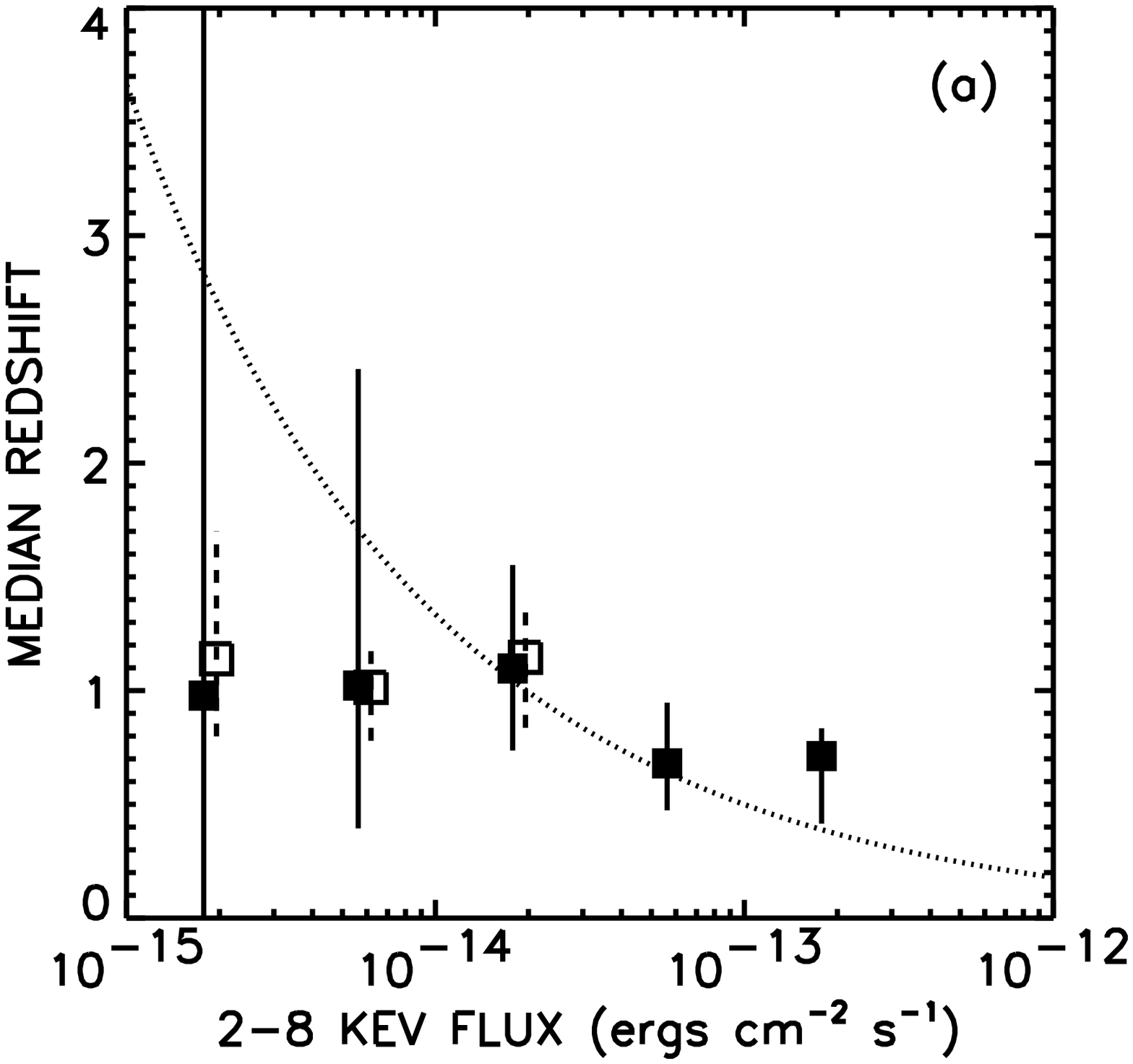,width=3.5in,angle=90}
\psfig{figure=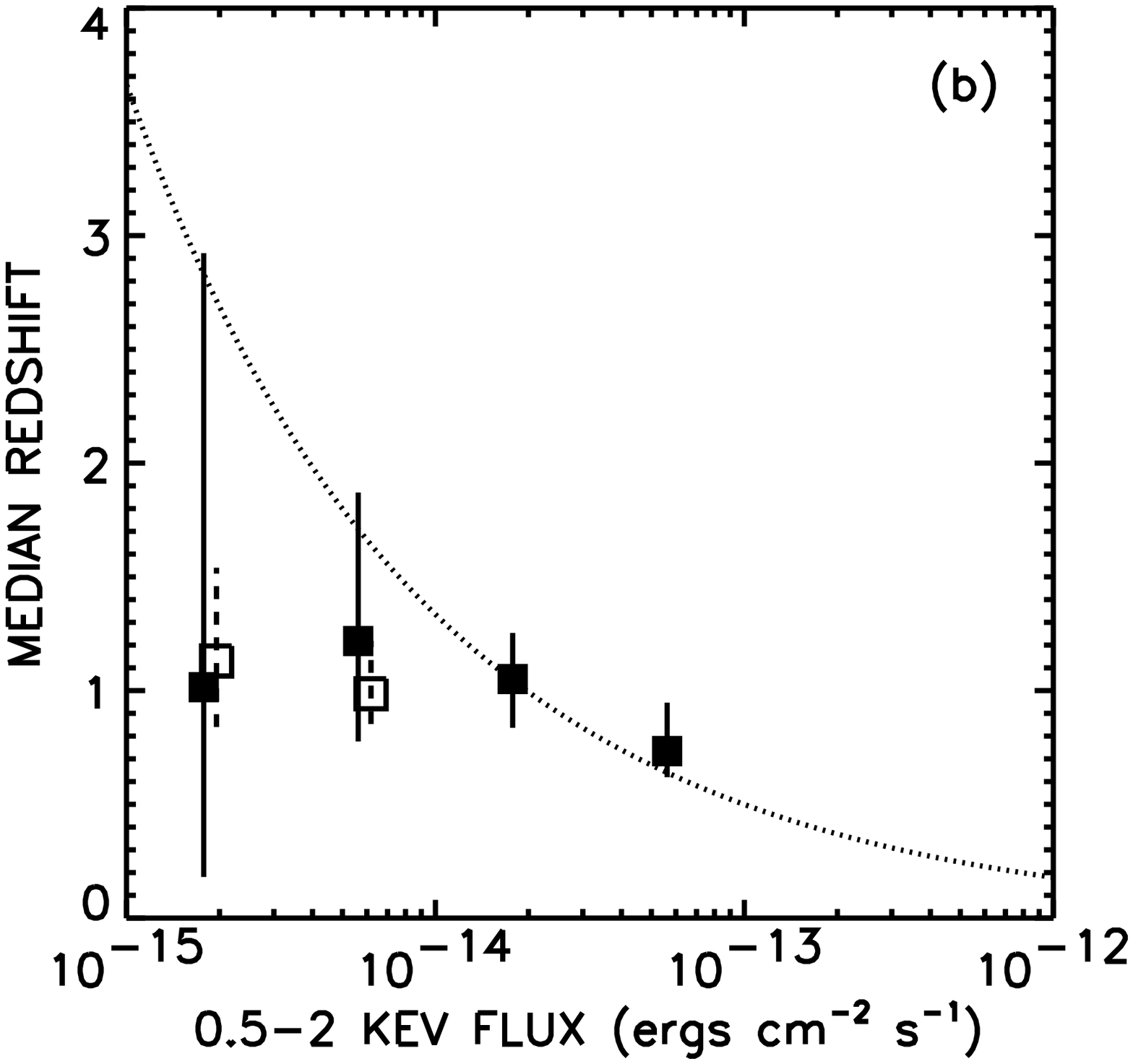,width=3.5in,angle=90}}
\figurenum{10}
\caption{
Median redshifts {\em (solid squares)\/}
with $1\sigma$ median redshift ranges {\em (solid bars)\/}
vs. (a) $2-8$~keV and (b) $0.5-2$~keV flux. The $1\sigma$ upper
and lower limits on the median redshifts were determined
by placing all of the unidentified sources at
arbitrarily high and low redshifts, respectively.
For the lower X-ray fluxes, we also show the median redshifts
{\em (open squares)\/} and $1\sigma$ ranges {\em (dashed bars)\/}
when we use only the CDF-N and CDF-S data and include the
photometric redshifts in the computation, placing sources
with neither spectroscopic nor photometric redshifts at
arbitrarily high and low redshifts, respectively.
These have been slightly offset in flux to distinguish
them from the spectroscopic points. Dotted curves show the
standard redshift-luminosity relation for a source with
rest-frame $2-8$~keV luminosity $L_X=10^{44}$~ergs~s$^{-1}$.
\label{figfxmedz}
}
\end{figure*}

We examine this issue in more detail in Figures~\ref{figfxmedz}a 
and \ref{figfxmedz}b, where we show the spectroscopic median 
redshifts {\em (solid squares)\/} and $1\sigma$ median redshift 
ranges {\em (solid bars)\/}, calculated as in Figure~\ref{figmedz},
versus $2-8$~keV and $0.5-2$ X-ray flux, respectively, for the
total hard and soft X-ray samples. At the lower X-ray fluxes, we 
also show the median redshifts {\em (open squares)\/} 
and $1\sigma$ ranges {\em (dotted bars)\/} for the spectroscopic
plus photometric CDF-N and CDF-S data. Again we see that the median 
redshifts of the hard and soft X-ray samples are very similar.
Moreover, with the tighter spectroscopic plus photometric median 
redshift ranges, we see that the median 
redshifts are indeed fairly constant with flux. The dotted curves show 
the standard redshift-luminosity relation for a source with 
rest-frame $2-8$~keV luminosity $L_X=10^{44}$~ergs~s$^{-1}$. The
lower X-ray flux sources are clearly dominated by sources with
luminosities less than this value and redshifts near one. 
We shall return to this point when we compute the hard X-ray 
luminosity functions in \S\ref{sechxlf}.

%
%

%
%

\section{Optical Spectral Classifications}
\label{secclass}

The optical spectra of the X-ray sources in our sample, 
while generally of high quality, span different
rest-frame wavelengths and suffer varying degrees of AGN
mixing with the host galaxy spectrum at different redshifts. 
It would therefore be quite hard to classify the X-ray sources 
in any uniform way using a conventional AGN classification scheme.
Recognizing this, we have instead only roughly classified the 
spectroscopically identified {\em Chandra\/} sources into four 
optical spectral classes. We call sources without any strong emission lines 
(EW([OII])$<3$~\AA\ or EW(H$\alpha+$NII)$<10$~\AA) {\em absorbers\/};
sources with strong Balmer lines and no broad or high-ionization
lines {\em star formers\/}; sources with [NeV] or CIV lines or strong
[OIII] (EW([OIII]~5007~\AA$)>3$~EW(H$\beta)$) {\em high-excitation (HEX) 
sources\/}; and, finally, sources with optical lines having FWHM line 
widths $>2000$~km~s$^{-1}$ {\em broad-line AGNs\/}. We have chosen 
these four classes to roughly match those used by Szokoly et al.\ (2004) 
for the CDF-S (they call the second category {\em low-excitation {\rm or} 
LEX\/} sources) in order to combine our own classifications of the 
CDF-N and CLASXS sources with their classifications of the
CDF-S sources. In this paper, we will sometimes combine the 
absorber and the star former classes into a {\em normal galaxy\/} 
class.

Table~\ref{tab1} gives the breakdown of the CDF-N, CDF-S, and CLASXS 
samples by spectral type. Note that four of the CDF-N redshifts
are from the literature (see Barger et al.\ 2003b for references)
and hence do not have spectral typings.
Hereafter, we call all of the sources that 
do not show broad-line (FWHM$>2000$~km~s$^{-1}$) signatures 
``optically-narrow'' AGNs. However, we note that there may be a few 
sources where our wavelength coverage is such that we are missing 
lines which would result in us defining the spectrum as broad-line.

For the CDF-N and CLASXS fields, we made a more quantitative
analysis by measuring the FWHM line widths 
for each of the CIV~1550~\AA, [CIII]~1909~\AA, MgII~2800~\AA, 
H$\beta$, and H$\alpha$ lines that were in the
observed spectra by fitting Gaussian profiles to the lines. 
Where more than one of these lines was within the spectrum, we 
took the maximum of the measured FWHMs to be the FWHM.
There were only significant differences in the measured widths 
of the various lines for a small number of cases where 
H$\beta$ was narrow and MgII~2800~\AA\ was wide. For a small number
of other spectra, none of the lines were in the observed range, and
hence we were not able to classify them. For the cases where the 
wavelengths of the lines were within our spectra, but no emission 
lines were visible, we set the widths to zero. 
Figure~\ref{figclwid} shows line width versus optical spectral class 
for the CDF-N and CLASXS sources.

The choice of 2000~km~s$^{-1}$ as the dividing line between
the high-excitation sources and the broad-line AGNs is
in itself rather arbitrary, and many of the sources lying
in the high-excitation category would rather naturally fall into the
narrow-line Seyfert~1 galaxy definition
(Osterbrock \& Pogge 1985;
Goodrich 1989). To investigate this further,
we ran the same classification scheme on the large optical
spectroscopic sample given in Cowie et al.\ (2004b)
and Wirth et al.\ (2004) for the ACS GOODS-North region of the 
CDF-N field. Figure~\ref{field_widths}
shows soft X-ray flux versus FWHM line width for these sources.
Of the 1718 sources in the GOODS region where we could measure
line widths, 20 had FWHM$>800$~km~s$^{-1}$, and 13 had
FWHM$>2000$~km~s$^{-1}$. All of these sources are X-ray sources. 

%
%
\begin{inlinefigure}
\psfig{figure=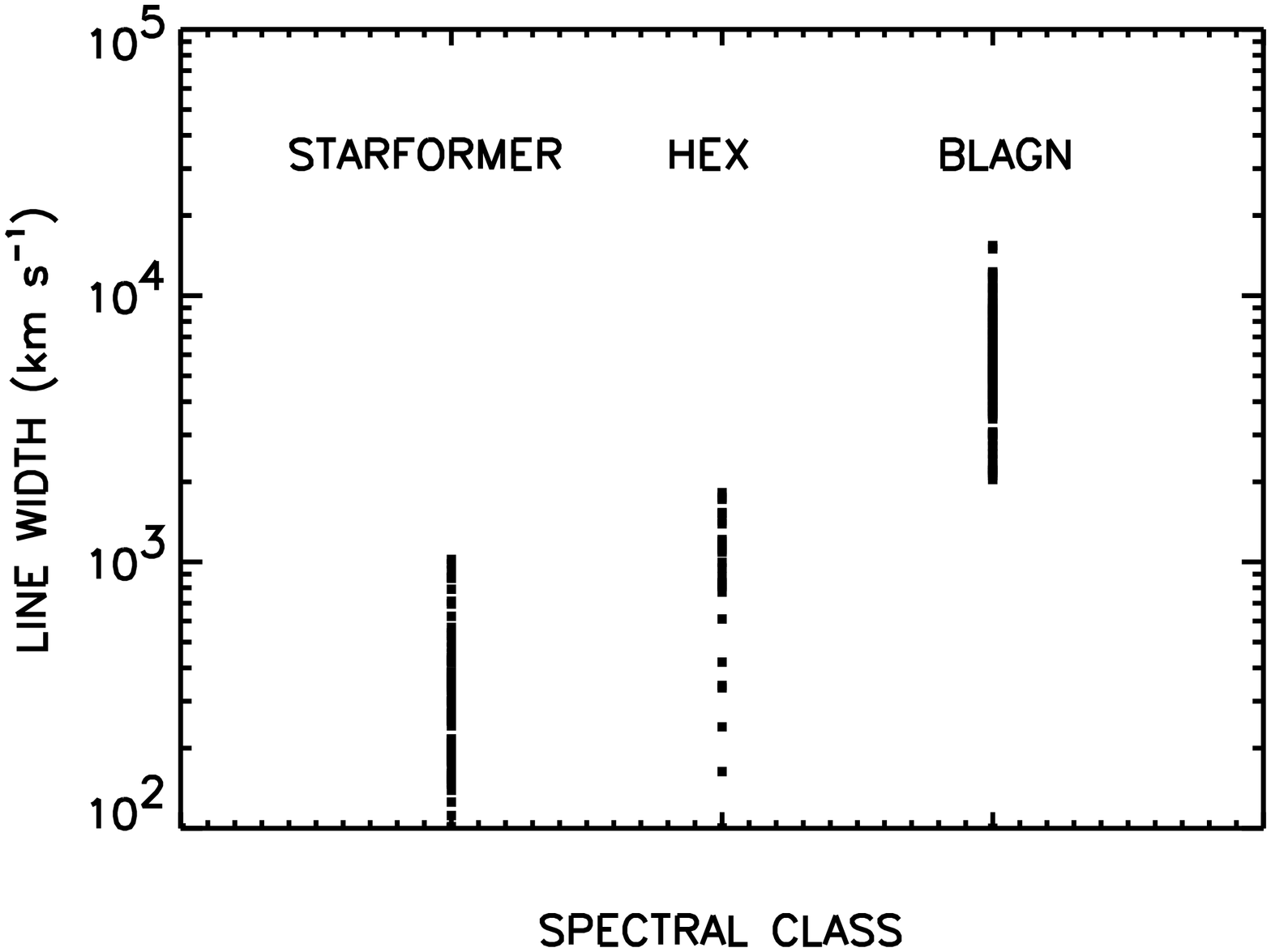,width=8.5cm,angle=90}
\figurenum{11}
\caption{
FWHM line width vs. optical spectral class for the CDF-N and CLASXS
sources. By definition, the broad-line AGNs lie above 2000~km~s$^{-1}$.
\label{figclwid}
}
\addtolength{\baselineskip}{10pt}
\end{inlinefigure}

We may make two points from this. First, X-ray--selected
samples at the flux limit of the CDF-N essentially find
all of the broad-line AGNs. However, some of the intermediate width
($800-2000$~km~s$^{-1}$) sources would not be found in a field 
with the CLASXS depth, though they are well above the on-axis 
CDF-N flux limit. Second, a split at 800~km~s$^{-1}$ might seem, 
in some ways, to be more objective than one at 2000~km~s$^{-1}$.

%
%
\begin{inlinefigure}
\psfig{figure=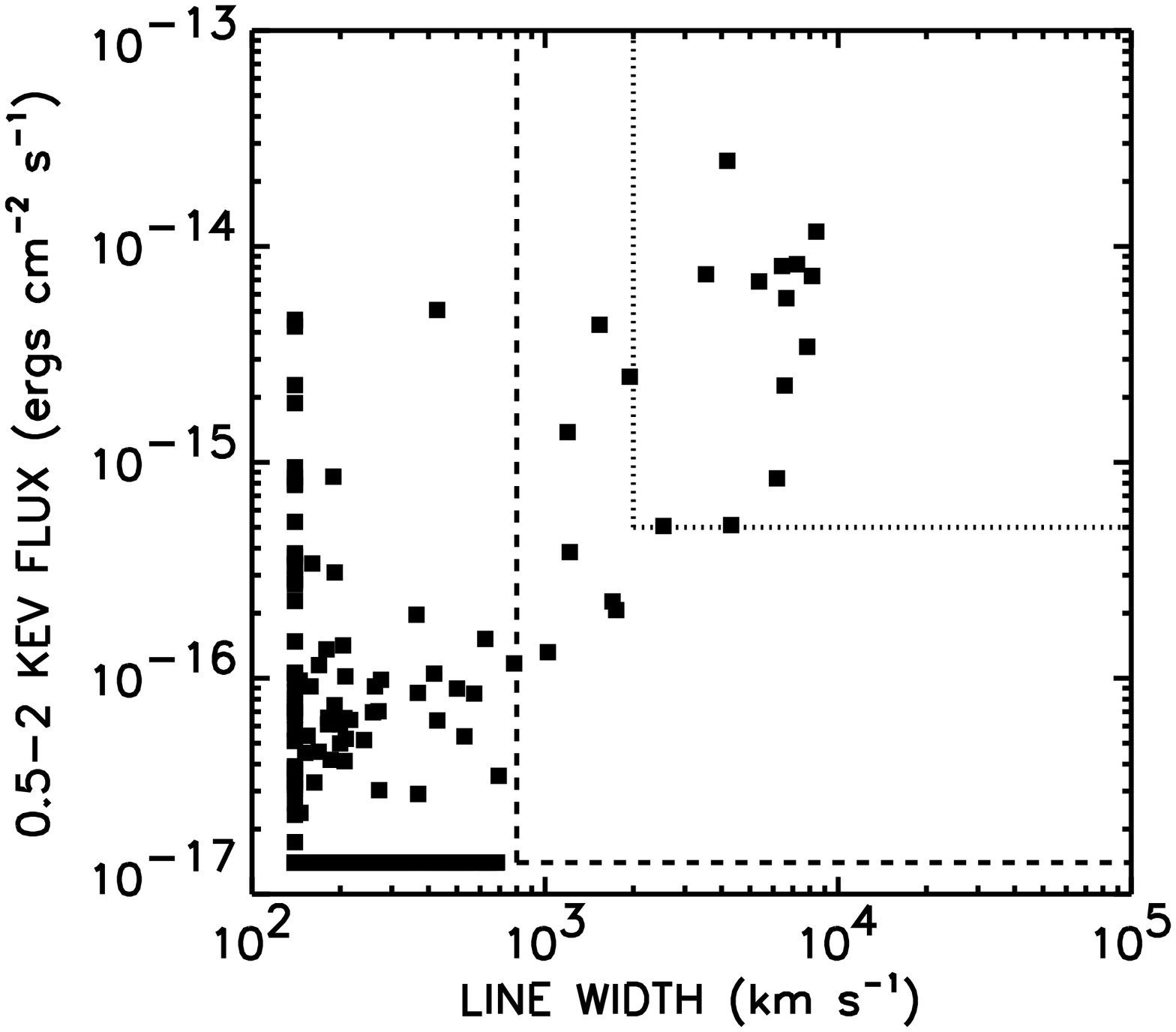,width=8.5cm,angle=90}
\figurenum{12}
\caption{
$0.5-2$~keV flux vs. FWHM line width for a large,
optically-selected sample in the CDF-N {\em (squares)\/}.
Dashed lines show a FWHM of 800~km~s$^{-1}$ {\em (vertical)\/}
and the CDF-N $0.5-2$~keV flux limit {\em (horizontal)\/}.
Dotted lines show the minimum FWHM of 2000~km~s$^{-1}$
for a broad-line source {\em (vertical)\/} and
the CLASXS $0.5-2$~keV flux limit {\em (horizontal)\/}.
\label{field_widths}
}
\addtolength{\baselineskip}{10pt}
\end{inlinefigure}

However, when we turn to the X-rays, it is clear that the
primary distinction is between the broad-line AGNs and the 
optically-narrow AGNs. Indeed, this is the only distinction 
that can easily be made from the X-ray colors or X-ray spectra 
(Szokoly et al.\ 2004). In Figure~\ref{softwid},
we show the $0.5-2$~keV to $2-8$~keV flux ratio for
the sample of sources with strong $2-8$~keV fluxes in the
CDF-N and CLASXS fields versus FWHM line width. 
Above 2000~km~s$^{-1}$, 
almost all of the sources are soft ($\Gamma=1.8$), while below 
this line width, there is a wide span of X-ray colors. 

The intermediate width sources show a very similar spread in their
X-ray colors relative to all of the remaining optically-narrow 
sources. This result is consistent with recent work by
Williams, Mathur, \& Pogge (2004), who
used {\em Chandra\/} observations of a sample of 
optically-selected, X-ray--weak, narrow-line Seyfert~1 galaxies 
to show that strong, ultrasoft X-ray emission is not a universal
characteristic of narrow-line Seyfert~1 galaxies, and, indeed, 
that many narrow-line Seyfert~1 galaxies have weak or hard 
X-ray emission.

%
%
\begin{inlinefigure}
\psfig{figure=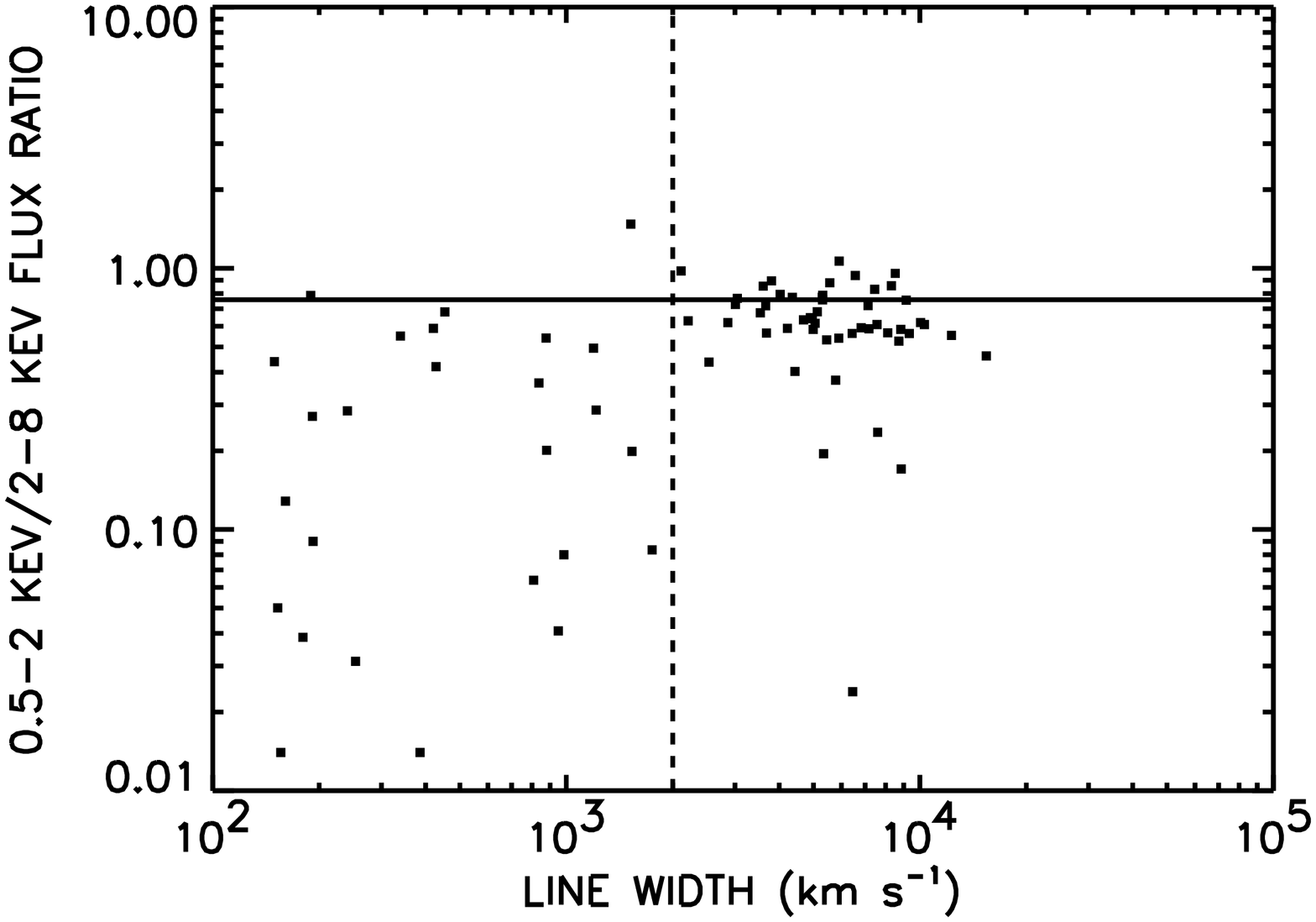,width=8.5cm,angle=90}
\figurenum{13}
\caption{
$0.5-2$~keV to $2-8$~keV flux ratio vs. FWHM for
the sample of sources with
$f_{2-8~{\rm keV}}>10^{-15}$~ergs~cm$^{-2}$~s$^{-1}$
that lie within 10 arcminutes of the CDF-N aim point, and for
the sample of sources with
$f_{2-8~{\rm keV}}>2\times 10^{-14}$~ergs~cm$^{-2}$~s$^{-1}$
in CLASXS. Horizontal line corresponds to $\Gamma=1.8$.
Vertical line corresponds to the mimimum FWHM of
2000~km~s$^{-1}$ for a broad-line source.
\label{softwid}
}
\addtolength{\baselineskip}{10pt}
\end{inlinefigure}

For the CDF-N sources, which have the deepest exposures, we can 
perform a more sophisticated color 
analysis. In Figure~\ref{figcol}, we show the color-color plot of
the $2-4$~keV to $4-8$~keV counts ratio versus the $0.5-2$~keV to
$2-8$~keV counts ratio for the $4-8$~keV {\em Chandra\/} sources that
have more than 100 counts in that band, separated by optical
spectral type (broad-line AGNs in Figure~\ref{figcol}a; optically-narrow AGNs
in Figure~\ref{figcol}b).
The broad-line AGNs are nearly all soft with mean photon indices
$\Gamma=1.8$ and essentially no visible absorption in X-rays.
By contrast, the optically-narrow AGNs are well described by a
simple model in which a power-law spectrum with $\Gamma=1.8$
is suppressed at low energies by photoelectric absorption spread
over a very wide range of absorbing column densities
(Barger et al.\ 2002;
Alexander et al.\ 2003b).
There is little dependence of the absorbing column densities
on optical spectral type (shown by the different symbols in
Figure~\ref{figcol}b) or on the line widths of the optically-narrow AGNs. 
Thus, the presence of optical Seyfert~2 or narrow-line Seyfert~1 
galaxy characteristics does not seem to be dependent on the absorbing 
column density to the X-ray source. 

From Figures~\ref{softwid} and \ref{figcol}, we see that it
is possible to separate roughly the broad-line AGNs and the
optically-narrow AGNs on the basis of the X-ray colors alone
(e.g., Szokoly et al.\ 2004),
without knowing the optical spectra.
However, there will be a small amount of contamination from stars,
from the small number of optically-narrow AGNs that have
soft X-ray colors, and from the small number of broad-line AGNs that
have hard X-ray colors. (Note that 
Perola et al.\ 2004 found that about 10\% of 
their broad-line AGNs showed some X-ray obscuration, with estimated
$N_H>10^{22}$~cm$^{-2}$.)

%
%
\begin{inlinefigure}
\psfig{figure=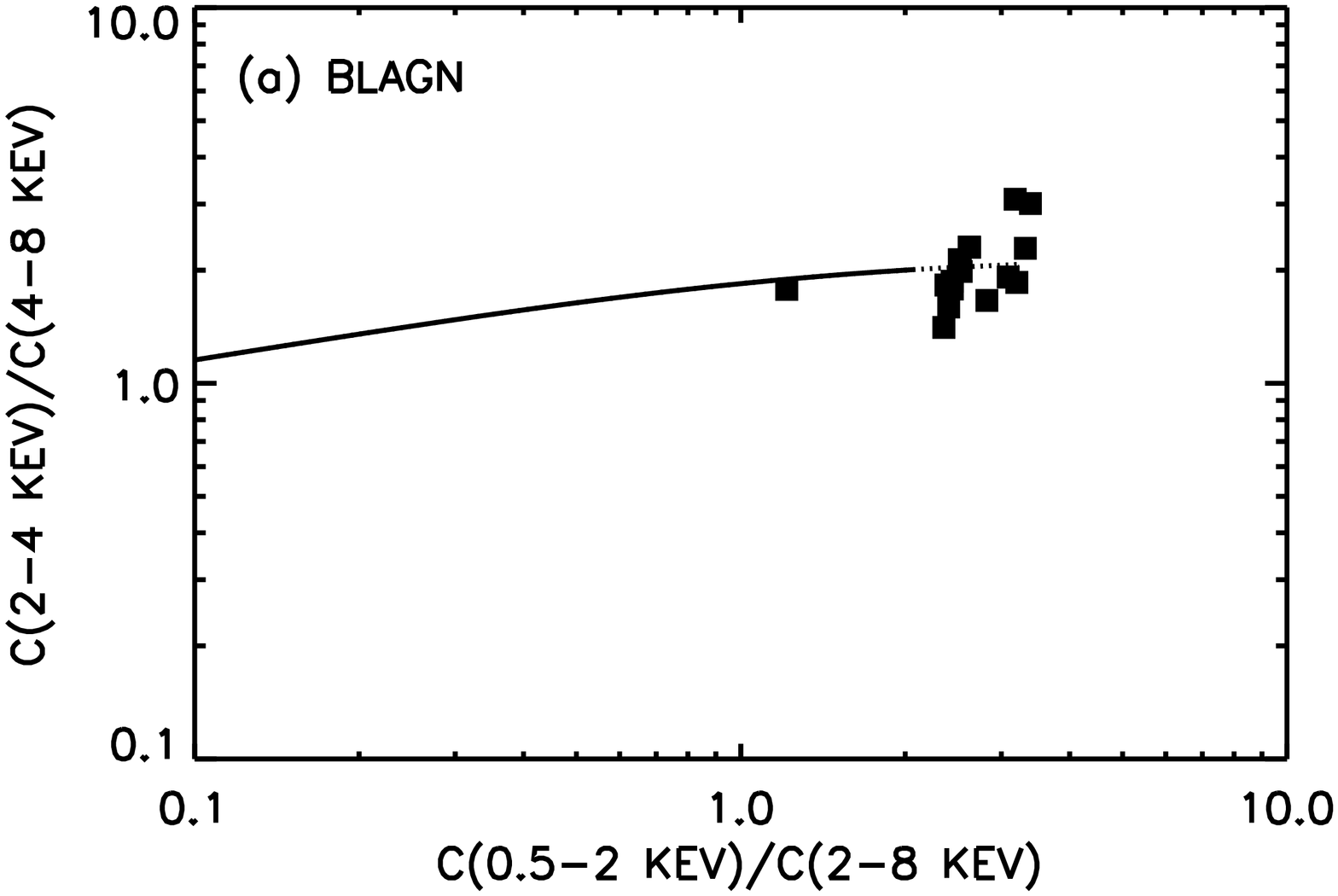,width=8.5cm,angle=90}
\psfig{figure=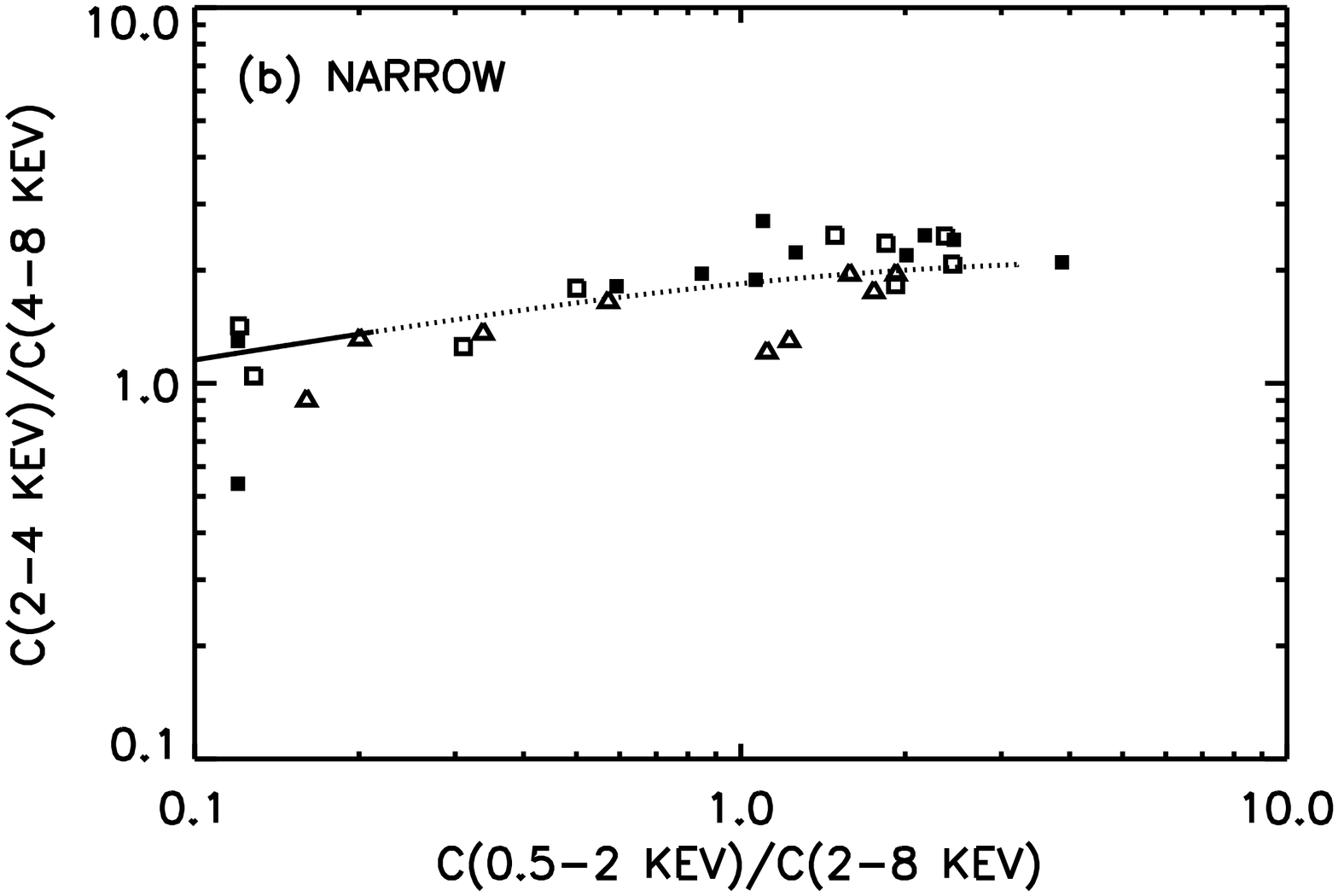,width=8.5cm,angle=90}
\figurenum{14}
\caption{
Color-color plot of the $2-4$~keV to $4-8$~keV counts ratio
vs. the $0.5-2$~keV to $2-8$~keV counts ratio for the $4-8$~keV
sources in the CDF-N 2~Ms exposure with more than 100 counts in the
$4-8$~keV band. Curves show a $\Gamma=1.8$ source with absorption from
varying column densities. (a) Sources with broad-line optical spectra
(``BLAGN'') {\em (large squares)\/} are closely grouped. All but one is
consistent with $\Gamma=1.8$ and an effective absorbing column
density below $2\times10^{21}$~cm$^{-2}$ {\em (curve is solid
above this column density and dotted below)}.
(b) Sources without broad-line optical spectra (``NARROW'')
({\em solid squares\/}---high-excitation signatures;
{\em open squares\/}---absorbers and star formers;
{\em triangles\/}---unidentified sources)
have a wide range of effective column densities, even up to
values above $3\times10^{22}$~cm$^{-2}$ {\em (curve is solid
above this column density and dotted below)}.
\label{figcol}
}
\addtolength{\baselineskip}{10pt}
\end{inlinefigure}

\section{Nuclear UV/Optical Magnitudes}
\label{secnuc}

To study the nuclear UV/optical properties of the $2-8$~keV--selected
sources, we use multicolor observations of the CDF-N taken
with the ACS camera on {\em HST\/} in
four bands (ACS F435W, F606W, F814W, and F850LP) as part
of GOODS (Giavalisco et al.\ 2004).
(Grogin et al.\ 2003 used earlier {\em HST\/}
observations of the CDF-S for this type of analysis.) Because of
the high spatial resolution of these data, we can separate the
nuclear component of each source from the host galaxy light, even
at the higher redshifts, in order to analyze the nuclear colors.
Thus, {\em HST\/} makes it possible for us to reproduce the
types of analyses that have been done for decades on low-redshift 
AGNs and optically-bright AGNs. 

\vskip 10cm

%
%
\begin{inlinefigure}
\figurenum{15}
\caption{
Thumbnail images of sources in the ACS GOODS-North region of the
CDF-N at three redshifts:
{\em (top panel)\/} $z=0.5$ (F485W),
{\em (middle panel)\/} $z=1.0$ (F606W), and
{\em (bottom panel)\/} $z=1.5$ (F814W).
Each of the three panels contains three separate rows of images:
{\em (top row)\/} nucleated, {\em (middle row)\/}
somewhat nucleated, {\em (bottom row)\/} no strong nucleus.
The sources with spectroscopic typings of broad-line AGN (``BROAD'')
or high-excitation (``HEX'') have been labeled as such.
\label{figimages}
}
\addtolength{\baselineskip}{10pt}
\end{inlinefigure}

Since the galaxies often have complex morphologies
(see Figure~\ref{figimages} for examples), we decided
to use a very simple prescription to separate the nuclear magnitudes
from the galaxy light. For each source, we first located the peak
of the optical light within a $2''$ radius around the X-ray position.
We then measured the magnitude in a $0.15''$ aperture
radius centered on this optical position. Based on measurements of
196 spectroscopically confirmed stars in the field, we found
that this radius corresponded to the 80 percent enclosed energy fraction.
(This is also consistent with the enclosed energy curves in the ACS handbook.)
Thus, the $0.15''$ aperture magnitude corrected for this enclosed
energy fraction provides a first estimate of the nuclear magnitude.

In order to provide an improved estimate of the nuclear magnitude
with the galaxy contribution removed, we followed an iterative
procedure. We first measured the light
within a $0.15''-0.3''$ annulus around the peak position of the
optical light. We then subtracted from this measurement the expected
contribution from
the nucleus (10\% of the $0.15''$ radius light, as determined from
the PSF) in order to leave only the galaxy contribution.
Based on measurements of objects without strong nuclei that otherwise
appear to have similar spectral characteristics,
we found that the value of the galaxy light in the central $0.15''$
radius relative to that in the $0.15''-0.3''$ annulus is $\sim 0.87$.
(We note that this number is rather rough and does not take into account
redshift effects.) Thus, we subtracted this estimate of the galaxy
light from the $0.15''$ aperture magnitude to obtain a revised estimate
of the nuclear magnitude. The whole process was then repeated until
the estimates converged.

%
%
\begin{inlinefigure}
\psfig{figure=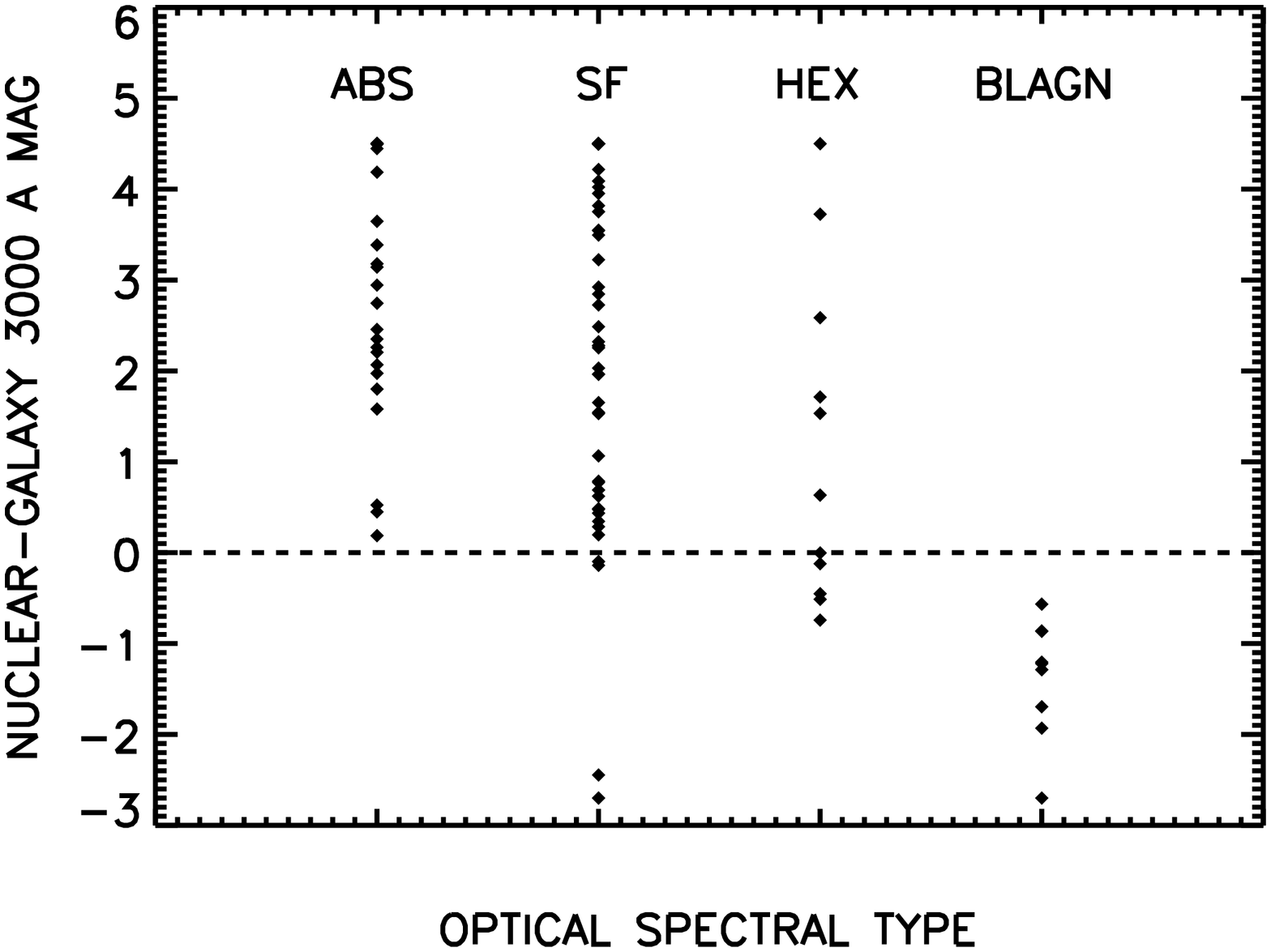,width=8.5cm,angle=90}
\figurenum{16}
\caption{
Rest-frame (nuclear -- galaxy) 3000~\AA\ magnitude
(computed by interpolating between the GOODS bandpasses) vs.
optical spectral type for sources in the ACS GOODS-North region of
the CDF-N that lie in the redshift range $z=0.5-2$ and have
$L_X\ge 10^{42}$~ergs~s$^{-1}$. For each of the four spectral classes,
we show the difference between the nuclear magnitude and the galaxy
magnitude of each source in that class {\em (solid diamonds)}.
Sources with (nuclear -- galaxy) magnitudes larger than 4.5 or less
than $-4.5$ are plotted at these values.
The four spectral classes are labeled ``ABS'' (absorbers), ``SF''
(star formers), ``HEX'' (high-excitation sources), and ``BLAGN''
(broad-line AGNs). All of the broad-line AGNs are strongly
nucleated, while almost none of the absorbers or star formers are.
Most of the high-excitation sources have comparable nuclear and
galaxy magnitudes.
\label{nuc_gal_type}
}
\addtolength{\baselineskip}{10pt}
\end{inlinefigure}

The division between the nucleated
sources and those without strong nuclei is generally quite
clear. We have illustrated this for three redshifts in
Figure~\ref{figimages}. Some objects (such as those with double
nuclei) can be problematic, but rather than correct these by hand,
we have chosen to stay with the objectively measured values.
We ran the above procedures on all four color bands in the ACS
sample, determining nuclear magnitudes for all 286 objects in the
CDF-N 2~Ms list that lie within the GOODS-North region. The galaxy
magnitudes are then just the isophotal ACS GOODS-North magnitudes
minus the nuclear magnitudes.

There is a good correlation between the fraction of nuclear UV
light in the galaxy and the optical spectral type, which gives us
confidence in the procedure, since these are completely independently
measured properties. In Figure~\ref{nuc_gal_type}, we show nuclear
UV magnitude minus galaxy magnitude for all of the X-ray sources
versus optical spectral type. The broad-line AGNs and the stars are all
strongly nucleated, while the absorbers and star formers have much
weaker nuclei, and the high-excitation sources lie at intermediate values.

Of the 286 sources measured, 130 have no detectable nucleus in
the $B$-band, and 126 have no detectable nucleus in the $z'$-band.
Only 4 of the sources without nuclei are spectrally classified 
as high-excitation sources, and none are classified as broad-line AGNs.

\section{Low-Redshift Hard X-ray Luminosity Functions}
\label{sechxlf}

With the advent of the {\em Chandra\/} and {\em XMM-Newton\/} data, 
there have been several computations of the evolution of the 
hard X-ray luminosity functions with redshift (e.g., Cowie et al.\ 2003;
Hasinger 2003;
Steffen et al.\ 2003;
Ueda et al.\ 2003;
Fiore et al.\ 2003).
It is a measure of how rapidly the science is evolving that
the {\em Chandra\/} spectroscopic samples are now much improved over 
those used in even these recent calculations. The {\em Chandra\/}
samples of Table~\ref{tab1} now contain 1165 spectroscopically 
observed sources and 715 spectroscopically identified sources. 

In calculating the hard X-ray luminosity functions, we use only 
the spectroscopically observed sources. This approach is strictly 
valid for CLASXS and the CDF-S, where the X-ray sources were 
randomly observed. For the CDF-N, there may be a small bias since
the spectroscopy does not come purely from targeted observations of 
the X-ray sample. However, as there are now only 52 X-ray sources 
in the CDF-N that have not been observed, it would make very little 
difference to the results if we were instead to use all of the 
sources.

There are a total of 698 spectroscopically observed sources
in the CDF-N and CDF-S samples. Of these, 601 have 
either spectroscopic or photometric redshifts. We can test for
the effects of incompleteness in our analysis by using the 
spectroscopic plus photometric redshifts in these two fields and 
the spectroscopic redshifts for the {\em ASCA\/} data, where only 
one source is unidentified.

Here we recompute the hard X-ray luminosity functions following
Cowie et al.\ (2003), who used the
traditional $1/V_a$ method of Felten (1977). 
We use the spectroscopically
observed samples of Table~\ref{tab1}. We define the hard X-ray 
luminosity function
versus rest-frame $2-8$~keV X-ray luminosity and redshift,
$d\Phi(L_X, z)/{d\log L_X}$, as the number of X-ray sources
per unit comoving volume per unit base 10 logarithmic luminosity
that lie in the redshift interval. 
We determine the solid angle covered by the observed sources at 
a given flux by comparing the observed number of sources versus
flux with the averaged number counts in the appropriate energy
band from Cowie et al.\ (2002) and 
Yang et al.\ (2004).
This method allows a simple treatment of the incompleteness
that was modeled in computing the counts. However, the counts
in Cowie et al.\ also include the low CDF-S counts, which may
affect the normalization at the 10\% level relative to the
CLASXS average (Yang et al.\ 2004). We consider
this to be a reasonable estimate of the systematic errors.
At $10^{-14}$~ergs~cm$^{-2}$~s$^{-1}$ ($2-8$~keV), where the
contribution to the $2-8$~keV X-ray background peaks (see \S\ref{seccomplete}), 
the solid angle is dominantly from CLASXS.
In Figure~\ref{figarea}, we show solid angle versus (a) $2-8$~keV 
and (b) $0.5-2$~keV flux for the observed hard and soft X-ray 
samples, respectively.

%
%
\begin{inlinefigure}
\psfig{figure=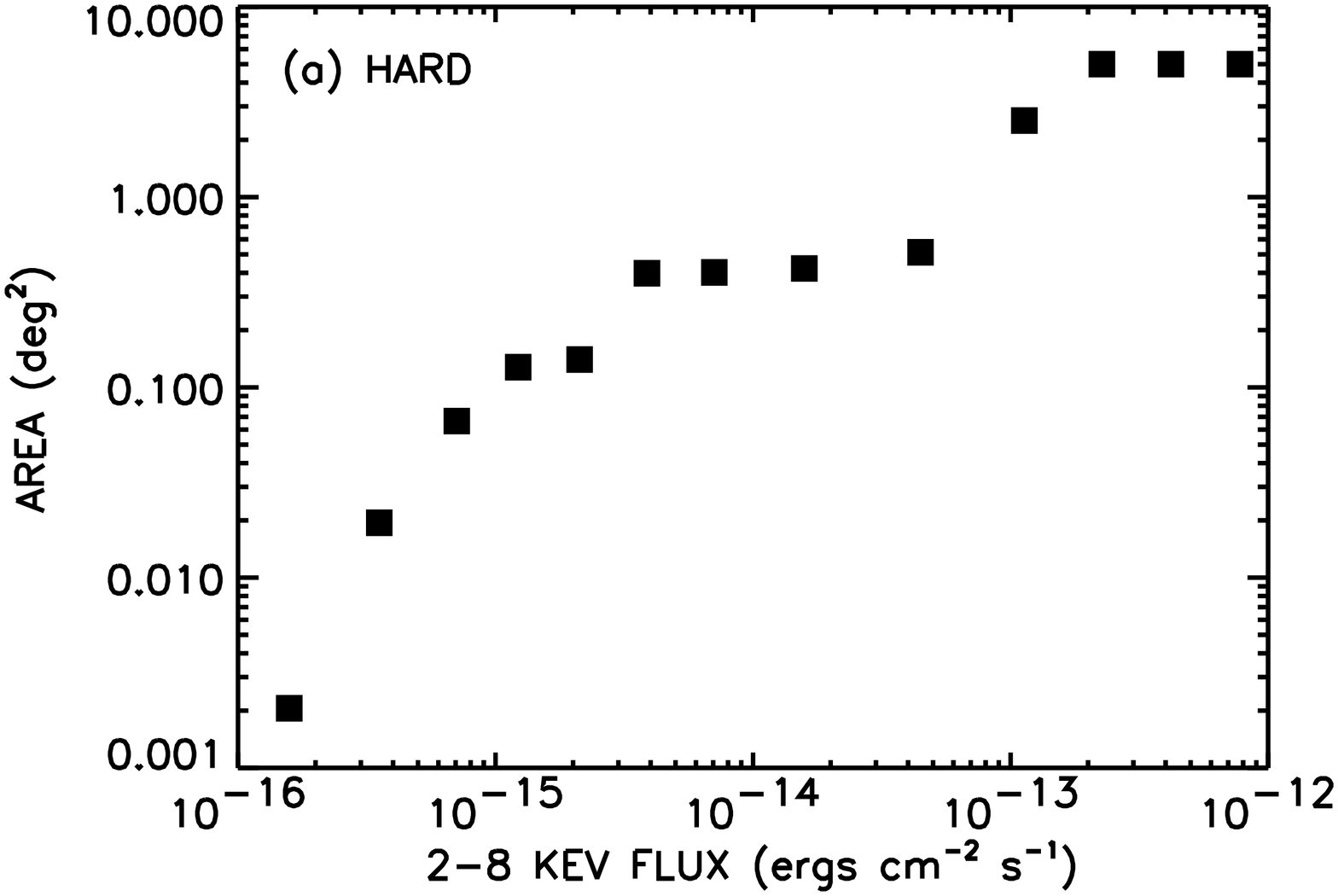,width=3.5in,angle=90}
\psfig{figure=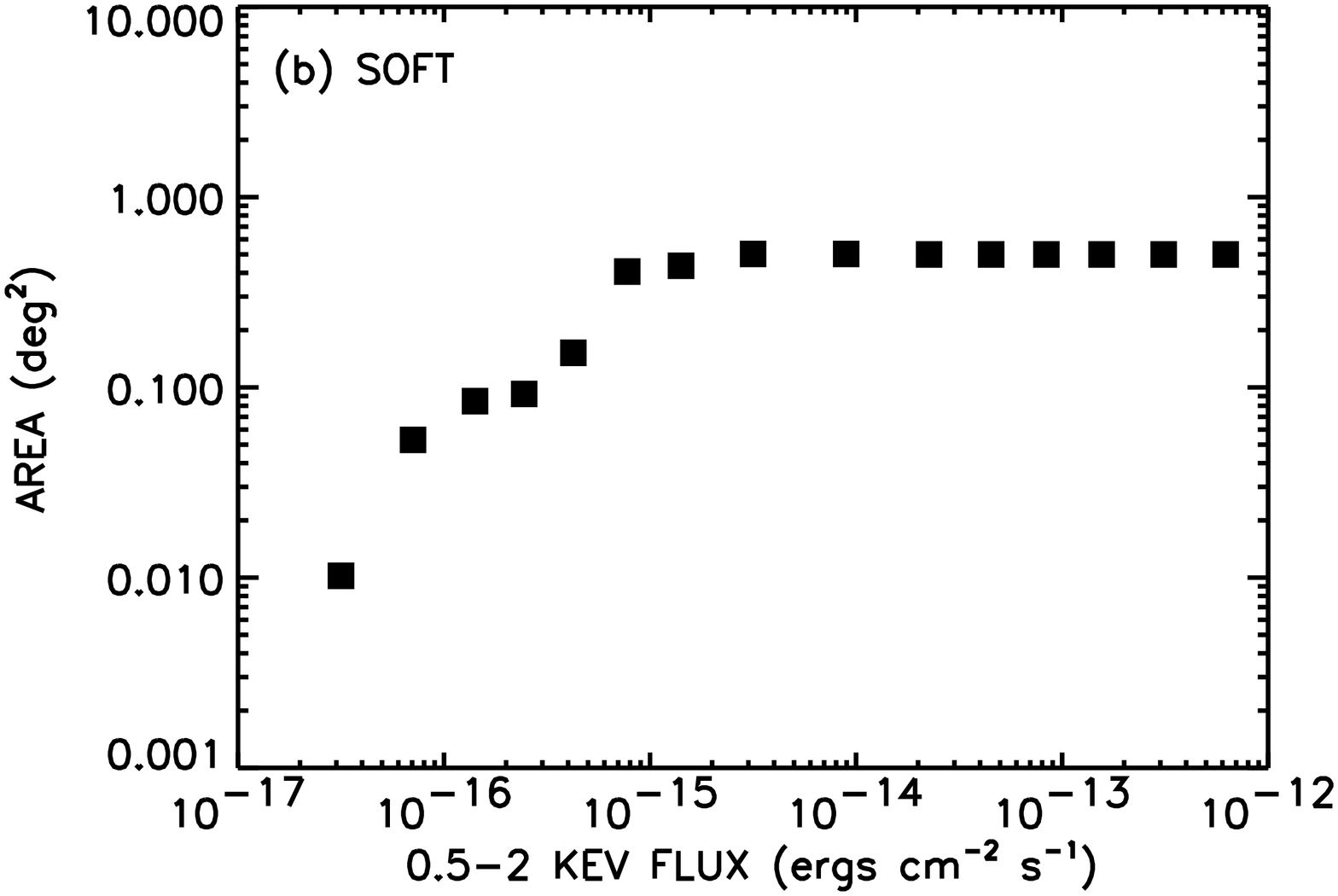,width=3.5in,angle=90}
\figurenum{17}
\caption{
Solid angle vs. (a) $2-8$~keV and (b) $0.5-2$~keV flux
for the observed hard and soft X-ray samples, respectively.
\label{figarea}
}
\addtolength{\baselineskip}{10pt}
\end{inlinefigure}

In Figures~\ref{figlowhxlf}a,
\ref{figlowhxlf}b, and \ref{figlowhxlf}c, we show the hard 
X-ray luminosity functions for all spectral types
({\em squares\/}; this includes all of the spectroscopically
identified X-ray sources with $L_X\ge 10^{42}$~ergs~s$^{-1}$, 
without regard to the optical spectroscopic classifications)
and for broad-line AGNs ({\em diamonds\/}; this includes all of
the spectroscopically identified broad-line AGNs)
for three low-redshift intervals. These hard X-ray luminosity 
functions were computed from the observed-frame $2-8$~keV 
measurements, assuming an intrinsic $\Gamma=1.8$.
The Poissonian $1\sigma$ uncertainties are based on the number
of galaxies in each luminosity bin.
At these low redshifts, there is relatively little uncertainty
from incompleteness, since the sources have X-ray fluxes lying
in the range where most sources are spectroscopically identified.
The triangles show the values computed for what we hereafter refer 
to as the ``spectroscopic plus photometric'' sample.
For this sample, we consider only the CDF-N, CDF-S, and 
{\em ASCA\/} spectroscopically observed sources and assign
CDF-N and CDF-S photometric redshifts, where available,
to those sources that are spectroscopically unidentified.
The resulting identified sample is therefore a combination
of spectroscopic and photometric redshifts.
Note that because this sample does not contain the 
CLASXS data, in some cases the values {\em (triangles)\/}
lie below the spectroscopic values {\em (squares)\/} due to
poor sampling by the deep fields of the low-redshift volume and 
of the high-luminosity sources. In Figure~\ref{figlowhxlf}d,
we show the local $3-20$~keV luminosity function from the 
{\em RXTE\/} analysis of Sazonov \& Revnivtsev (2004).

The forms of the hard X-ray luminosity functions for all spectral 
types and for broad-line AGNs are clearly very different. The hard 
X-ray luminosity functions for broad-line AGNs peak at a 
characteristic luminosity, such that the dominant population 
at the higher X-ray luminosities is broad-line AGNs, while the 
dominant population at the lower X-ray luminosities is
optically-narrow AGNs. This reproduces the Steffen 
effect discussed in \S\ref{secxlum}. We show this more clearly
in Figure~\ref{figtype}, where we plot the ratio {\em (squares)\/}
of the hard X-ray luminosity functions for all spectral types to 
the hard X-ray luminosity functions for broad-line AGNs. 
Above $L_X=10^{44}$~ergs~s$^{-1}$, the
fraction of sources that are broad-line AGNs is very high, while below 
$L_X=10^{43}$~ergs~s$^{-1}$, the fraction of sources that are 
broad-line AGNs is very low. This result still holds when we include
in the hard X-ray luminosity functions for all spectral types
all of the spectroscopically unidentified sources. This is done by
placing the unidentified sources at the center of each 
redshift interval and then including them in the luminosity bin
where their luminosities at that redshift would put them 
{\em (dashed curves)}. Thus, we have clearly detected a luminosity 
dependence in optical spectral type.

In Figures~\ref{figlowhxlf}e--g, we show the division between the 
optically normal galaxies (our absorber and star former 
classes) and the sources which show clear AGN signatures in 
their optical spectra (our high-excitation and broad-line AGN 
classes). Including the  
high-excitation sources slightly raises the normalization of the 
broad-line AGN luminosity function, but it does not change its shape 
much. At low X-ray luminosities, the optically normal galaxies
comprise nearly the entire population. The cross-over
point lies around $L_X<10^{43}$~ergs~s$^{-1}$ at $z=0.25$ and 
rises to $L_X<10^{44}$~ergs~s$^{-1}$ near $z=1$. All of the
curves in Figure~\ref{figlowhxlf} show pure luminosity 
evolution in both the luminosity functions and the spectral type 
mix. This must be understood in any model that seeks to
explain the difference between the optically normal 
galaxies and those with AGN characteristics in their
optical spectra. This type of parallel evolution is probably 
most easily understood for a luminosity dependent unified model 
(see \S\ref{seclum}).

\section{Evolution of the Low-Redshift Hard X-ray luminosity functions}
\label{secevol}

The low-redshift hard X-ray luminosity functions are well represented
({\it solid and dashed curves\/}; Figure~\ref{figlowhxlf}) 
by a conventional double power-law fit 
(Piccinotti et al.\ 1982) of the form
\begin{equation}
{{d\Phi(L_X, z)}\over {d\log L_X}}={a\over {(L/L_\ast)^{g1}+(L/L_\ast)^{g2}}}
\,.
\end{equation}
Parameterizing the redshift evolution as 
\begin{equation}
L_\ast=L_0 \Bigl({{1+z}\over 2}\Bigr)^A \,
\end{equation}
and
\begin{equation}
a=a_0 \Bigl({{1+z}\over 2}\Bigr)^B \,,
\end{equation}
we determined the six parameters and their $1\sigma$ 
uncertainties for both the hard X-ray luminosity function for 
all spectral types and the hard X-ray luminosity function for 
broad-line AGNs using maximum likelihood fits
over the redshift range $z=0-1.2$ (Cash 1979;
Marshall et al.\ 1983).
The parameters are summarized in Table~\ref{tab2}. The
evolution of both hard X-ray luminosity functions is
consistent with pure luminosity 
evolution---as is the evolution of the hard X-ray 
luminosity function for optically-selected broad-line AGNs 
(Boyle et al.\ 2000)---where the values of  
$g1$, $g2$, and $a$ remain constant and only $L_\ast$ evolves. 
The fits are acceptable throughout the redshift range, as shown in
Figures~\ref{figlowhxlf}a--d.

%
%
\begin{deluxetable}{lcc}
\tablecaption{Maximum Likelihood Fit Parameters for
All Spectral Types and for Broad-line AGNs}
\tablehead{Parameter & All & Broad-line}
\startdata
$\log L_0$ & $44.11\pm 0.08$  &  $43.81\pm 0.12$ \cr
$\log a_0$ & $-4.42\pm 0.07$  &  $-4.44\pm 0.14$ \cr
$g_1$      & $0.42\pm 0.06$   &  $-0.9\pm 0.5$   \cr
$g_2$      & $2.2\pm 0.5$     &  $1.6\pm 0.3$    \cr
$A$        & $3.2\pm 0.8$     &  $3.0\pm 1.0$    \cr
$B$        & $-0.1\pm 0.7$    &  $-0.1\pm 0.6$   \cr
\enddata
\label{tab2}
\end{deluxetable}

The above fit to the hard X-ray luminosity function for all 
spectral types agrees well with that of Ueda et al.\ (2003)
at higher luminosities but is shallower below the break,
where Ueda et al.\ find $g1=0.82\pm 0.13$ for their pure
luminosity evolution model. In fact, the Ueda et al.\ slope 
overpredicts our low-luminosity counts throughout the $z=0-1.2$ 
redshift range.  Our data sample is much larger than the sample
used in Ueda et al., particularly in this redshift range, because
of our inclusion of the wide-field CLASXS data; correspondingly,
our fit should be more robust.

Our $L_\ast=(1+z)^{3.2\pm0.8}$ evolution at $z<1.2$
is steeper than the $L_\ast\sim (1+z)^{2.7}$ evolution
found by Ueda et al.\ (2003), though
consistent within the uncertainties.
Sazonov \& Revnivtsev (2004) found that the
Ueda et al.\ (2003) model lay significantly above their local
determination. In contrast, our steeper luminosity 
evolution matches the Sazonov \& Revnivtsev (2004)
local determination within the uncertainties in the determination 
of the $A$ parameter (see Figure~\ref{figlowhxlf}d).

Thus, AGNs drop in luminosity by almost an order of magnitude
over the $z=0-1$ redshift range. This drop applies equally to 
all AGNs, regardless of optical spectral type. 
We shall return to the question of whether this is an
evolution of the accretion rate or of the active population
of supermassive black holes in \S\ref{secsmbh}.

%
%
\begin{figure*}
\centerline{\psfig{figure=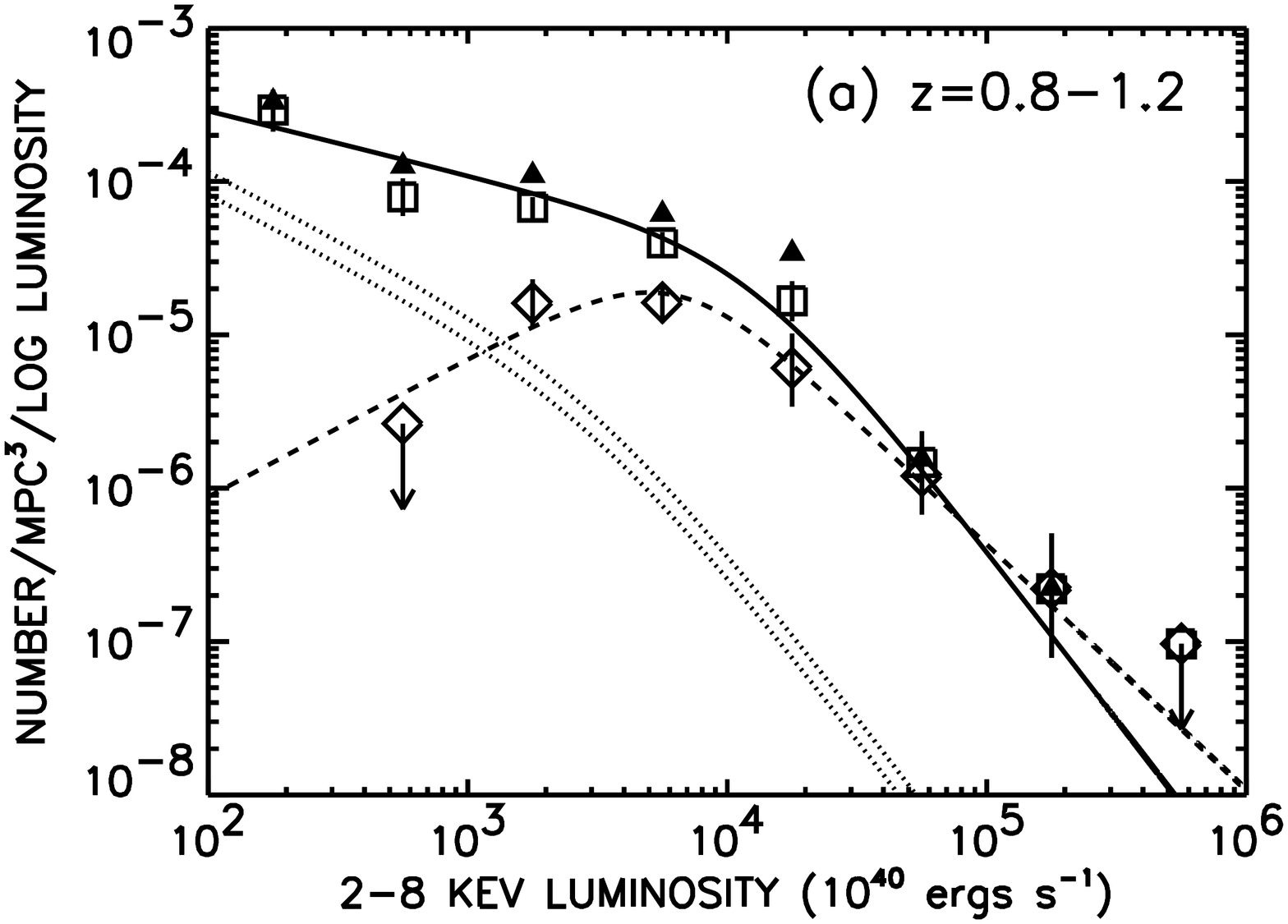,width=8.5cm,angle=90}
\psfig{figure=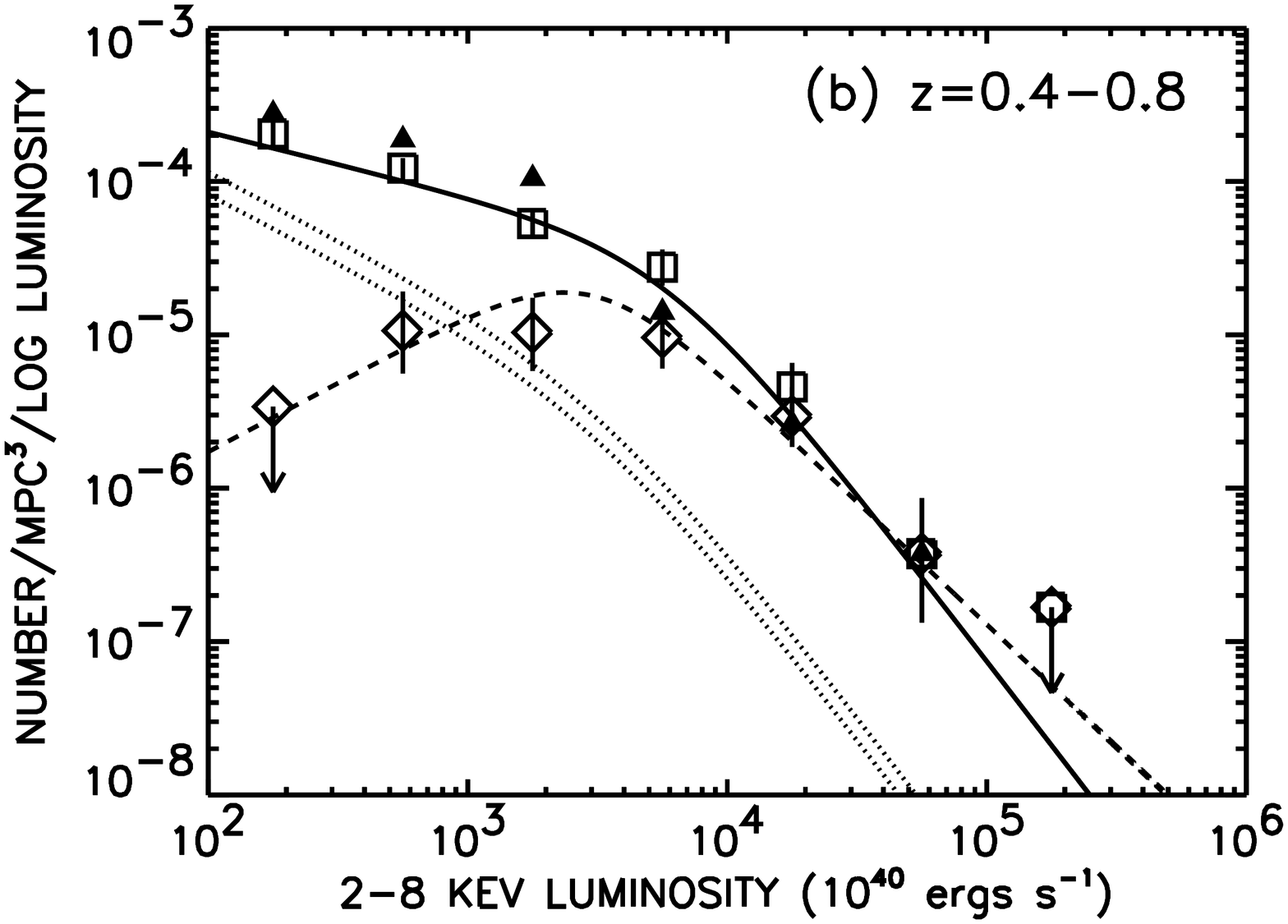,width=8.5cm,angle=90}}
\centerline{\psfig{figure=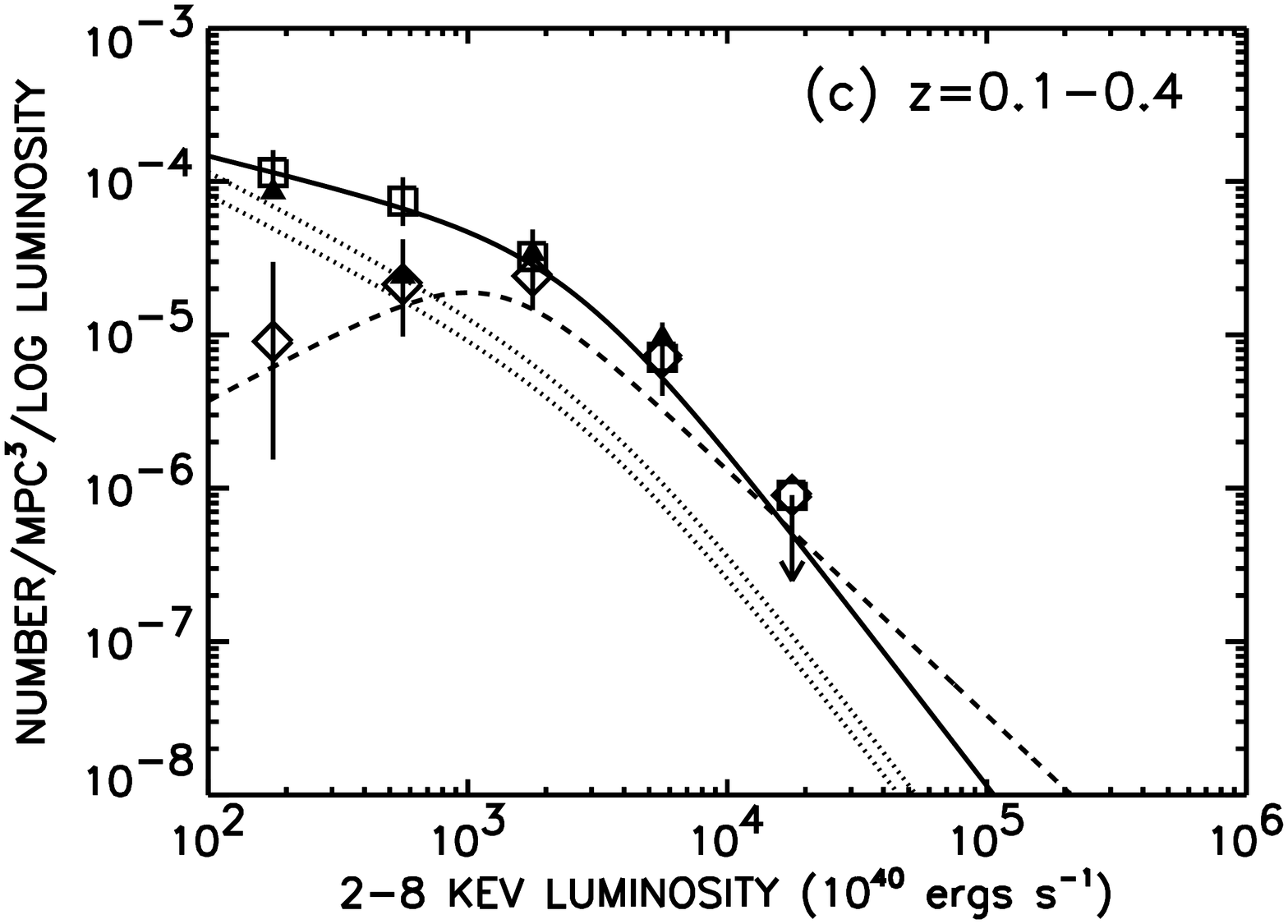,width=8.5cm,angle=90}
\psfig{figure=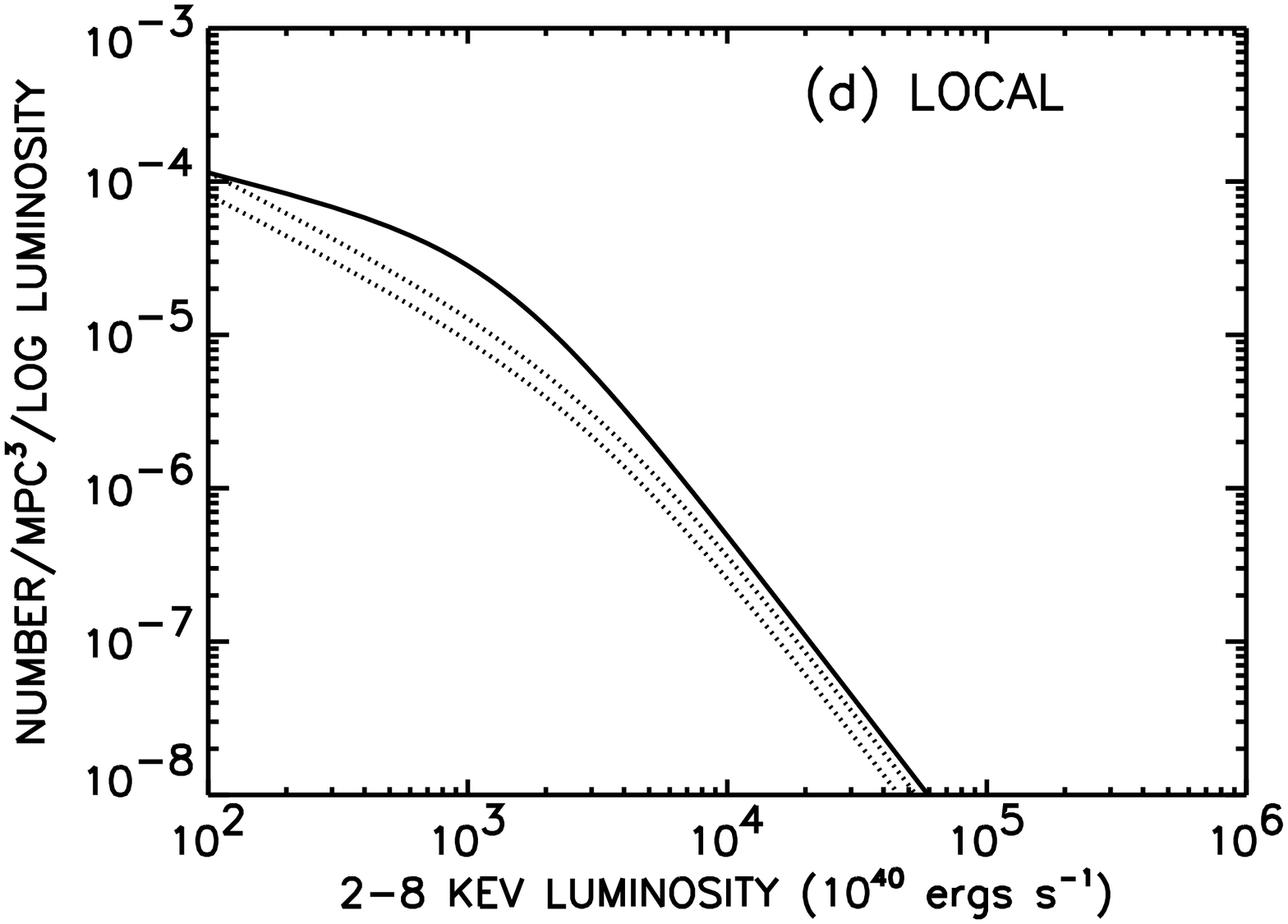,width=8.5cm,angle=90}}
\vspace{6pt}
\figurenum{18}
\caption{
Rest-frame $2-8$~keV luminosity function
per unit (base 10) logarithmic luminosity in the redshift intervals
(a) $z=0.8-1.2$, (b) $z=0.4-0.8$, and (c) $z=0.1-0.4$.
Hard X-ray luminosity functions for all spectral types
(broad-line AGNs) are denoted by squares (diamonds) and were
computed from the observed-frame $2-8$~keV data.
We assumed an intrinsic $\Gamma=1.8$, for which
there is only a small differential $K$-correction to
rest-frame $2-8$~keV. Poissonian $1\sigma$ uncertainties
are based on the number of sources in each luminosity bin.
Triangles denote the
spectroscopic plus photometric hard X-ray luminosity functions,
which use only the CDF-N, CDF-S, and {\em ASCA\/} data (see text
for details). (d) Local hard X-ray luminosity function determined
by Sazonov \& Revnivtsev (2004)
using {\em RXTE\/} data, shown with ({\em upper dotted curve\/})
and without ({\em lower dotted curve\/}) their assumed
incompleteness correction of 1.4. These curves are also shown
in the other panels. Solid (dashed) curves show the
maximum likelihood fit to the hard X-ray luminosity function for
all spectral types (broad-line AGNs) computed at the geometric
mean for each redshift interval. Only the maximum likelihood fit
to the hard X-ray luminosity function for all spectral
types computed at $z=0$ is shown in (d).
\label{figlowhxlf}
}
\end{figure*}

%
%
\begin{figure*}
\centerline{\psfig{figure=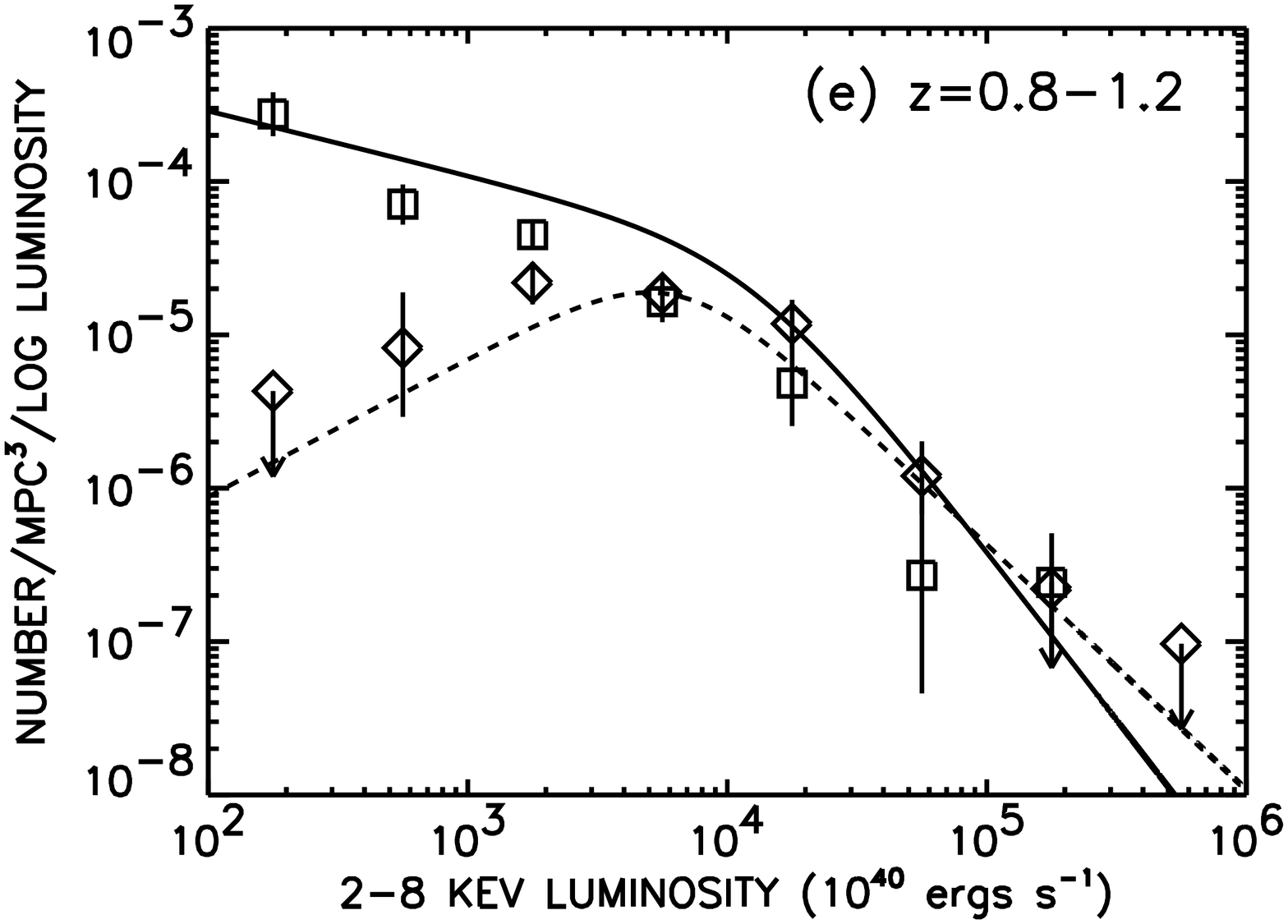,width=8.5cm,angle=90}
\psfig{figure=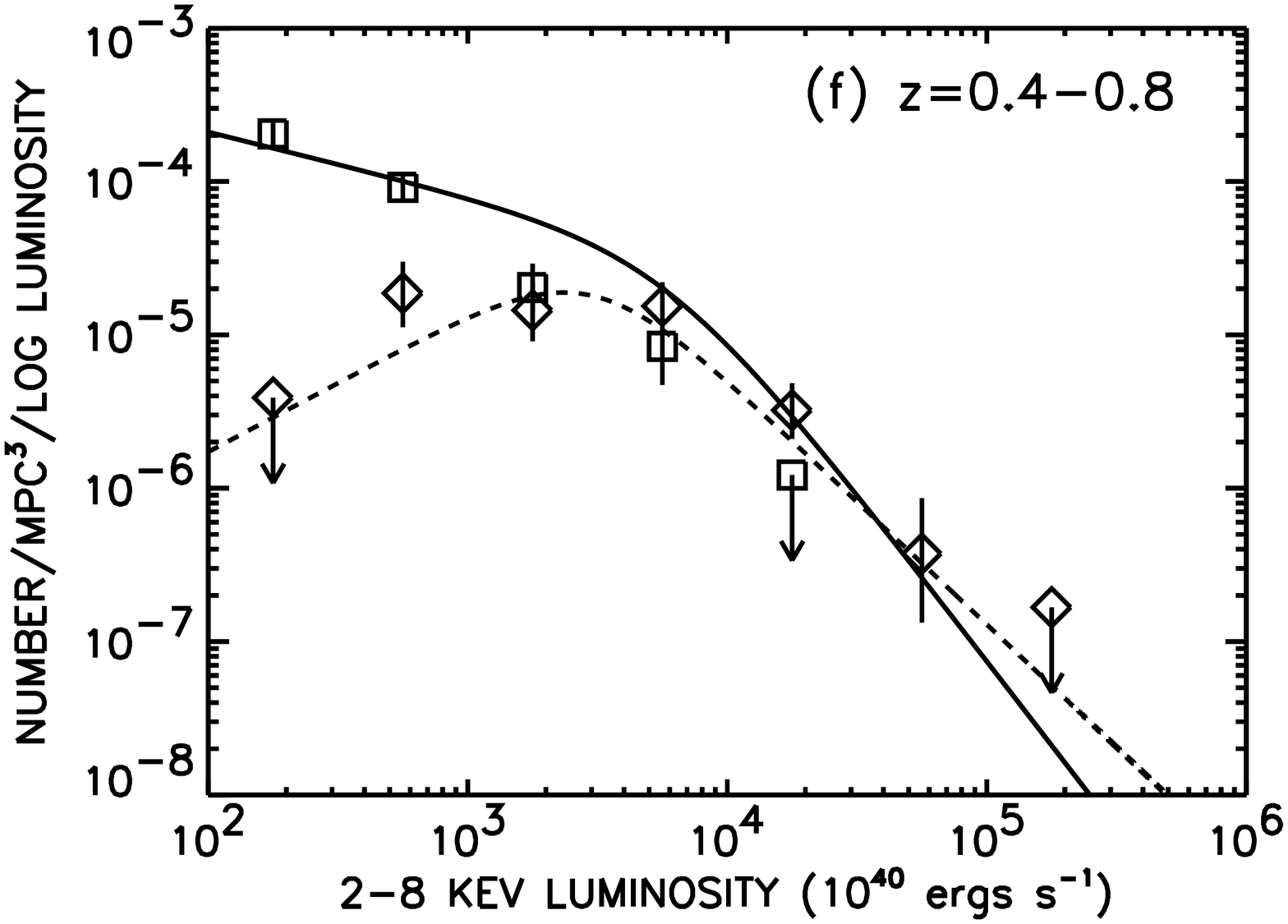,width=8.5cm,angle=90}}
\centerline{\psfig{figure=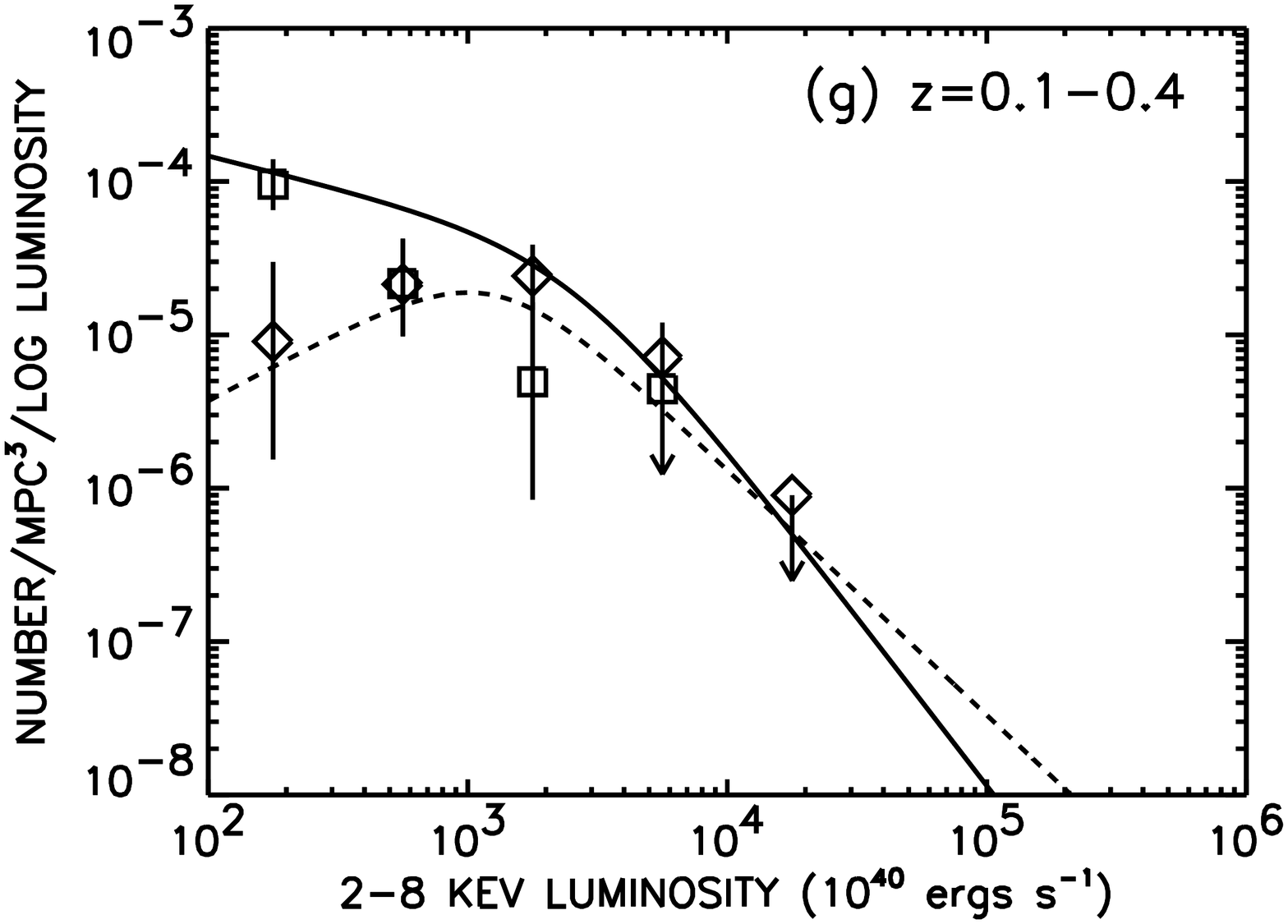,width=8.5cm,angle=90}}
\vspace{6pt}
\figurenum{18}
\caption{(continued) Parts (e)--(g) show the same redshift 
intervals as (a)--(c), but
now the open squares denote optically normal galaxies alone,
and the open diamonds denote high-excitation sources and broad-line
AGNs together. The solid and dashed lines are the same as in
(a)--(c). The optically normal galaxies dominate the
populations at low X-ray luminosities, while the inclusion of
the high-excitation sources only slightly increases the broad-line
AGN luminosity functions.
}
\end{figure*}

%
%
\begin{figure*}
\centerline{\psfig{figure=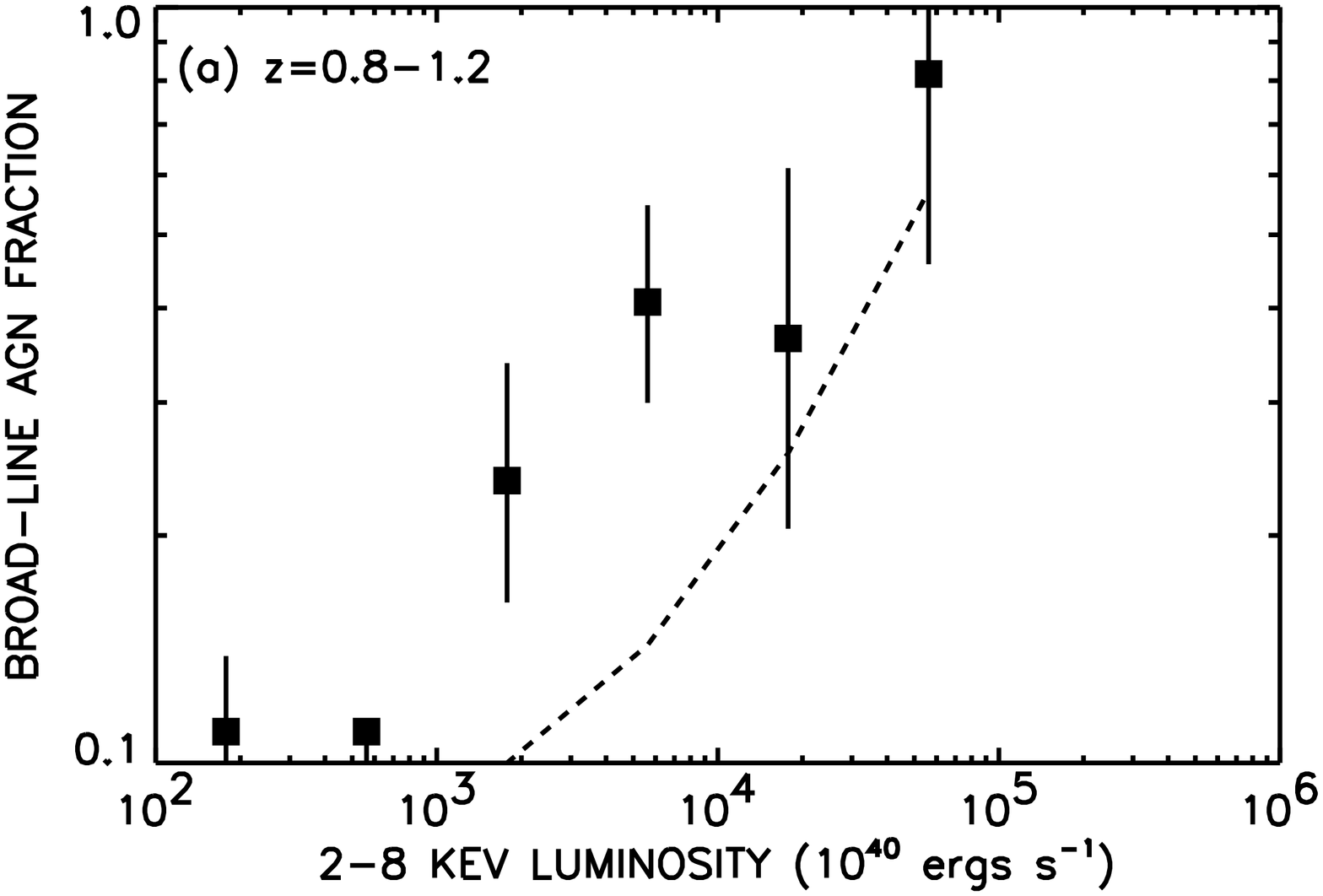,width=8.5cm,angle=90}
\psfig{figure=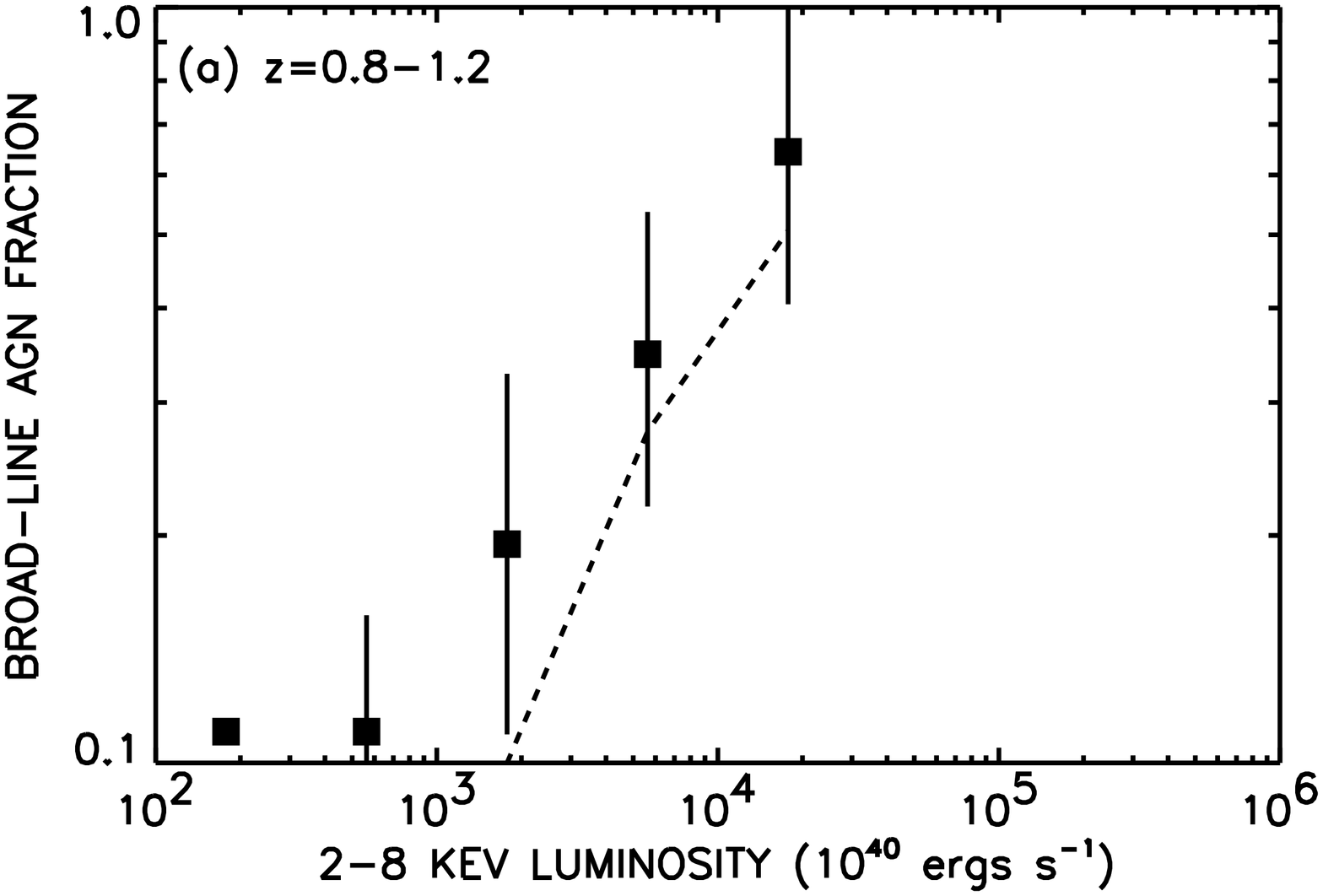,width=8.5cm,angle=90}}
\centerline{\psfig{figure=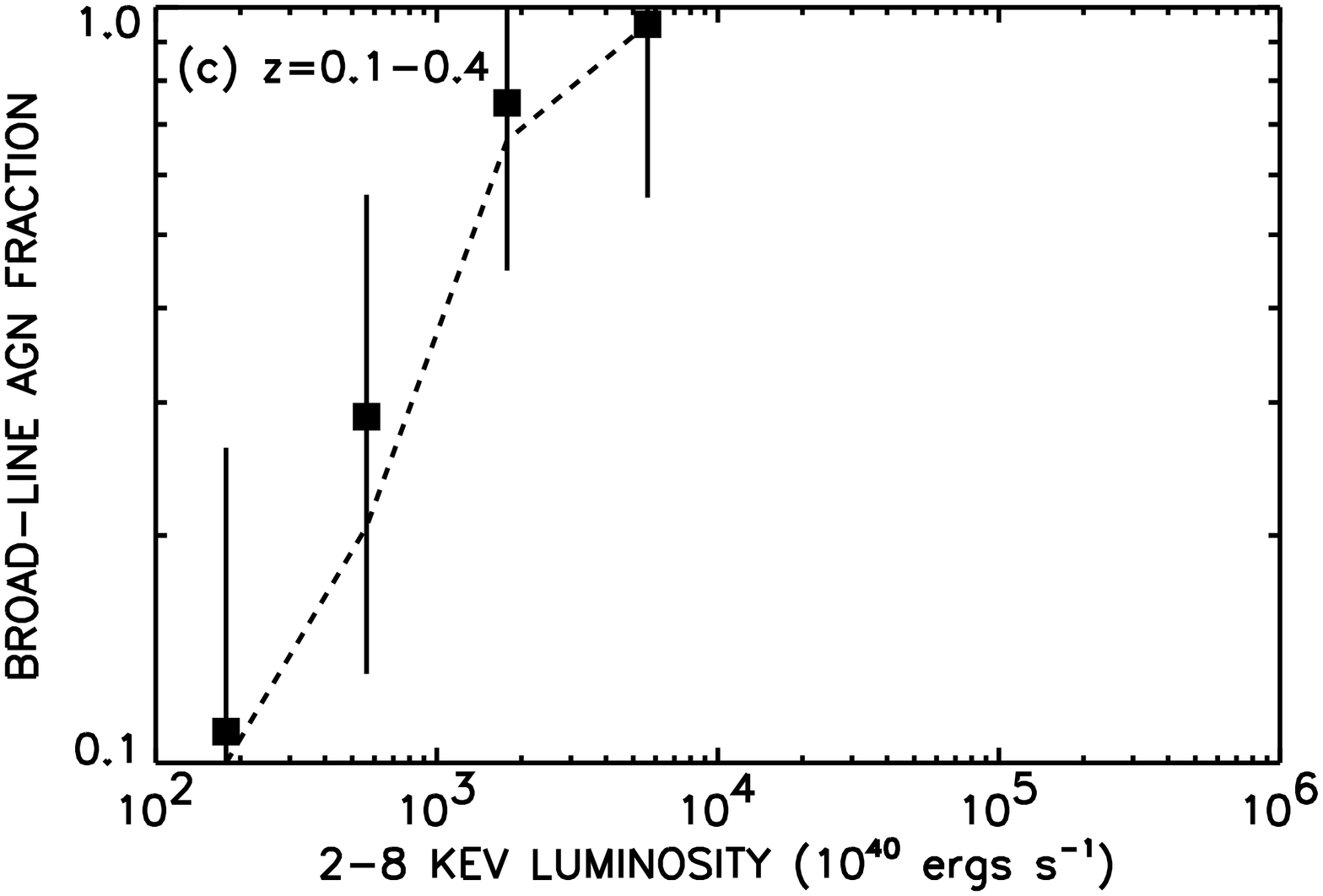,width=8.5cm,angle=90}}
\vspace{6pt}
\figurenum{19}
\caption{
Ratio {\em (squares)\/} of the rest-frame $2-8$~keV
luminosity function per unit logarithmic luminosity in the redshift
intervals (a) $z=0.8-1.2$, (b) $z=0.4-0.8$, and (c) $z=0.1-0.4$
for all spectral types (squares from Figures~18a--c) relative
to that for broad-line AGNs (diamonds from Figures~18a--c). Dashed
curves show the same ratio, but with all of the spectroscopically
unidentified sources in the spectroscopically observed sample
included in the hard X-ray luminosity functions for all spectral types.
This was done by placing the unidentified sources at the center of
each redshift interval and then including them in the luminosity bin
where their luminosities at that redshift would put them.
Uncertainties are $1\sigma$ based on the number of sources in each bin.
\label{figtype}
}
\end{figure*}

%
%
\begin{figure*}
\centerline{\psfig{figure=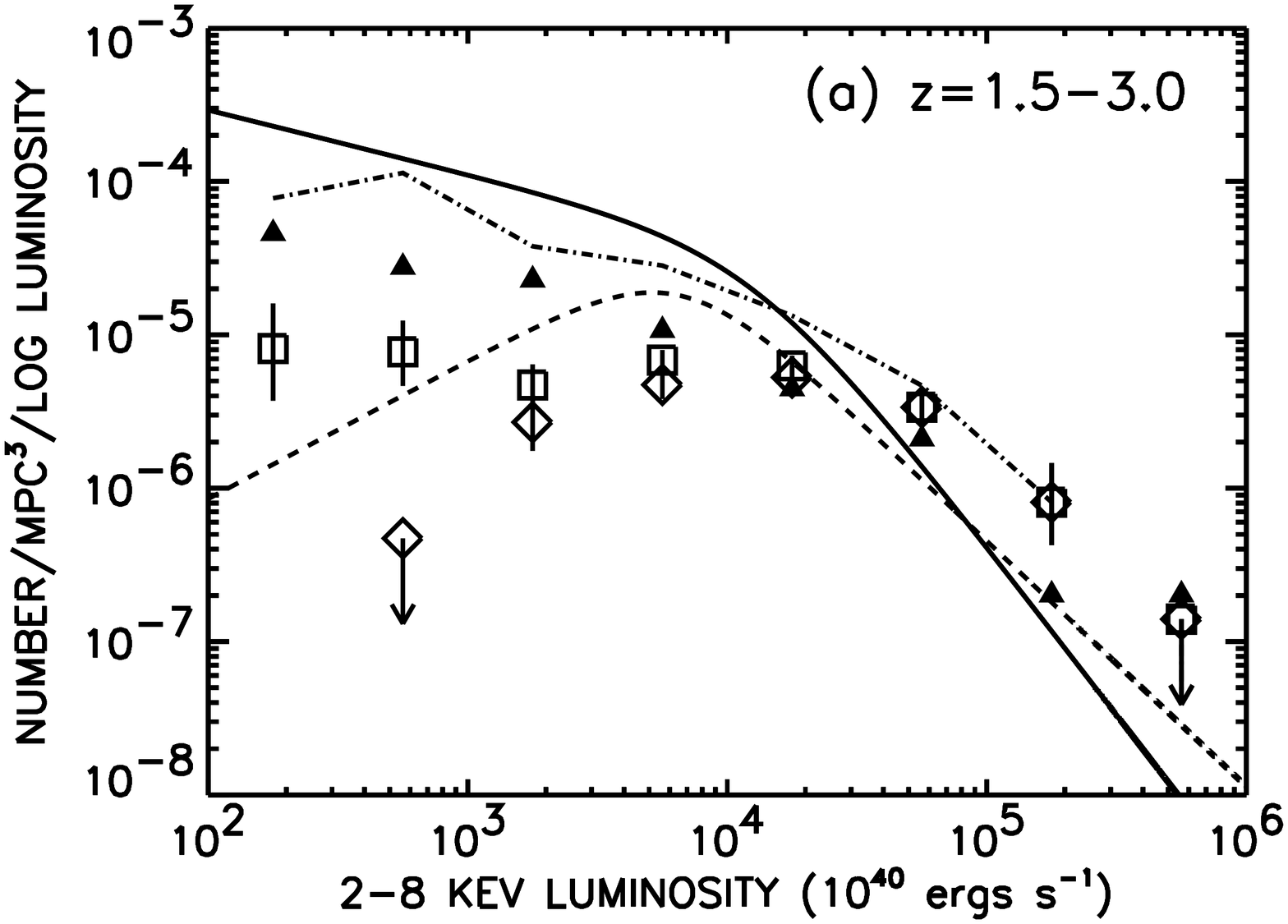,width=8.5cm,angle=90}
\psfig{figure=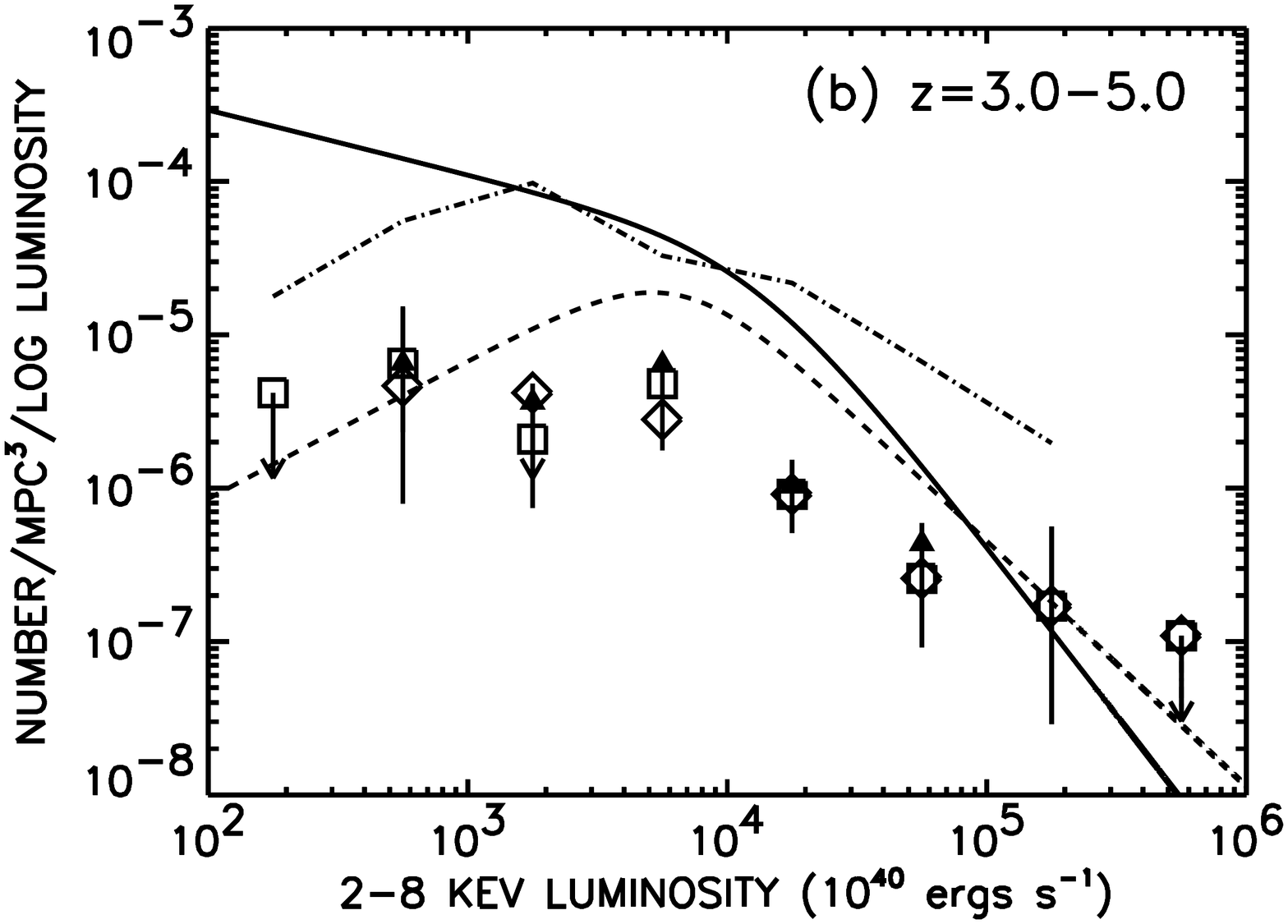,width=8.5cm,angle=90}}
\figurenum{20}
\vspace{6pt}
\caption{Rest-frame $2-8$~keV luminosity function per unit
logarithmic luminosity in the redshift intervals (a) $z=1.5-3$
and (b) $z=3-5$. Hard X-ray luminosity functions for all spectral
types (broad-line AGNs) are denoted by squares (diamonds)
and were computed from the observed-frame $0.5-2$~keV data.
Poissonian $1\sigma$ uncertainties are based on the number of
sources in each luminosity bin. Dot-dashed curves are the maximal
hard X-ray luminosity functions for all spectral types in each
redshift interval. These were found by assigning
redshifts at the center of each redshift interval to all of the
spectroscopically unidentified sources in the spectroscopically
observed sample. Triangles denote the spectroscopic plus
photometric hard X-ray luminosity functions, which use only the
CDF-N, CDF-S, and {\em ASCA\/} data (see text for details).
Solid (dashed) curves denote the maximum likelihood fit to the
$z=0-1.2$ hard X-ray luminosity function for all spectral types
(broad-line AGNs) computed at $z=1$.
\label{fighihxlf}
}
\end{figure*}

\section{High-Redshift Hard X-ray Luminosity Functions}
\label{sechizhxlf}

We also determined the hard X-ray luminosity functions for 
two high-redshift intervals,
$z=1.5-3$ and $z=3.0-5.0$, using a $0.5-2$~keV observed-frame
sample to provide the best possible match to the lower redshift
data. Figures~\ref{fighihxlf}a and \ref{fighihxlf}b show these
measured hard X-ray luminosity functions 
({\em squares\/}---all spectral types;
{\em diamonds\/}---broad-line AGNs). It is very difficult to measure 
host galaxy redshifts for sources that lie in the high-redshift 
intervals, so incompleteness is potentially a large source of 
error. We therefore computed maximal hard X-ray luminosity 
functions for all spectral 
types by assigning redshifts at the center of each redshift 
interval to all of the spectroscopically unidentified sources
(i.e., we included all of the unidentified sources
in both redshift intervals). We show these as dot-dashed
curves in the figures. It is important to keep in mind 
that because all of the unidentified sources have been included 
in both redshift intervals (provided that the sources, when 
assigned those redshifts, are at $L_X\ge 10^{42}$~ergs~s$^{-1}$), 
the curves are not consistent with one another.
Because the spectroscopic
identifications are much more complete at higher X-ray fluxes,
the associated systematic uncertainties are larger at lower $L_X$.
(Note that any low X-ray flux source assigned to a given redshift
interval will have a low $L_X$ in that redshift interval.)
We also show as triangles the 
spectroscopic plus photometric hard X-ray luminosity 
functions determined from the CDF-N, CDF-S, and {\em ASCA\/} data 
only. In the $z=1.5-3$ redshift interval, the 
spectroscopic plus photometric hard X-ray 
luminosity function is much closer to the maximal hard X-ray 
luminosity function than is the spectroscopic hard X-ray luminosity 
function, whereas in the $z=3-5$ redshift interval, the 
spectroscopic plus photometric 
hard X-ray luminosity function is very similar to the spectroscopic 
hard X-ray luminosity function. This suggests that many of the 
spectroscopically unidentified sources lie in the $z=1.5-3$ interval 
rather than at higher redshifts.

Even with our maximal incompleteness corrections, at
$L_X<10^{44}$~ergs~s$^{-1}$, the hard X-ray luminosity functions 
for all spectral types in the two high-redshift intervals lie below 
the maximum likelihood fits to the $z=0-1.2$ hard X-ray luminosity 
function computed at $z=1$ ({\em solid curve\/}), suggesting a peak 
in the universal AGN energy density production rate near $z=1$. 
However, at $L_X>10^{44}$~ergs~s$^{-1}$, 
the maximal hard X-ray luminosity functions in both high-redshift 
intervals appear to continue the increasing trend with redshift. 

Before proceeding, we want to make sure that we would not
have gotten different results if we had used the $2-8$~keV 
observed-frame sample instead of the $0.5-2$~keV observed-frame 
sample in determining the high-redshift hard X-ray luminosity functions. 
We performed this check by computing the $z=1.5-3$ hard X-ray 
luminosity function for broad-line AGNs 
using the $0.5-2$~keV observed-frame sample and also
using the $2-8$~keV observed-frame sample. In Figure~\ref{figshcomp}, 
we show that the agreement between these two computations is very good. 

%
%
\begin{inlinefigure}
\psfig{figure=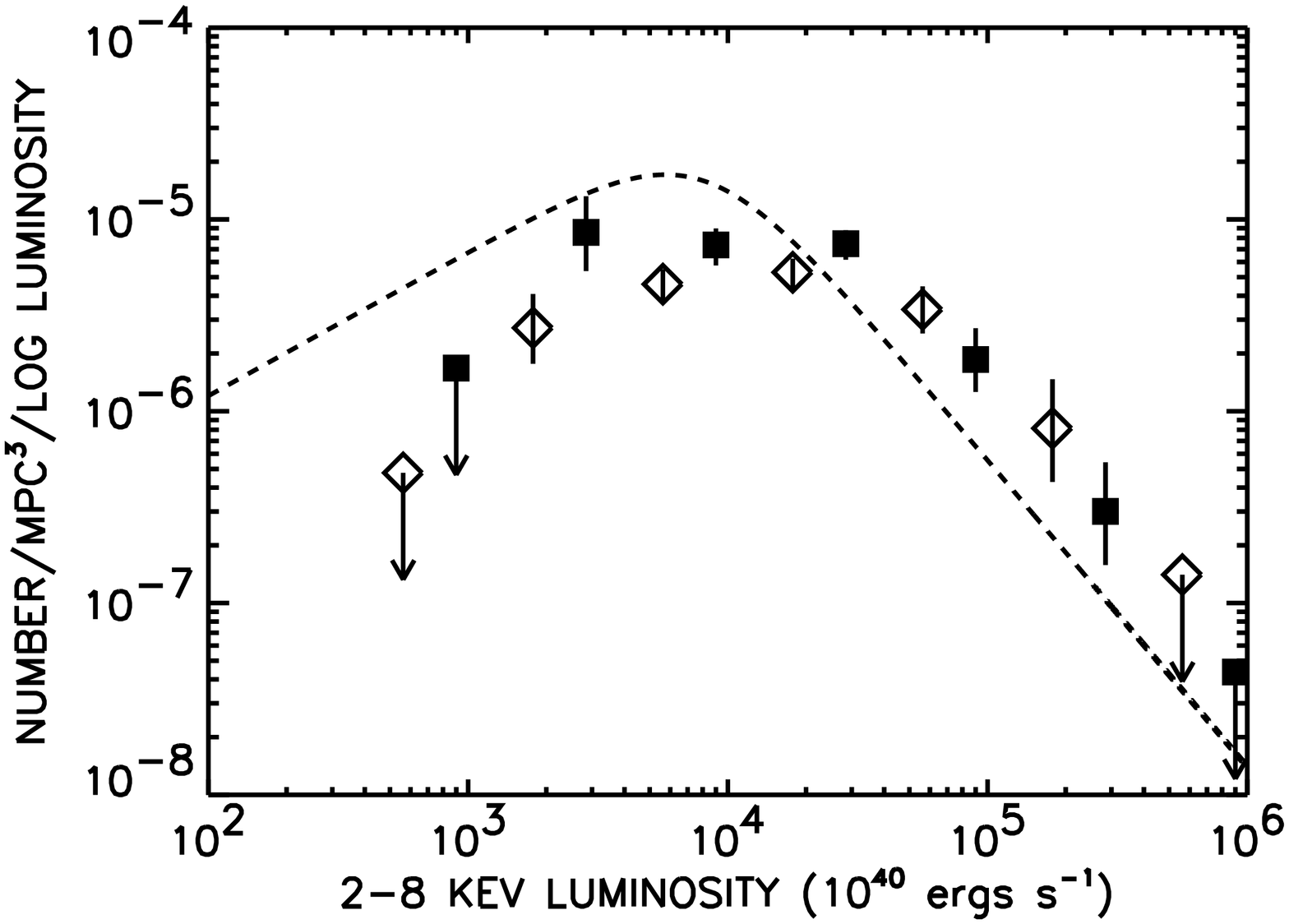,width=3.5in,angle=90}
\figurenum{21}
\caption{
Hard X-ray luminosity function for broad-line AGNs at $z=1.5-3$
computed from the observed-frame
$0.5-2$~keV {\em (diamonds)\/} and observed-frame
$2-8$~keV {\em (squares)\/} samples. Dashed curve denotes
the maximum likelihood fit to the $z=0-1.2$ hard X-ray
luminosity function for broad-line AGNs computed at $z=1$.
\label{figshcomp}
}
\addtolength{\baselineskip}{10pt}
\end{inlinefigure}

\section{Comparison with Optical QSO Luminosity Functions}
\label{seclf}

\begin{figure*}
\centerline{\psfig{figure=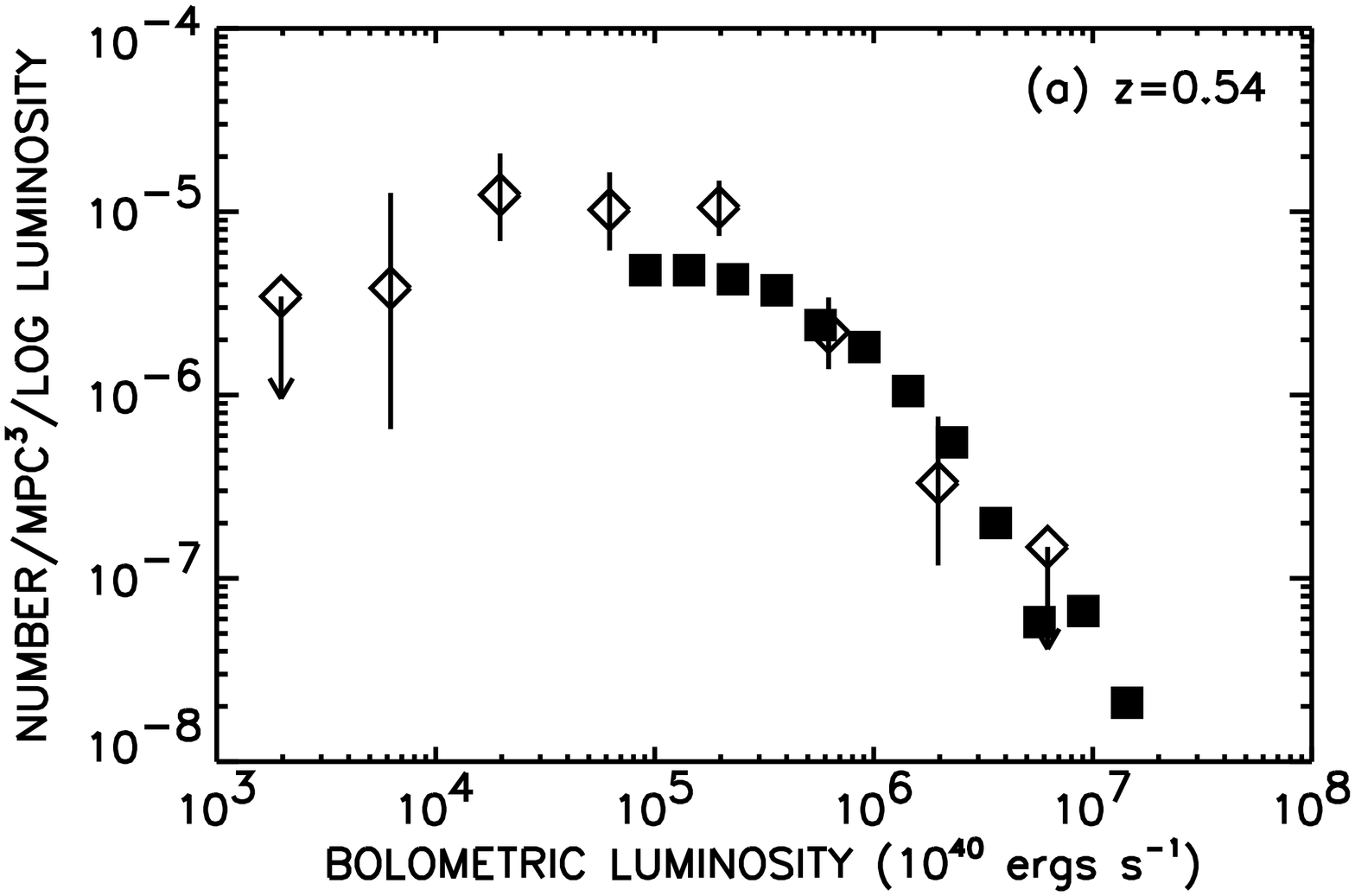,width=8.5cm,angle=90}
\psfig{figure=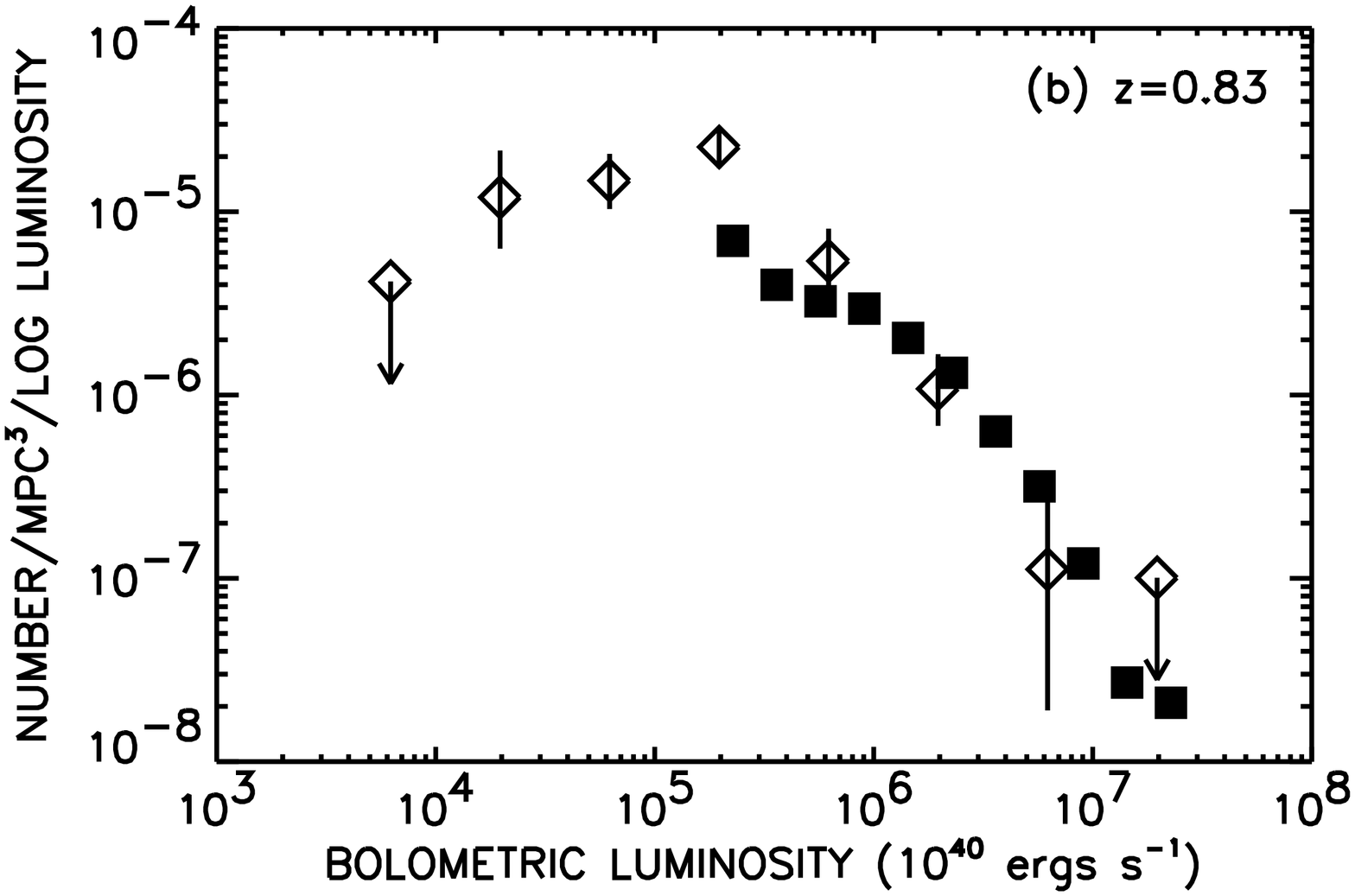,width=8.5cm,angle=90}}
\centerline{\psfig{figure=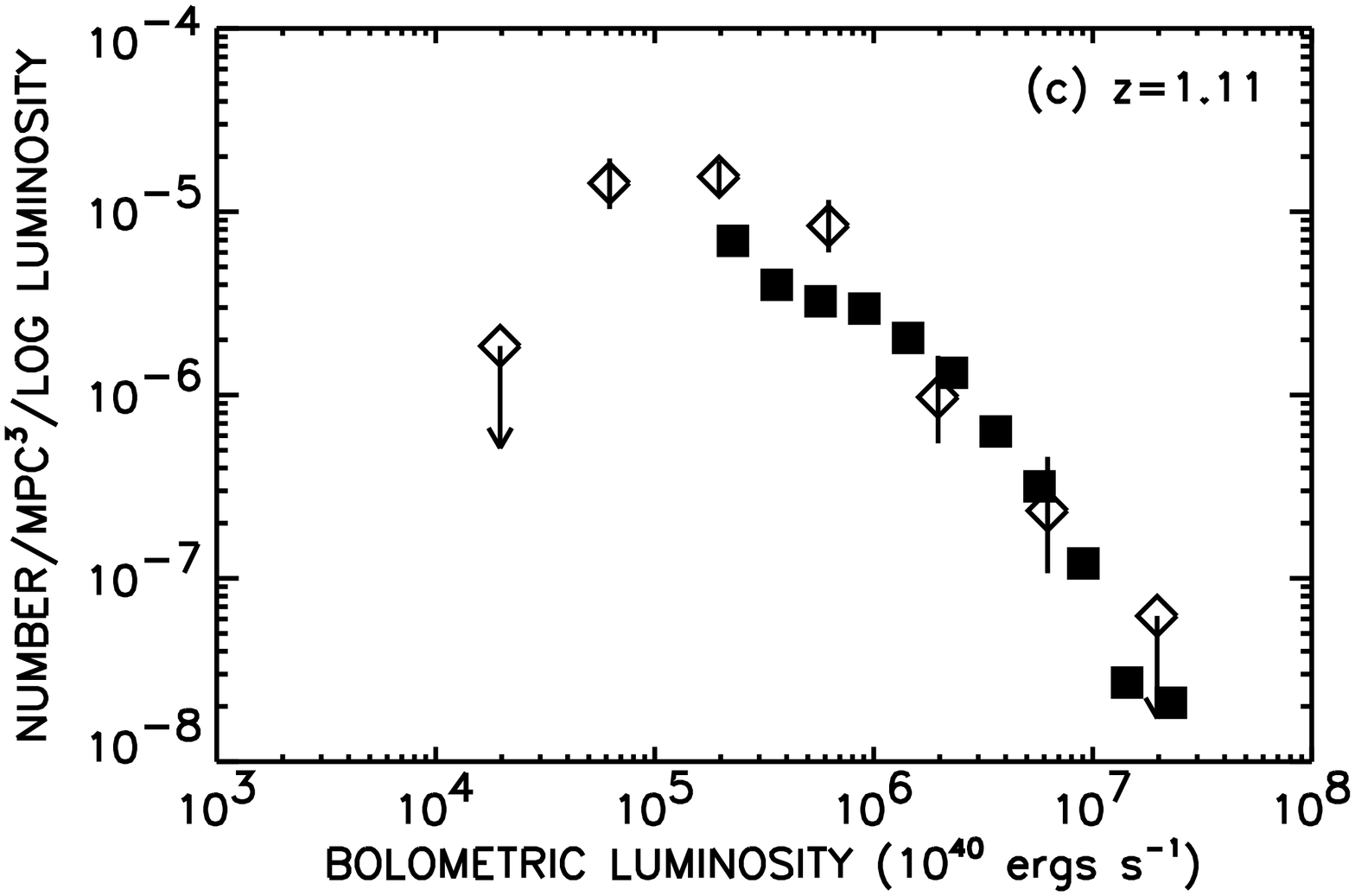,width=8.5cm,angle=90}
\psfig{figure=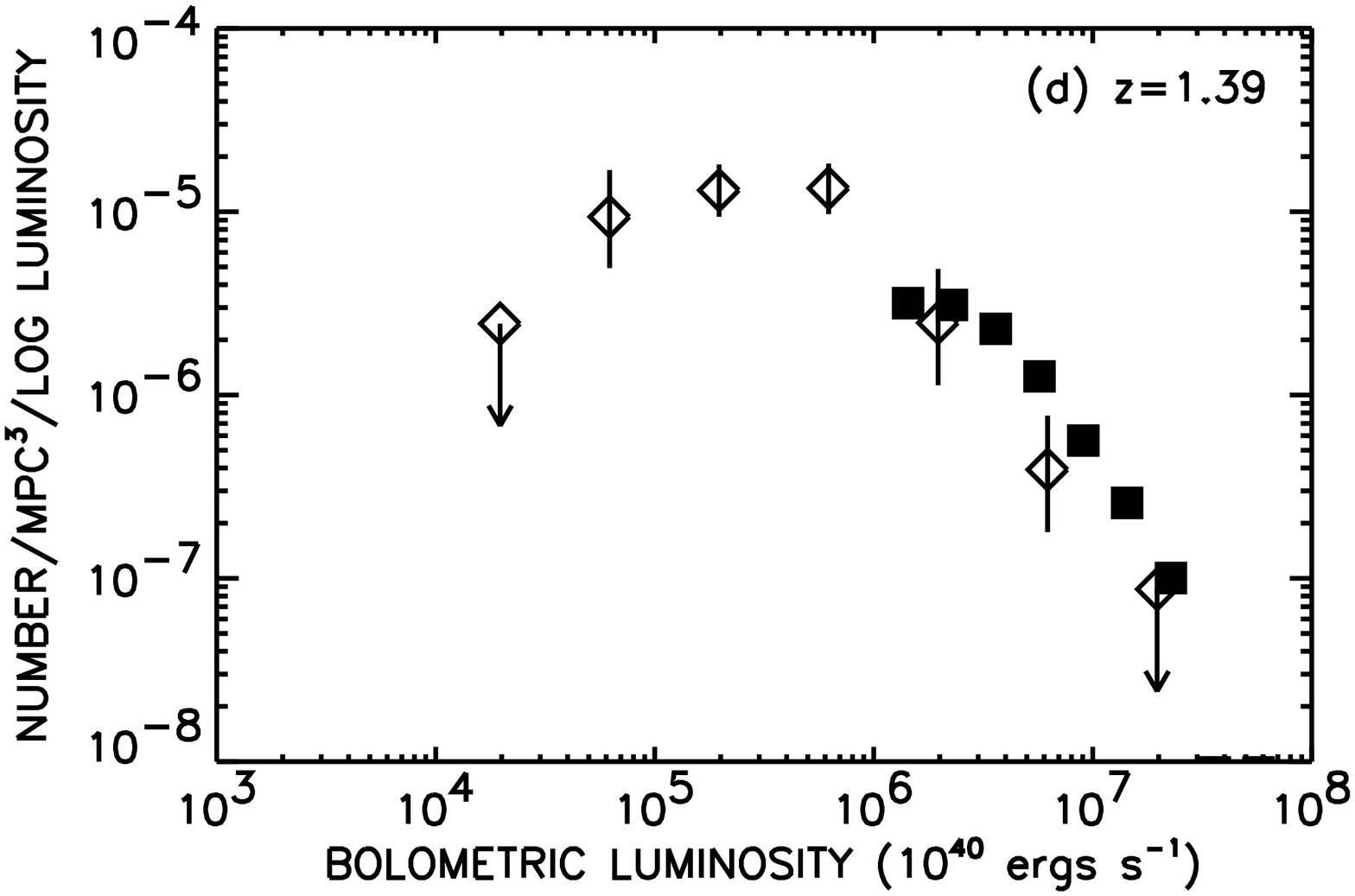,width=8.5cm,angle=90}}
\centerline{\psfig{figure=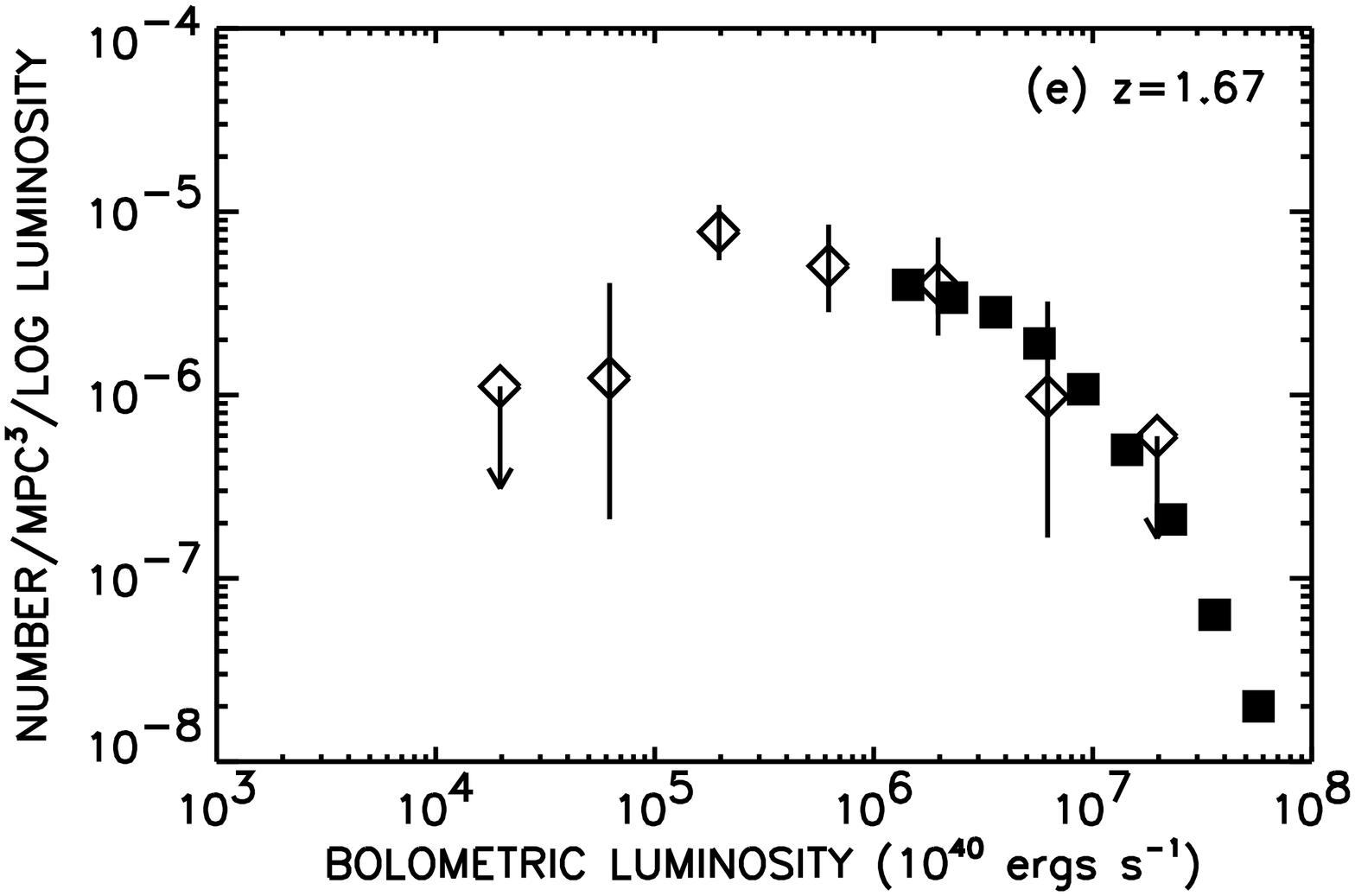,width=8.5cm,angle=90}
\psfig{figure=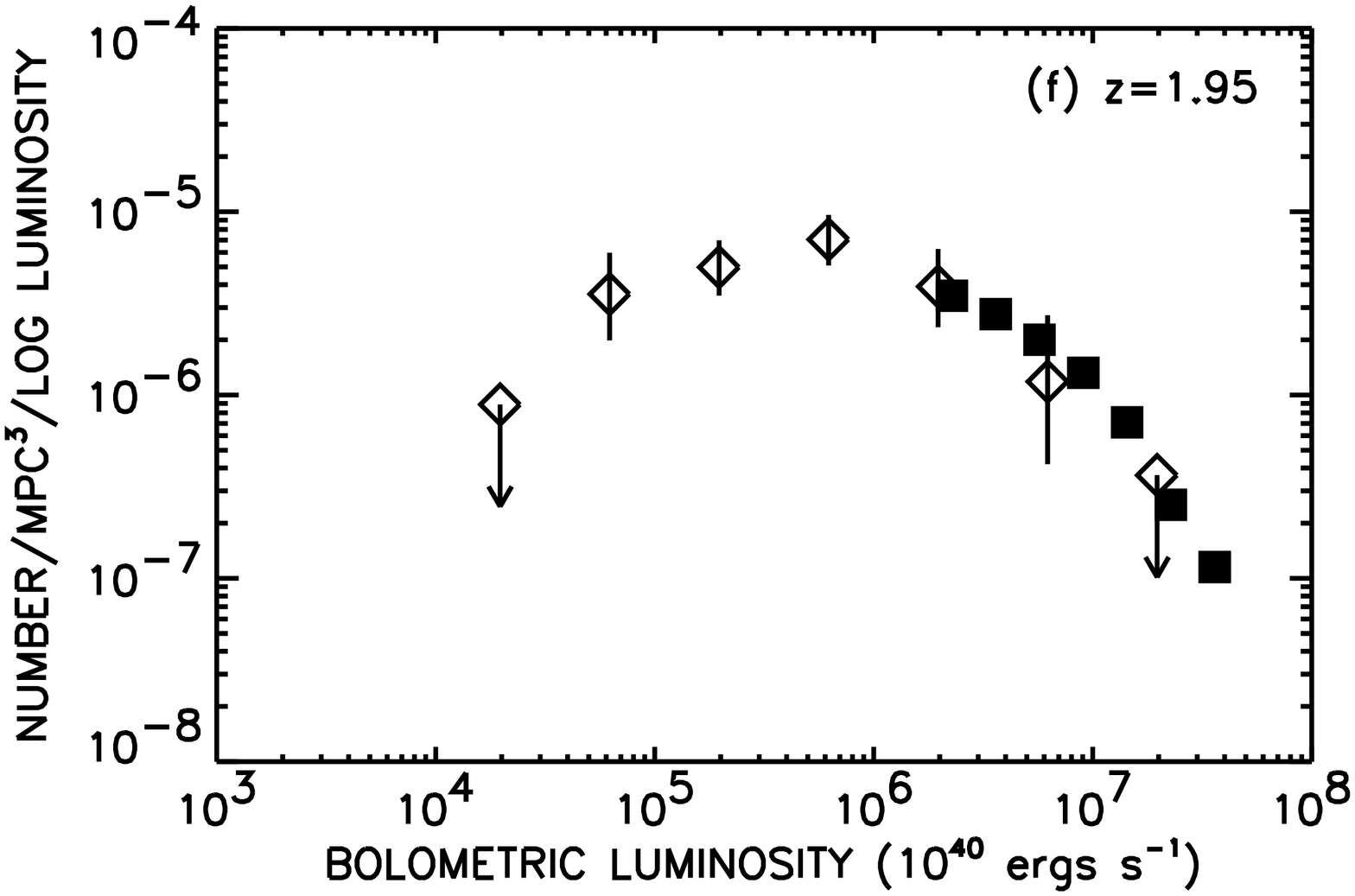,width=8.5cm,angle=90}}
\vspace{6pt}
\figurenum{22}
\caption{
Open diamonds show the rest-frame $2-8$~keV luminosity function
per unit (base 10) logarithmic luminosity for the broad-line AGNs
vs. bolometric luminosity ($L_{BOL}=35\times L_X$) in the redshift
intervals (a) $z=0.28-0.8$, (b) $z=0.6-1.06$, (c) $z=0.8-1.42$,
(d) $z=1.2-1.58$, (e) $z=1.51-1.83$, and (f) $z=1.7-2.21$.
Solid squares show the rest-frame $B$-band
luminosity function of Croom et al.\ (2004) vs. bolometric luminosity
($L_{BOL}=11.8 \nu_B L_\nu(B)$) in the redshift intervals
(a) $z=0.40-0.68$, (b) $z=0.68-0.97$, (c) $z=0.97-1.25$,
(d) $z=1.25-1.53$, (e) $z=1.25-1.53$, and (f) $z=1.81-2.10$
adapted from their Table~4.
\label{blagn_opt}
}
\end{figure*}

Since the X-ray--selected broad-line AGNs recover essentially all 
of the optically-selected type~1 AGNs (see \S\ref{secclass}), we 
should be able to compare directly the broad-line AGN hard 
X-ray luminosity functions computed above with the optical QSO 
luminosity functions. In order to compare the two, we 
calculate the bolometric luminosities using the bolometric 
corrections determined by Elvis et al.\ (1994) 
(i.e., $L_{BOL}=35\times L_X$ for the broad-line AGN hard X-ray 
luminosity functions, and $L_{BOL}=11.8 \nu_B L_{\nu_B}$ for the 
QSO luminosity functions). In Figure~\ref{blagn_opt}, we show for 
six redshift intervals the broad-line AGN hard X-ray luminosity 
functions {\em (open diamonds)\/} and the rest-frame $B$-band QSO 
luminosity functions (Croom et al.\ 2004; {\em solid squares\/}) 
versus the calculated bolometric luminosities. The Croom et al.\ 
luminosity functions have been renormalized to our assumed geometry.

The bright end luminosity functions agree extremely well at all 
redshifts, confirming that the two methodologies are measuring 
the same sample and that the bolometric corrections used are 
appropriate. The optical QSO luminosity functions do not probe 
faint enough to see the downturn in the broad-line AGN hard X-ray 
luminosity functions. Moreover, they may be missing some 
sources at the very lowest luminosities to which they probe, 
as can be seen from the lowest redshift panels.

The optical QSO luminosity function in the $M_{B} = -23$ to $-26$ 
range is well described by pure luminosity evolution 
over the redshift range 
$z=0.3-2.1$ (Boyle et al. 2000). Croom et al.\ (2004)
parameterize this as $L$ evolving as $10^{k1 z + k2 z^{2}}$,
where $k1=1.39$ and $k2=-0.29$. In Figure~\ref{lumevol_comp},
we compare the Croom et al.\ evolution with the $(1+z)^{3.0}$ law 
we determined in \S\ref{secevol} for the broad-line AGNs 
(see Table~\ref{tab2}). 
The Croom et al.\ evolution is slightly steeper, though part of 
the difference may lie in the adopted functional forms and in
the different redshift ranges over which the laws have been 
fitted. However, within the uncertainties, the two determinations 
are consistent over the $z=0-1.2$ redshift interval.

We further investigate the pure luminosity evolution model for 
broad-line AGNs by using the Croom et al.\ (2004) 
evolution law (which was fit over a wider redshift range,
$z=0.3-2.1$, than our maximum likelihood fit was, $z=0-1.2$) 
to correct all of the X-ray luminosities to their values at 
$z=1$. We then compute the hard X-ray luminosity functions 
for broad-line AGNs over the wide redshift ranges $z=0.2-0.7$, 
$z=0.7-1.5$, and $z=1.5-2.5$. The resulting hard X-ray luminosity 
functions are compared in Figure~\ref{delum_comp}. The lower 
redshift functions {\em (open triangles and open diamonds)\/} 
match each other and our maximum likelihood fit computed at 
$z=1$ {\em (dashed curve)\/}
throughout the luminosity range, while the highest redshift 
function {\em (solid squares)\/} matches the lower redshift 
functions and our maximum likelihood fit only at the bright end,
where the optical QSO determinations are made. It is clear that there 
are fewer intermediate luminosity sources in the highest redshift 
interval, and hence that the pure luminosity evolution 
model cannot in fact be carried 
reliably to the higher redshifts.
This is another statement that the lower luminosity sources
peak in number at lower redshifts than the high-luminosity
sources (e.g., Cowie et al.\ 2003;
Hasinger 2003; Barger et al.\ 2003a; Steffen et al.\ 2003;
Fiore et al.\ 2003; Ueda et al.\ 2003).

%
%

%
%
\begin{inlinefigure}
\psfig{figure=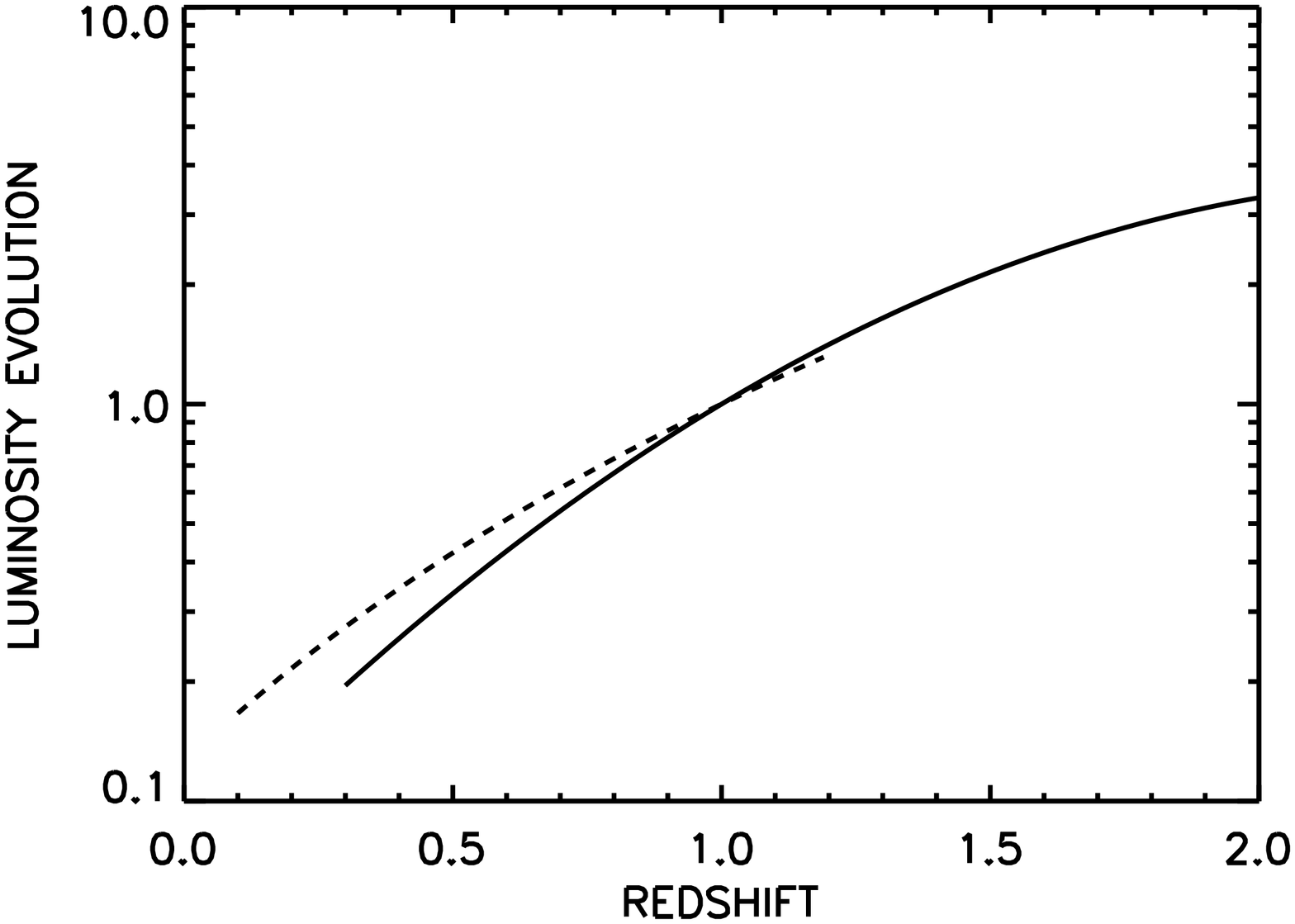,width=3.5in,angle=90}
\figurenum{23}
\caption{
Evolution with redshift of the normalizing luminosity in the
pure luminosity evolution
models for the broad-line AGN hard X-ray luminosity function
{\em (dashed line)\/} and the optical QSO luminosity function
{\em (solid line)\/}. Both have been normalized to unity at $z=1$
and are shown only over the fitted ranges.
\label{lumevol_comp}
}
\addtolength{\baselineskip}{10pt}
\end{inlinefigure}

%
%
\begin{inlinefigure}
\psfig{figure=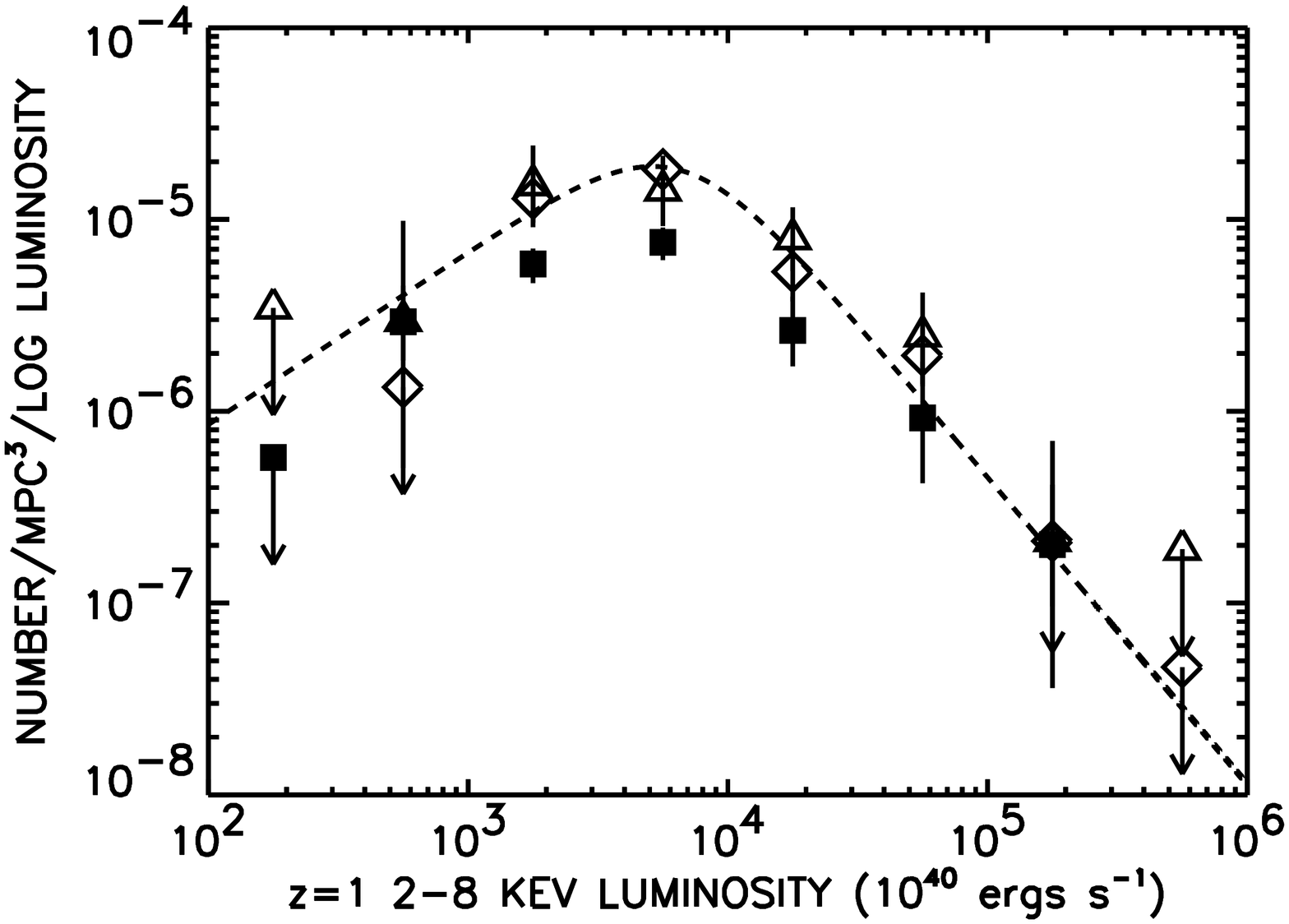,width=3.5in,angle=90}
\figurenum{24}
\caption{
Broad-line AGN hard X-ray luminosity functions
({\em open triangles\/}---$z=0.2-0.7$;
{\em open diamonds\/}---$z=0.7-1.5$;
{\em solid squares\/}---$z=1.5-2.5$) with luminosities
normalized to the values at $z=1$ using the pure luminosity evolution
model of Croom et al.\ (2004). Dashed curve shows the maximum
likelihood fit to the $z=0-1.2$ hard X-ray luminosity function
for broad-line AGNs (see \S6) computed at $z=1$.
\label{delum_comp}
}
\addtolength{\baselineskip}{10pt}
\end{inlinefigure}

\section{Hard X-ray Energy Density Production Rate}
\label{secenergy}

In Figure~\ref{figevol}, we show the evolution with
redshift of the $2-8$~keV comoving energy density production
rate, ${\dot\lambda_X}$, of all spectroscopically identified
$L_X\ge 10^{42}$~ergs~s$^{-1}$ AGNs {\em (solid squares)\/}
and broad-line AGNs alone {\em (open diamonds)\/}.
The production rate rises rapidly from $z=0$ to $z=1$.
The solid and dashed curves show the $(1+z)^{3.2}$ and
$(1+z)^{3.0}$ laws found for all spectral types and for
broad-line AGNs alone, respectively, in \S\ref{secevol}. 
There is no reason why the
evolution should be well fitted by a power of $(1+z)$, and, 
indeed, at $z>1$, the evolution is turning over in redshift, 
at least in part due to incompleteness. If we had instead
fitted over a lower redshift range (say $z=0$ to $z=1$),
then we would have obtained a slightly steeper relation.

Such an extremely rapid evolution with 
redshift bears a striking resemblance to the overall redshift 
evolution of the star formation rate density in $z<1$ galaxies.
The UV luminosity density in small star-forming galaxies
is falling as $(1+z)^{1.7\pm 1.0}$ 
(Wilson et al.\ 2002; note that there
have been a whole slew of relations determined, including
some that are steeper). However, AGNs are more likely to be
related to the bulge luminosity of galaxies, and hence
to the submillimeter and ultraluminous infrared galaxy
$(L_{FIR}>10^{12}~{\rm L}_\odot)$ populations. 
The evolution of the star-forming 
luminosity function at $z=0.5-1.5$, as determined from radio 
sources with the equivalent radio power of at least
$L_{FIR}=10^{12}~{\rm L}_\odot$, is steeper and reasonably well 
described by a $(1+z)^{3.8}$ evolution
(Cowie et al.\ 2004a).

%
%
\begin{inlinefigure}
\psfig{figure=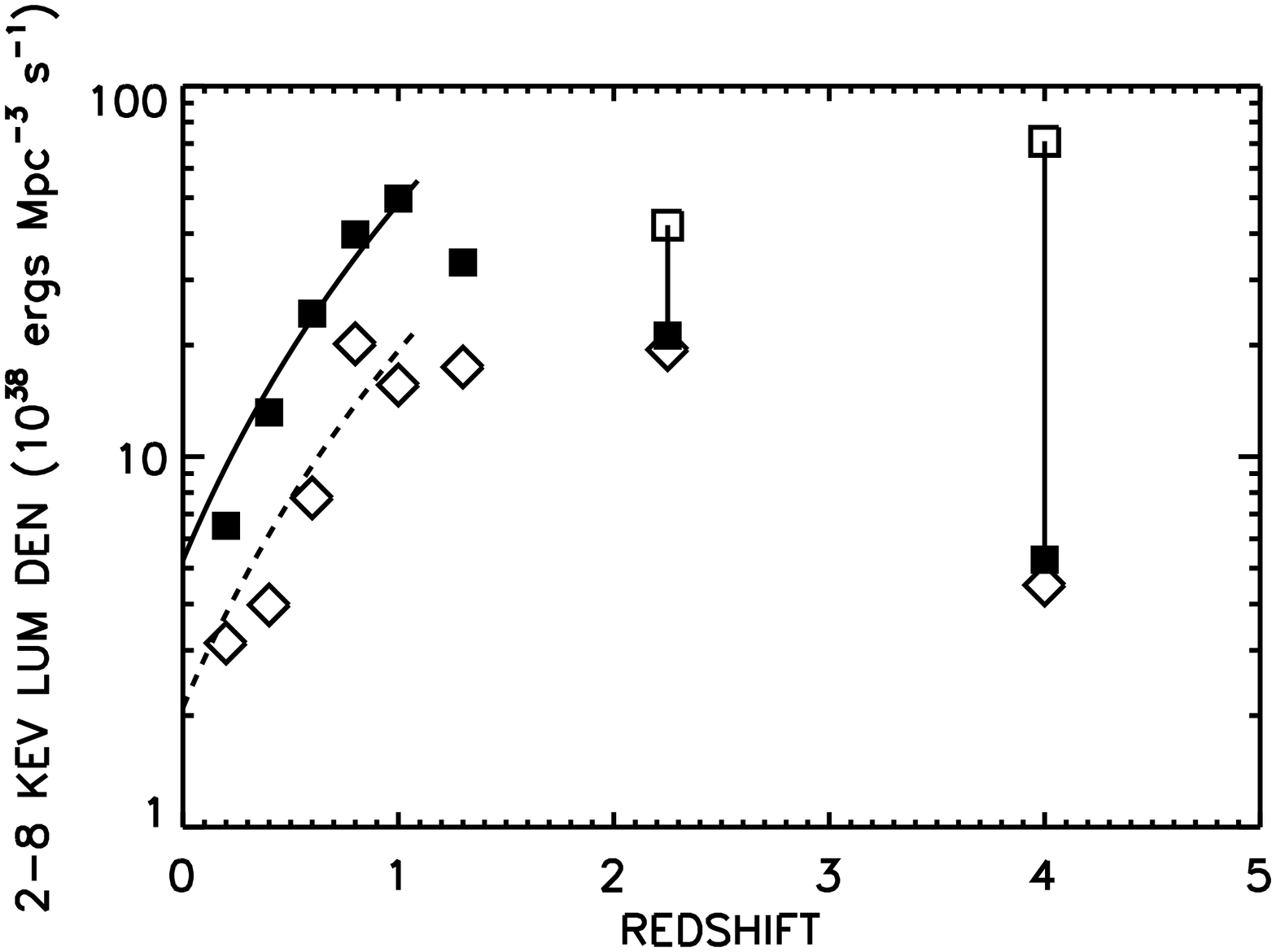,width=3.5in,angle=90}
\figurenum{25}
\caption{
Evolution with redshift of the rest-frame $2-8$~keV
comoving energy density production rate, ${\dot\lambda_X}$,
of all spectroscopically identified $L_X\ge 10^{42}$~ergs~s$^{-1}$
AGNs {\em (solid squares)\/}. Open diamonds show the values for
broad-line AGNs. Open squares show the maximal incompleteness found by
assigning redshifts at the center of each high-redshift interval
to all of the spectroscopically unidentified sources in the
spectroscopically observed samples. Maximal incompleteness values
are not consistent with one another, because all of the unidentified
sources are included in both redshift intervals, provided that
the sources, when placed at those redshifts, lie at
$L_X\ge 10^{42}$~ergs~s$^{-1}$. Curves show the pure luminosity
evolution maximum likelihood fits for all spectral types {\em (solid)\/}
and for broad-line AGNs alone {\em (dashed)\/} over the range $z=0-1.2$,
where $L_\ast=(1+z)^{3.2}$ and $L_\ast=(1+z)^{3.0}$, respectively.
\label{figevol}
}
\addtolength{\baselineskip}{10pt}
\end{inlinefigure}

The peak of the hard X-ray energy density production rate of all
spectroscopically identified $L_X\ge 10^{42}$~ergs~s$^{-1}$
AGNs is in the interval $z=0.8-1.2$ and is about
$4.6\times 10^{39}$~ergs~s$^{-1}$~Mpc$^{-3}$.
About one-third of this production arises in broad-line AGNs and
about two-thirds in optically-narrow AGNs.
Half of this production arises in
sources more luminous than $8.2\times 10^{43}$~ergs~s$^{-1}$,
where the majority of the sources are broad-line AGNs.

The above results are based on direct summation over the
individual sources, but integration of the power-law fits 
to the hard X-ray luminosity functions give similar answers.
The largest uncertainty is the redshift distribution of the
spectroscopically unidentified sources, not the small effects 
of extrapolation outside of the observed luminosity range.

In Figure~\ref{figevol}, we assume that most of the 
spectroscopically unidentified sources in the spectroscopically
observed samples lie at $z>1.5$ (we examine this in more detail 
below). We then determine the maximal incompleteness for the 
two high-redshift intervals {\em (open squares)\/} by using the 
sources that have a measured spectroscopic redshift in these intervals 
and by assigning redshifts at the center of each interval to all 
of the spectroscopically unidentified sources in the spectroscopically 
observed samples. Note that the maximal incompleteness values are 
not consistent with one another, because all of the unidentified
sources are included in both redshift intervals, provided that
the sources, when placed at those redshifts, lie at 
$L_X\ge 10^{42}$~ergs~s$^{-1}$.
Even with the incompleteness uncertainty,
${\dot\lambda_X}$ is at most flat beyond $z=1$.

To explore more carefully the incompleteness at all redshifts,
in Figure~\ref{figlightden}, we show the production rate
for the spectroscopic samples in the CDF-N, CDF-S, and {\em ASCA\/} 
fields only {\em (solid squares)\/}, and, at $z<1.5$, the production 
rate for the spectroscopic plus photometric samples in these fields 
{\em (triangles)\/}. We see that the only substantial difference 
at $z<1.5$ is in the $z=1.1-1.5$ redshift interval. 
At $z=1$, the symbols are a bit higher than the solid
square was in Figure~\ref{figevol} (the solid curve from 
Figure~\ref{figevol} has been replotted in Figure~\ref{figlightden}
to illustrate this) because of the excess in the CDF-N at that 
redshift, which gets smoothed out when the CLASXS data are 
included. We compute the maximal incompleteness for the
two high-redshift intervals {\em (open squares)\/} by using the 
sources with either a spectroscopic or photometric redshift
in these intervals and by assigning redshifts at the center of 
each interval to all of the remaining unidentified sources 
in the CDF-N, CDF-S, and {\em ASCA\/} fields. The inclusion of 
the photometric data does not change the conclusion that at $z>1$,
${\dot\lambda_X}$ is flat or, more realistically, slightly falling.

We can go one step further in studying the evolution of the energy 
density production rate. In Figure~\ref{figevolblagn}, we
separate out the optically-narrow AGNs {\em (solid circles)\/},
incompleteness correct their production rates (see below),
and then compare the evolution of their corrected production 
rates with the evolution of the production rates of the broad-line 
AGNs ({\em open diamonds\/}; values from Figure~\ref{figevol}). 
Over the redshift range $z=0-1.5$, the incompleteness corrections
were obtained by multiplying the spectroscopically identified 
optically-narrow AGNs by the ratios of the spectroscopic plus photometric 
values in Figure~\ref{figlightden} (the triangles in that figure) 
to the spectroscopic values in Figure~\ref{figlightden} (the solid 
squares in that figure). 
At $z<1.5$, the only place where the incompleteness corrections 
are significant is in the redshift interval $z=1.1-1.5$. 

For $z=1.5-3$, we cannot incompleteness correct
so simply. Instead, we compute the possible range of values
({\em vertical bar\/}) by subtracting the value 
given in Figure~\ref{figevol} for broad-line AGNs (the diamond at $z=1.5-3$)
from the value given in Figure~\ref{figevol} for all spectral types
(the solid square at $z=1.5-3$) to get
the lower value {\em (solid circle)\/}, and by subtracting
the same value for broad-line AGNs from the value given in Figure~\ref{figevol} 
for the maximal incompleteness (the open square at $z=1.5-3$)
to get the upper value {\em (open circle)\/}.

From Figure~\ref{figevolblagn}, we conclude that at $z<1$, most of 
the hard X-ray energy density production is coming from optically-narrow
AGNs, while at higher redshifts, this reverses, and the broad-line AGNs
dominate the production, even with the maximal incompleteness.
The broad-line AGN production is fairly well represented by
a constant at higher redshifts (see discussion in \S\ref{secacc}), 
as illustrated by the dashed curve at $z>1$ in the figure.
(At $z<1$, the dashed curve shows the pure luminosity evolution 
maximum likelihood fit for broad-line AGNs over the 
redshift range $z=0-1.2$ from Figure~\ref{figevol}.)

%
%
\begin{inlinefigure}
\psfig{figure=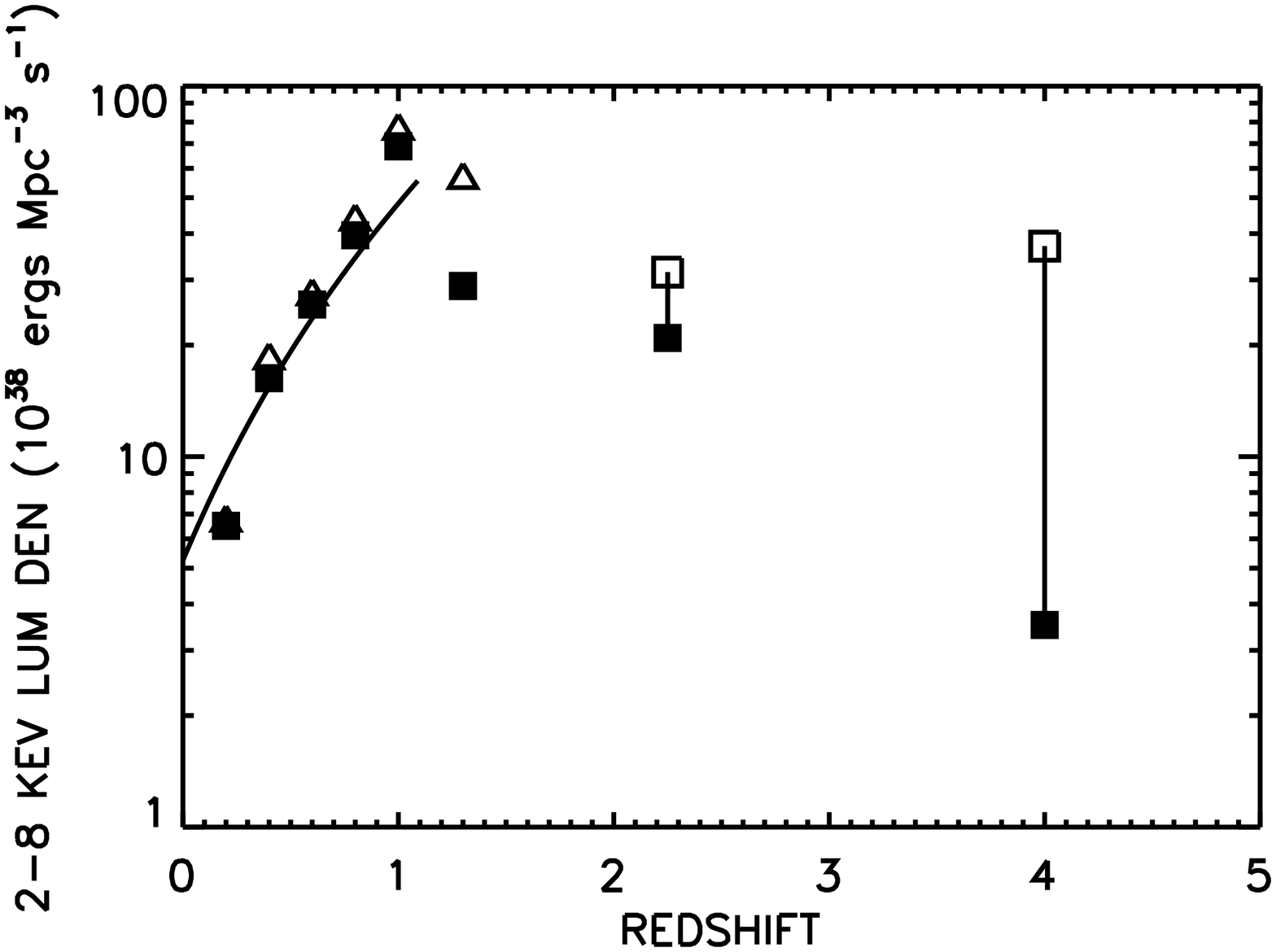,width=3.5in,angle=90}
\figurenum{26}
\caption{
Evolution with redshift of the rest-frame $2-8$~keV
comoving energy density production rate, ${\dot\lambda_X}$,
of $L_X\ge 10^{42}$~ergs~s$^{-1}$ AGNs. Solid squares show the
measured values for the spectroscopically identified
samples in the CDF-N, CDF-S, and {\em ASCA\/} fields only.
Open triangles at $z<1.5$ show the measured values for the
spectroscopic plus photometric
samples in the three fields. Open squares show the maximal
incompleteness found by using the sources with either a
measured spectroscopic or photometric redshift in the two high-redshift
intervals and by assigning redshifts at the center of each
interval to all of the remaining unidentified sources
in the CDF-N, CDF-S, and {\em ASCA\/} fields.
Maximal incompleteness values are not
consistent with one another, because all of the unidentified
sources are included in both redshift intervals, provided that
the sources, when placed at those redshifts, lie at
$L_X\ge 10^{42}$~ergs~s$^{-1}$. Solid curve shows the
pure luminosity evolution
maximum likelihood fit for all spectral types over the range
$z=0-1.2$, where $L_\ast=(1+z)^{3.2}$.
\label{figlightden}
}
\addtolength{\baselineskip}{10pt}
\end{inlinefigure}

%
%
\begin{inlinefigure}
\psfig{figure=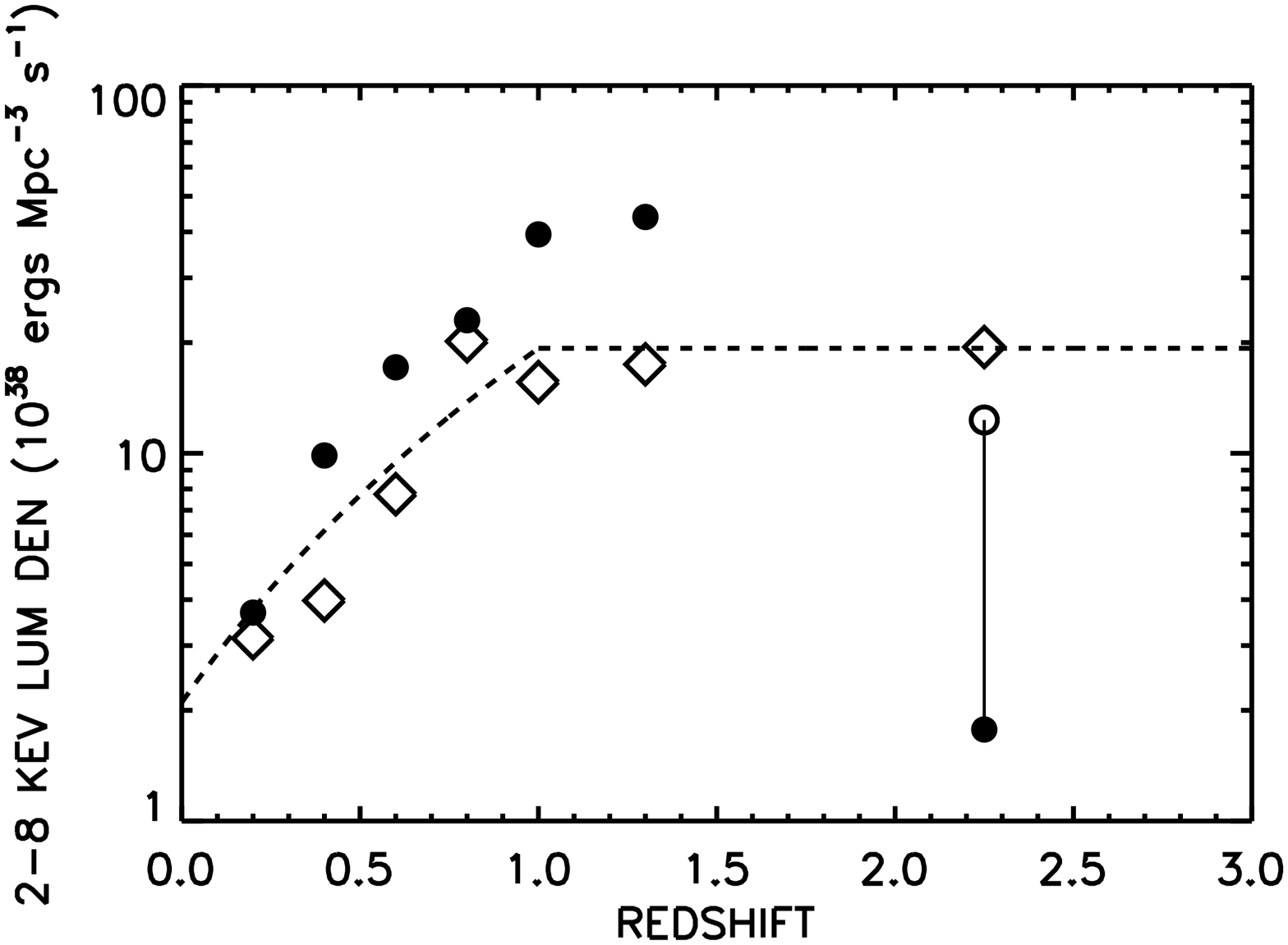,width=3.5in,angle=90}
\figurenum{27}
\caption{
Incompleteness corrected evolution with redshift
of the rest-frame $2-8$~keV comoving energy density production
rate, ${\dot\lambda_X}$, of $L_X\ge 10^{42}$~ergs~s$^{-1}$
optically-narrow AGNs {\em (solid circles)\/}. Open diamonds
show the evolution of the broad-line AGNs from Figure~25.
Vertical bar in the $z=1.5-3$
redshift interval shows the range from the spectroscopically
measured value for the optically-narrow AGNs {\em (solid circle)\/}
to the maximally incompleteness corrected value ({\em open circle\/};
see text for details). Dashed curve shows the pure luminosity
evolution maximum likelihood fit for broad-line AGNs
over the range $z=0-1.2$ (dashed curve from Figure~25)
and a flat line at $z>1$.
\label{figevolblagn}
}
\addtolength{\baselineskip}{10pt}
\end{inlinefigure}

\section{Downsizing of the Host Galaxies and Black Hole Masses}
\label{secsmbh}

The rapid drop in the luminosities of the individual black holes
(both broad-line AGNs and optically-narrow AGNs)
with decreasing redshift between $z=1$ and $z=0$,
as well as the drop in the 
overall energy density production by the AGNs, could arise in two ways. 
First, the accretion rate of all supermassive black holes, regardless 
of mass, could be declining. We refer to this as ``mass starvation''.
Second, the more massive black holes could be being
preferentially starved, while the less massive black holes continue 
to accrete. This would parallel the behavior of star formation,
which is dominated by lower mass sources at lower redshifts.
Following the naming introduced by Cowie et al.\ (1996) 
for that process, we refer to this possibility as ``AGN downsizing''.

For the broad-line AGNs, we can use the line widths and nuclear optical
luminosities to estimate supermassive black hole masses
(Wandel, Peterson, \& Malkan 1999;
Kaspi et al.\ 2000). Given the redshifts of the
sources, we only have a substantial sample of uniform line widths
for the MgII 2800~\AA\ line (measured from sources in the ACS 
GOODS-North region of the CDF-N). We use the 
McLure \& Dunlop (2004) calibration
based on this line and the rest-frame 3000~\AA\
luminosity of the nucleus to estimate the black hole masses for those
sources with MgII line widths greater than 2000~km~s$^{-1}$.
We note, however, that the sample is small (7 sources between 
redshifts of $z=0.5$ and 1.5), 
and the simple fitting technique that we use to measure the line 
widths may result in some contamination by the 
FeII line complex (see McLure \& Dunlop 2004). 
There are also many uncertainties in the method, so the presently 
derived values should only be considered as a rough measure
(accurate to factors of a few) of the black hole masses.

%
%
\begin{inlinefigure}
\psfig{figure=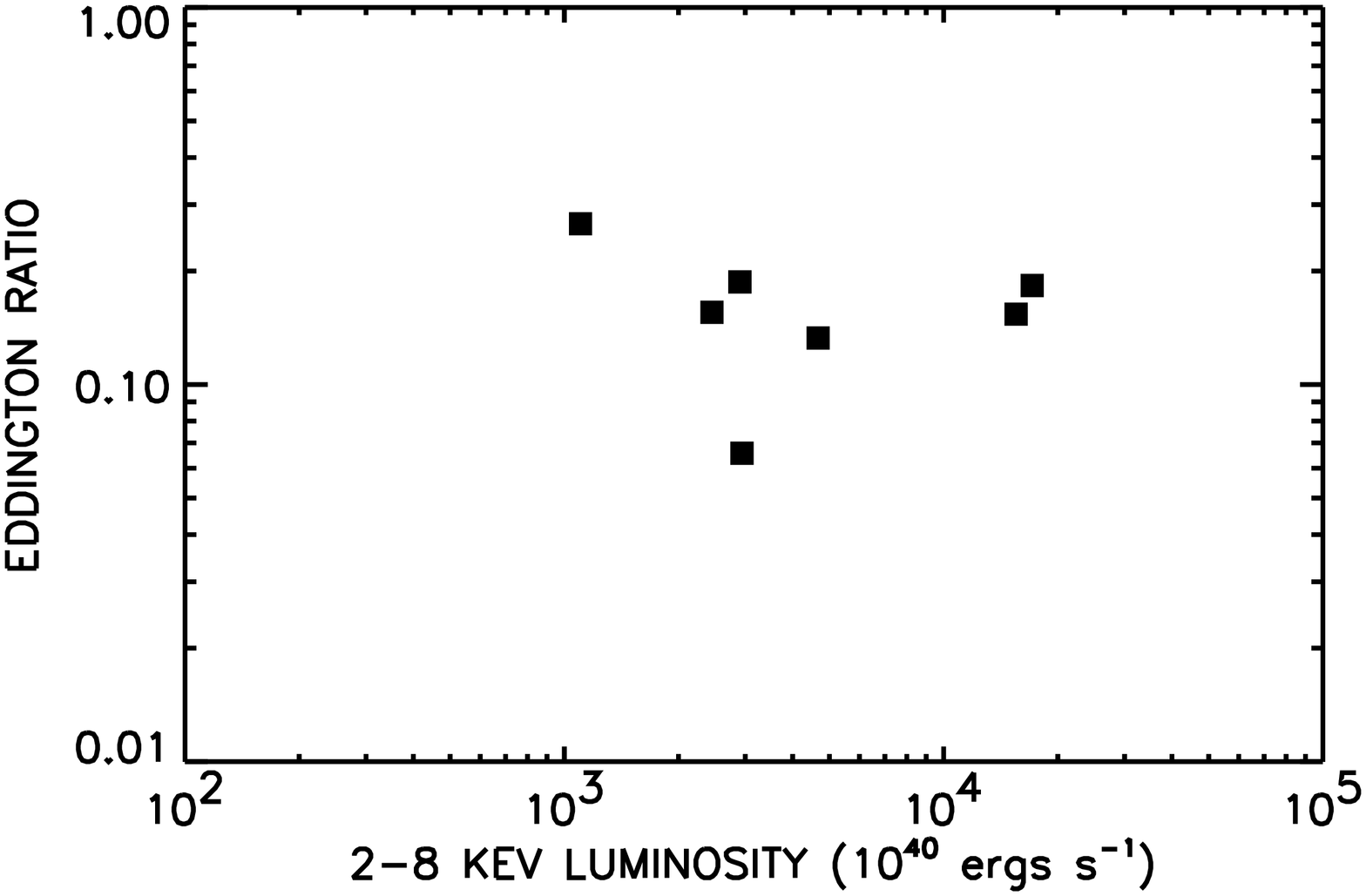,width=8.5cm,angle=90}
\figurenum{28}
\caption{
Ratio of bolometric luminosity ($L_{BOL}=35\times L_X$)
to Eddington luminosity vs. $L_X$ for the 7 sources with
measured ${\rm FWHM}>2000$~km~s$^{-1}$ in the MgII 2800~\AA\ line.
The sources lie between $z=0.5$ and $z=1.5$.
\label{edd_plot}
}
\addtolength{\baselineskip}{10pt}
\end{inlinefigure} 

In Figure~\ref{edd_plot}, we show the ratios of the bolometric
luminosities ($L_{BOL}=35\times L_X$; the bolometric correction
is from Elvis et al.\ 1994 and
Kuraszkiewicz et al.\ 2003 for broad-line AGNs) 
to the Eddington luminosities that correspond to the estimated
black hole masses. Over nearly two orders of magnitude in $L_X$, 
there is no obvious change in these Eddington ratios, 
with individual values varying by less than 
a factor of two about a median value of 0.16. 
(We note, however, that the X-ray selection means that we are 
only seeing the high-end values of the Eddington ratio; e.g., 
see Alonso-Herrero et al.\ 2002 and
Merloni et al.\ 2003.) This is partly 
a consequence of the tight relation between $L_X$ and $L_{OPT}$
(see Woo \& Urry 2002 for a discussion), 
as $L_{OPT}$ enters as roughly the square root in the black hole 
mass calculation (e.g., Wandel et al.\ 1999; 
McLure \& Dunlop 2004),
but the relation is much tighter than would be expected solely 
from this effect. Indeed, the Eddington ratio
depends on the line width squared, and the line widths of the
sources vary by nearly a factor of four. This suggests that at
$z=1$, there is a fairly tight correlation between $L_X$ and
black hole mass,
\begin{equation}
M_\bullet = 1.7\times 10^8 (L_X/10^{44}~{\rm ergs~s}^{-1}) {\rm M}_\odot \,.
\label{eq4}
\end{equation}

For a source accreting at the median value of $16\%$ of the 
Eddington luminosity measured for the seven sources, the 
mass doubling time 
is about $3\times 10^{8}$ years. (Note that the mass growth rate 
of a black hole accreting at the Eddington rate is exponential with 
an e-folding timescale of $t=4.5\times 10^7$~yr, the Salpeter time;
see Natarajan 2004 for a summary of the
basic definitions used in the accretion paradigm.)
Since a few percent of luminous galaxies are active at any
time (Barger et al.\ 2001a, 2003b;
Cowie et al.\ 2004b), we would expect
that the whole population of supermassive black holes would have significantly 
grown over timescales similar to that of the Hubble time at 
$z=1$. This result is consistent with the ongoing formation 
of the supermassive black hole population at $z=1$ that we found in \S\ref{secenergy}.

In Figure~\ref{galevol}, we show for the sources in the ACS GOODS-North 
region of the CDF-N the host galaxy luminosities (excluding the nuclear 
contributions) computed at rest-frame 5000~\AA\ versus redshift. 
The luminosities are shown for three intervals in $L_X$
({\em triangles\/}---$L_X>10^{44}$~ergs~s$^{-1}$;
{\em diamonds\/}---$L_X=10^{43}-10^{44}$~ergs~s$^{-1}$;
{\em squares\/}---$L_X=10^{42}-10^{43}$~ergs~s$^{-1}$). 
We chose rest-frame 5000~\AA\ because it is the largest rest 
wavelength that still corresponds to an observed optical
wavelength at $z<1$. We wanted to choose the reddest
rest-frame wavelength possible to give the best representation 
of the mass and to minimize the susceptibility to evolution
in the star formation.

From the figure, we can see that the most luminous X-ray sources 
lie in the most optically-luminous host galaxies. The median absolute 
magnitudes in the three X-ray luminosity ranges listed above
are $-22.8$, $-22.0$, and $-21.1$, respectively, while the median 
X-ray luminosities are $1.2\times 10^{44}$~ergs~s$^{-1}$,
$2.1\times 10^{43}$~ergs~s$^{-1}$, and
$3.4\times 10^{42}$~ergs~s$^{-1}$. 
This seems to be in accord with what we would expect from 
the well-known bulge luminosity--black hole mass relation 
(Kormendy \& Richstone 1995; Magorrian et al.\ 1998); that is,
the highest mass supermassive black holes (the accretion onto which, as we found
from our sample of seven, produces the highest X-ray luminosities) 
should lie in the most optically-luminous galaxies.

More importantly for the present discussion, the 
host galaxy optical luminosities for a given interval in $L_X$ 
do not appear to be changing much with redshift (i.e., the
X-ray sources at low redshifts are occurring in the same
host galaxies as they are at high redshifts), and the most 
optically-luminous host galaxies that existed at high redshifts 
have completely disappeared at low redshifts.
If the accretion rates of all supermassive black holes, regardless of mass, 
were just declining, then we would see a different picture:
the host galaxy optical luminosities for a given interval in 
$L_X$ would be higher at low redshifts than at high 
redshifts, and the most optically-luminous host galaxies 
would still contain the most X-ray luminous sources, even though 
the $L_X$ values of these sources would be lower. The most
optically-luminous host galaxies would not just cease to exist 
as X-ray sources. 

%
%
\begin{inlinefigure}
\psfig{figure=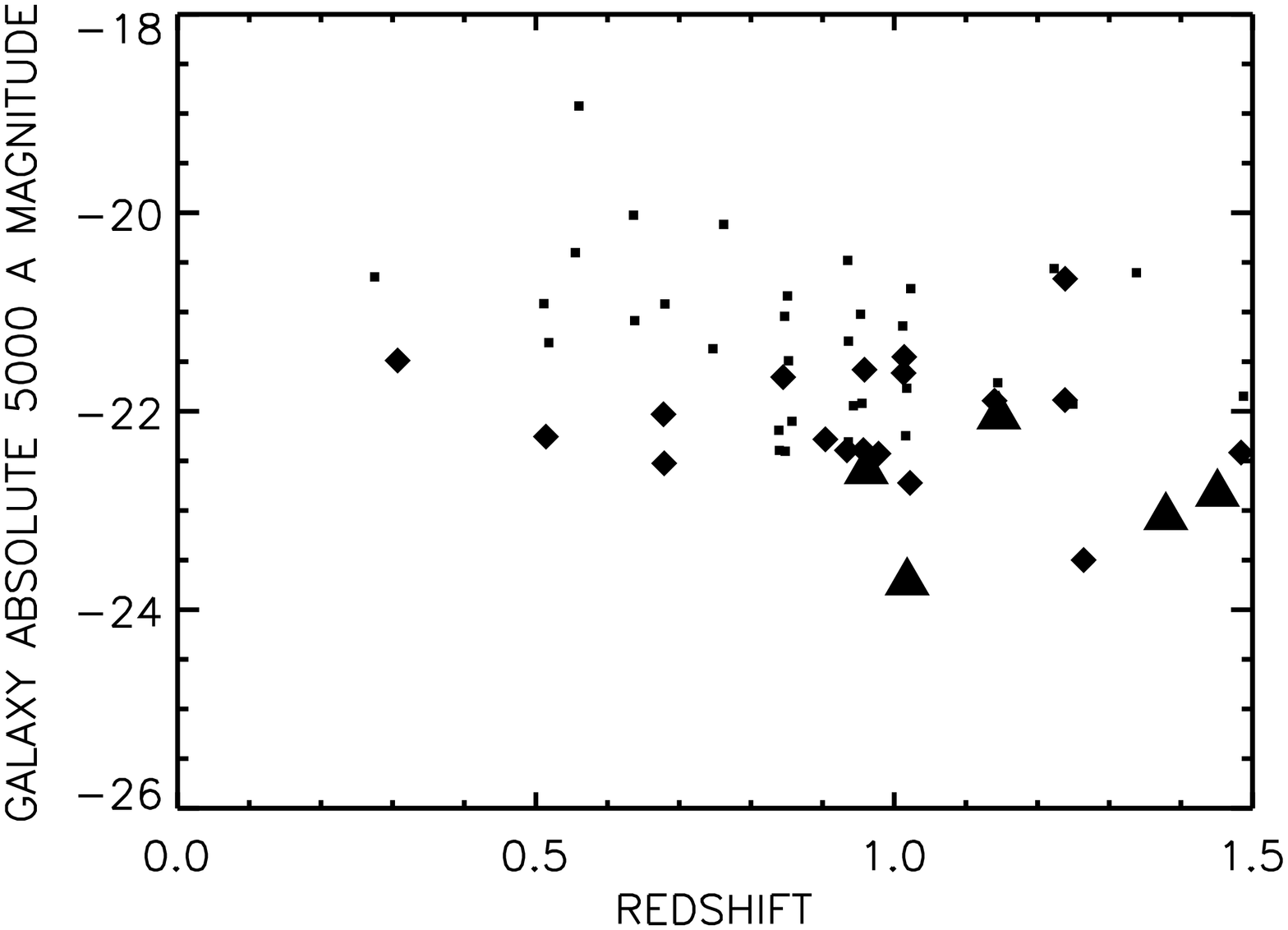,width=8.5cm,angle=90}
\figurenum{29}
\caption{
Host galaxy luminosity at rest-frame 5000~\AA\ vs. redshift
for sources in the ACS GOODS-North region of the CDF-N shown
for three X-ray luminosity intervals
({\em triangles\/}---$L_X>10^{44}$~ergs~s$^{-1}$;
{\em diamonds\/}---$L_X=10^{43}-10^{44}$~ergs~s$^{-1}$;
{\em squares\/}---$L_X=10^{42}-10^{43}$~ergs~s$^{-1}$). 
\label{galevol}
}
\addtolength{\baselineskip}{10pt}
\end{inlinefigure}

Thus, Figure~\ref{galevol} suggests that 
the loss of the highest $L_X$ sources is indeed an AGN downsizing 
effect. The highest mass supermassive black holes in the most 
optically-luminous host galaxies are switching off between $z=1$ 
and the present time, leaving only the lower mass supermassive 
black holes in the less optically-luminous 
galaxies active. This is consistent with the work of
Heckman et al.\ (2004), who found from
their analysis of the Sloan Digital Sky Survey data that most 
of the present-day accretion is occurring onto low-mass black 
holes ($M_\bullet<{\rm few}\times 10^7~{\rm M}_\odot$), with
massive black holes currently experiencing very little
additional growth.

\section{Failure of the Simple Unified Model}
\label{secfailure}

One possible way to account for at least part of the Steffen effect,
i.e., the absence of broad-line AGNs at low X-ray luminosities, 
would be if the nuclear optical light of the optically-narrow AGNs
were being swamped by the host galaxy light, 
thereby washing out the broad-line signal
(Moran, Filippenko, \& Chornock 2002; Moran 2004).
In the present spectroscopic obsertvations, this effect would be 
less pronounced at lower redshifts, where the 
nuclear light is a larger part of the light entering the spectrograph 
slit than the extended galaxy.
In this section, we explore the possibility of galaxy dilution
by looking at the nuclear UV/optical properties of the 
{\em Chandra\/} X-ray sources and determining how these properties 
relate to the spectral characteristics of the sources.
If the dilution hypothesis is correct, then we would expect the 
optically-narrow AGNs to have the same optical to X-ray ratios
as the broad-line AGNs when we isolate their nuclear UV/optical 
light.

In \S\ref{secnuc}, we found that the nuclear dominated sources 
are mostly observed as broad-line AGNs, while the galaxy dominated sources 
are not. At first glance, this result could be seen as consistent 
with the dilution hypothesis. However, if we look more closely
at the nuclear UV/optical magnitudes of the sources and compare
them with the X-ray fluxes, a different picture emerges.

It is well known (e.g., Zamorani et al.\ 1981)
that there is a strong correlation between the nuclear UV
magnitudes and the X-ray fluxes for broad-line AGNs. We can see this
correlation for the broad-line AGNs in our sample in Figure~\ref{figfs_nuc}. 
However, the correlation does not continue to hold for the 
narrow-line AGNs. In fact, from Figure~\ref{figfs_nuc}, we see that 
the nuclei of the narrow-line AGNs are much weaker relative
to their X-ray light than would be expected
if they were similar to the broad-line AGNs.

%
%
\begin{inlinefigure}
\psfig{figure=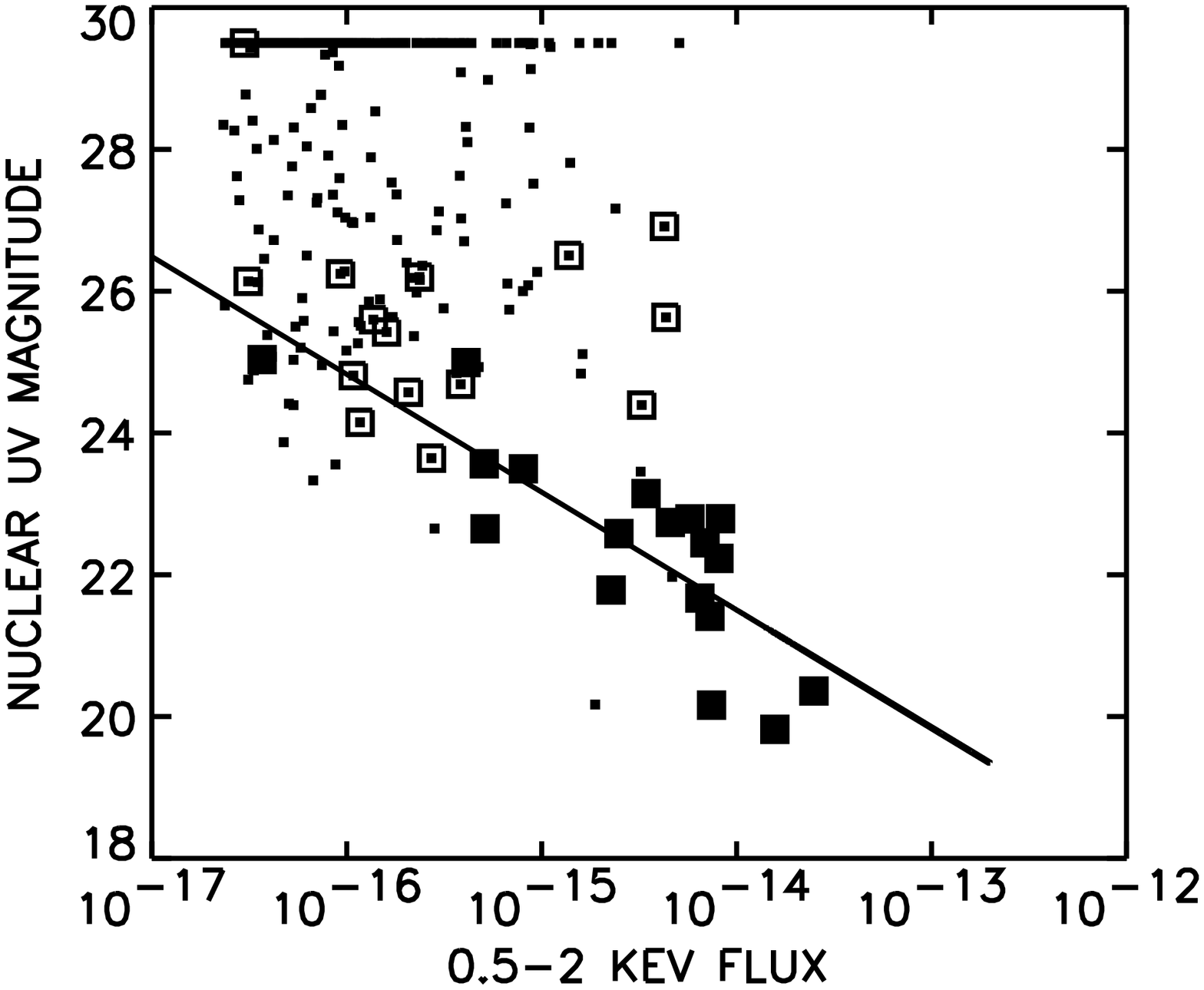,width=3.5in,angle=90}
\figurenum{30}
\caption{
Nuclear UV magnitude vs. $0.5-2$~keV flux, showing that
there is a strong correlation between the two for broad-line AGNs.
Broad-line AGNs (optically-narrow AGNs) are denoted by large
(small) squares. High-excitation sources are enclosed in large
open squares.
Sources with no measurable nuclei (all but one of these are
spectrally classified as normal galaxies) are shown at a nominal
magnitude of 29.5, which is roughly the $2\sigma$ limit for the
small $0.3''$ diameter aperture used to measure the
nuclear magnitude. Solid line shows a linear relation fitted to
the broad-line AGNs.
\label{figfs_nuc}
}
\addtolength{\baselineskip}{10pt}
\end{inlinefigure}

In Figure~\ref{nuc_flux_z1}, we plot nuclear rest-frame
3000~\AA\ flux versus rest-frame $2-8$~keV flux. We
restrict to the redshift range $z=0.6-1.4$, where the
deconvolution procedure should be similar for all of the
sources to provide a highly uniform sample, but the result
holds at all redshifts. In general, the optically-narrow AGNs 
fall well below the relation for the broad-line AGNs, with the 
high-excitation sources lying closer to the relation, and the 
absorbers and star formers lying well below the relation, on average.

%
%
\begin{inlinefigure}
\psfig{figure=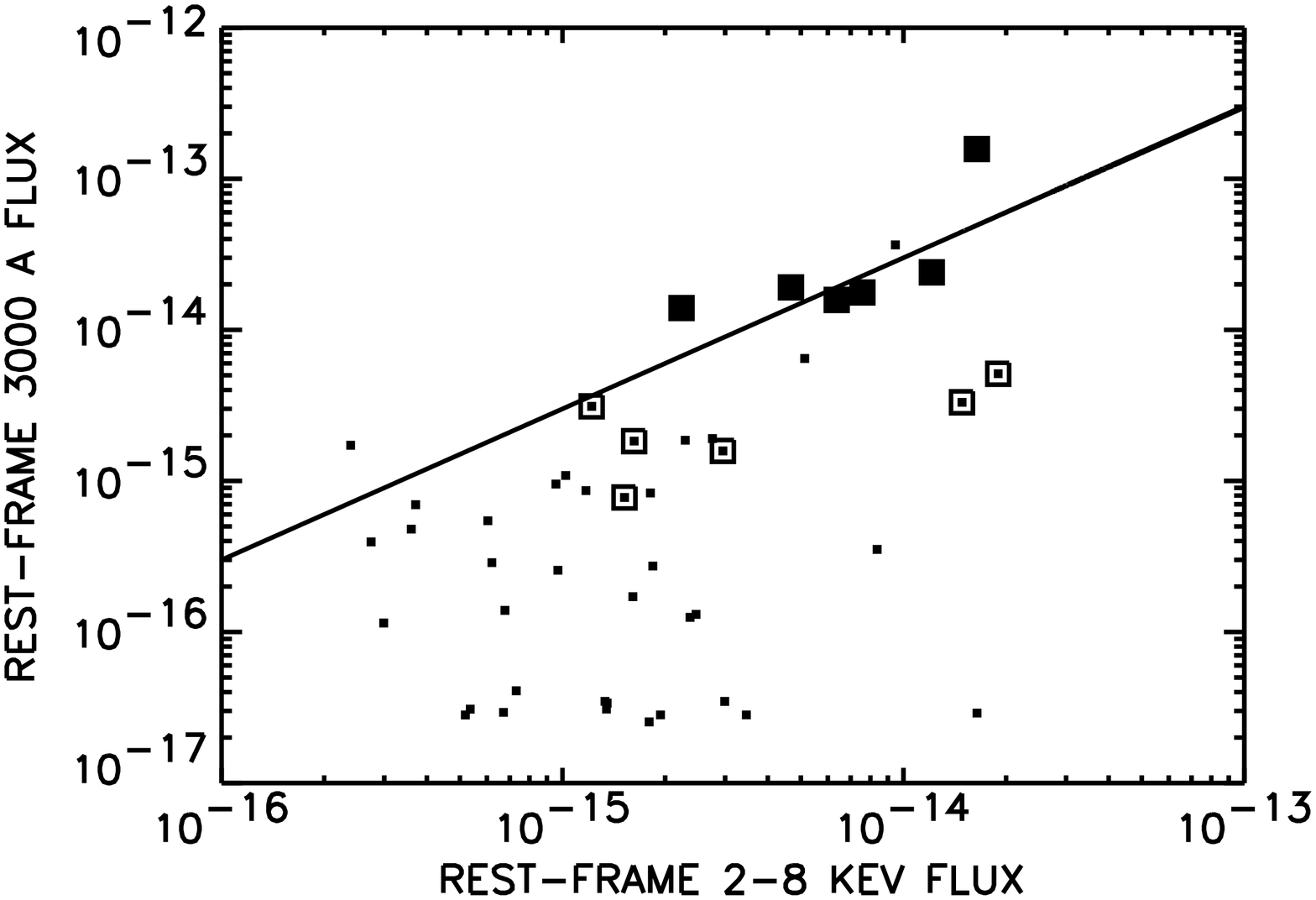,width=8.5cm,angle=90}
\figurenum{31}
\caption{
Rest-frame nuclear 3000~\AA\ flux computed by interpolating
between the GOODS bandpasses vs. rest-frame $2-8$~keV flux
for sources in the redshift range $z=0.6-1.4$ with
$L_X\ge 10^{42}$~ergs~s$^{-1}$. Broad-line AGNs (optically-narrow
AGNs) are denoted by large (small) squares.
High-excitation sources are enclosed in large open squares.
Solid line shows a linear relation normalized to the median
value of the 3000~\AA\ to $2-8$~keV ratio for the broad-line AGNs.
\label{nuc_flux_z1}
}
\addtolength{\baselineskip}{10pt}
\end{inlinefigure}

From the above results, we are led to conclude that 
the absence of broad-line AGNs at low X-ray luminosities is not a dilution 
effect, but rather that the optically-narrow AGNs really have 
weaker UV/optical nuclei relative to the X-rays. This leaves open the
question of whether the lower UV/optical fluxes are a consequence
of obscuration (as would be suggested by Figure~\ref{figcol} and
by the observational evidence supporting the unified
model; e.g., Antonucci 1993) or a
result of intrinsic differences in the source populations
(e.g., the optically-narrow AGNs intrinsically have less UV emission 
and no broad-line region or perhaps are at a different stage in their 
evolutionary sequence). 
At least having ruled out dilution as a partial explanation
for the Steffen effect, we can state firmly that the simple unified
model has failed because of the existence of the Steffen effect.

\section{The Luminosity Dependent Unified Model}
\label{seclum}

Probably the simplest interpretation of the Steffen 
effect is that the covering fraction of obscuration in the 
AGNs is extremely small at high X-ray luminosities (so all 
we see at these high luminosities are broad-line AGNs) and extremely 
large (near unity) at low X-ray luminosities (so we do not see
broad-line AGNs at these low luminosities). We shall hereafter 
refer to this as the luminosity dependent unified model.
If the luminosity depedent unified model is the correct 
explanation of the Steffen effect, then there are important 
implications for the interpretation of the spectral energy 
distributions of the X-ray sources (see \S\ref{sechxlum}).

In any unified scheme, where the absorption is due to the
orientation of the AGN relative to the line of sight, any FIR 
emission powered by the nucleus or by star formation will be 
emitted isotropically; thus, the FIR/submillimeter
properties of the absorbed sources should be the same as those of the
unabsorbed sources. (See Barvainas et al.\ (1995) for a nice example
that illustrates this point.) For the luminosity dependent
unified model, the FIR/submillimeter properties 
of the absorbed and unabsorbed sources should be the
same {\em at a given total bolometric luminosity}. We assume
that the hard X-ray luminosity is an approximate tracer of
the total bolometric luminosity for the present sources. Then 
if the luminosity dependent unified model holds, high X-ray 
luminosity sources, with their low covering
fractions, would not be expected to radiate as much in the FIR as 
low X-ray luminosity sources, with their high covering fractions.
This provides a good prediction that can be tested directly.

\subsection{High-Redshift Quasar Studies of the Luminosity
Dependent Unified Model}

One observational study may already argue against the
luminosity dependent unified model.
Page et al.\ (2001) measured 850$\mu$m fluxes for a sample of 
eight $z>1$ X-ray absorbed, radio-quiet quasars 
(with column densities of $21.4<\log N_H<22.5$ based on the
X-ray absorption; these sources
show little or no obscuration in their broad emission lines and 
ultraviolet continua) discovered serendipitously in 
archival X-ray data, while Page et al.\ (2004) 
measured 850$\mu$m fluxes for a comparison sample of 20 unabsorbed 
quasars. The latter sources were drawn from a combination of X-ray 
surveys and were matched in luminosity and redshift to the absorbed 
sample. Four of the eight absorbed sources were detected above 
5~mJy at $5\sigma$ (the other 
four were not detected at all), while none of the 20 unabsorbed
sources were detected above 5~mJy at $5\sigma$ (only one of the
unabsorbed sources was significantly detected at all, i.e., 
at $>3\sigma$).

From the level of segregation in the submillimeter properties
of their absorbed and unabsorbed samples,
Page et al.\ (2004) ruled out
orientation as the cause of the X-ray absorption. They proposed
that the two types of sources were at different stages in
an evolutionary sequence, with the absorbed quasars representing
an earlier phase. 

To put this on a more quantitative footing, we
used the published submillimeter fluxes from Page et al.\ (2001, 2004)
to calculate an error-weighted mean flux of $0.70\pm 0.20$~mJy for 
their unabsorbed sample at a mean soft X-ray luminosity of 
$9.5\times 10^{44}$~ergs~s$^{-1}$, and an error-weighted mean flux
of $4.50\pm 0.50$~mJy for their absorbed sample at a mean soft X-ray 
luminosity (corrected for instrinsic absorption) of 
$7.2\times 10^{44}$~ergs~s$^{-1}$. 
Both samples are significantly detected at $>3\sigma$. However,
in agreement with the Page et al.\ (2004) conclusions, the mean 
$850\mu$m flux of the absorbed AGNs is much larger (about a factor 
of 6) than that of the unabsorbed AGNs at roughly the same mean 
X-ray luminosity. This is not what would be expected in the
luminosity dependent unified model.
However, given the small sample size, larger hard 
X-ray--selected samples are needed to make the results definitive.

Other submillimeter studies have targeted large samples of 
optically-luminous quasars, and although it is harder to compare 
with these studies because of the lack of X-ray information,
Yuan et al.\ (1998) have shown that the great 
majority of optically-selected quasars have little or no X-ray absorption.
Thus, these optically-selected samples are most likely similar to the 
Page et al.\ (2001) unabsorbed sample in their X-ray properties, and 
we can try to make a comparison.

Priddey et al.\ (2003)
targeted a homogeneous sample of 57 radio-quiet, optically-luminous
quasars in the range $1.5<z<3.0$, detecting nine $7-17$~mJy
sources at greater than $3\sigma$ significance. Using
their published submillimeter fluxes, we calculate an
error-weighted mean flux of $3.22\pm 0.36$~mJy for the sample.
A previous, higher redshift ($z>4$) survey of 38 radio-quiet 
quasars by Isaak et al.\ (2002) detected eight 
$>10$~mJy sources at greater than $3\sigma$ sigificance. Again
using their published submillimeter fluxes, we calculate an 
error-weighted mean flux of $5.29\pm 0.45$~mJy for the
sample. In both of these studies, the mean 
$850\mu$m fluxes at X-ray luminosities that are probably fairly 
similar to, or perhaps somewhat higher than (based on the optical 
luminosities), those of the Page et al.\ (2001, 2004)
studies turn out to be similar to the mean $850\mu$m flux of the
Page et al.\ (2004) absorbed sample. This is what would be
expected in the luminosity dependent unified model. 
Thus, it may be premature to unambiguously conclude 
that the absorbed and unabsorbed sources at these high luminosities 
are intrinsically different populations.

\section{Determining Bolometric Corrections}
\label{sechow}

A major issue for mapping the accretion history of the universe
with redshift is knowing what bolometric corrections  
to use to correct $2-8$~keV 
luminosities to bolometric luminosities.
In the simple unified model, where both the covering fractions
and the intrinsic spectral energy distributions are invariant with luminosity (and redshift),
it is straightforward to determine the bolometric corrections:
since the FIR radiation is the absorbed UV through soft X-ray light 
reradiated isotropically, integrating a broad-line AGN spectral energy distribution, 
taking care to exclude the FIR to avoid double counting, would give
the bolometric correction. (Note that since the line of sight to the broad-line AGN would 
be unabsorbed, the observed light would account
for all of the radiation from the AGN.)
An alternative approach would be to take a large, homogeneous sample
that covered all possible orientations and sum the spectral energy distributions of 
all the sources in the sample, regardless of type, to create an average 
spectral energy distribution that could be integrated. This would then give the bolometric correction
of an unobscured source, which should have the same value as the
bolometric correction obtained from the broad-line AGN spectral energy distribution. 

Since we know the bolometric correction obtained by integrating a local broad-line AGN 
spectral energy distribution is approximately 35 (Elvis et al.\ 1994;
Kuraszkiewicz et al.\ 2003), if the simple
unified model were right, we would already have a determination 
of the invariant bolometric correction. However, we now know that the simple unified
model is not right, and we need to consider the possibility that 
the bolometric correction is not invariant and may in fact be spectral type
or X-ray luminosity dependent. 

In \S\ref{secenergy}, we found that
a large fraction of the low-redshift, hard X-ray energy density 
production is in optically-narrow AGNs, so we really need 
to determine the bolometric corrections for these sources if we 
are to map the 
accretion history with redshift. Given the fact that we do not 
know what the origin of the Steffen effect is (i.e., whether the
different spectral types are intrinsically different or whether 
the luminosity dependent unified model is correct), 
in \S\ref{secspectype} and \S\ref{sechxlum},
we construct the composite spectral energy distributions that 
are needed to calculate the 
bolometric corrections (see \S\ref{secbolcorrtype} 
and \S\ref{secbolcorrlum}) as a 
function of both spectral type and X-ray luminosity to cover
both possibilities. Either way, we need to include 
long-wavelength data. In \S\ref{seclongwave}, we describe the 
long-wavelength imaging data that we have available for our study.

\section{Long-Wavelength Imaging Data}
\label{seclongwave}

Any enhancement in the MIR/FIR radiation due to absorption and 
reradiation by gas and dust surrounding the central AGN should 
be directly observable at MIR/FIR wavelengths.
The observational situation in the MIR/FIR may be expected to
improve rapidly with the {\em Spitzer Space Telescope\/}
(e.g., Alonso-Herrero et al.\ 2004; Rigby et al.\ 2004;
Lonsdale et al.\ 2004; Dole et al.\ 2004).
However, we may already use observations with the ISOCAM 
and ISOPHOT instruments on the {\em ISO\/} satellite
and with the SCUBA bolometer array on the JCMT
to determine the MIR and FIR/submillimeter properties
versus optical spectral type.
There are deep ISOCAM observations of the CDF-N region
(Aussel et al.\ 1999), as well as
moderately deep SCUBA observations over much
of the CDF-N region (Wang, Cowie, \& Barger 2004).
The deepest ISOPHOT observations are of two Lockman Hole fields
(Kawara et al.\ 2004), one of which is the CLASXS field.

\subsection{Submillimeter Data}
\label{secsmm}

The submillimeter properties of the X-ray sample in the CDF-N
are discussed in Barger et al.\ (2001c)
and Alexander et al.\ (2003a) using
earlier, more restricted SCUBA observations, while analyses
on other fields are discussed in
Almaini et al.\ (2003) and
Waskett et al.\ (2003).
In all of these fields, only a small number of X-ray sources
are directly detected as submillimeter sources, and, given the
poor resolution of the submillimeter observations, some of
these may be chance projections. (A more detailed discussion may
be found in Wang et al.\ 2004, who
report that the fraction of $S_{850\mu{\rm m}}>6$~mJy and S/N$>4$
sources with X-ray counterparts in the CDF-N is $\sim 50$\%.
This fraction decreases to $\sim 35$\% once the number of random
overlaps has been taken into account.)

However, the error-weighted mean submillimeter fluxes of the 
ensemble of hard X-ray sources may be measured.
For the total ensemble of 61 $L_{X}\ge 10^{42}$~ergs~s$^{-1}$
sources lying in the submillimeter mapped region of the CDF-N
where the 850$\mu$m flux errors are less than 2.7~mJy,
we find an error-weighted mean 850$\mu$m flux of $0.77\pm 0.17$~mJy,
consistent with previous measurements.

If the optically-narrow AGNs are reprocessing their missing
UV/optical light to the MIR/FIR more so than the broad-line AGNs,
possibly due to these sources being at a dustier stage of the
evolutionary sequence than broad-line AGNs (if optically-narrow AGNs 
are fundamentally different than broad-line AGNs), or possibly
due to these, on average, lower $L_X$ sources having a higher 
covering factor (if the luminosity dependent unified model holds), 
then most of the above 
submillimeter signal may be expected to be from the optically-narrow
AGNs. Morever, because of the strong $K$-correction, which arises
from the steep spectral slope at submillimeter wavelengths, 
the signal may be expected to be much stronger in the high-redshift
sources, where the rest-frame corresponding to 850$\mu$m lies
much closer to the blackbody peak. We find that this is
marginally the case: for the 20 optically-narrow AGNs at $z>1$
that satisfy the conditions $L_{X}\ge 10^{42}$~ergs~s$^{-1}$ and 
850$\mu$m flux errors less than 2.7~mJy, we measure an error-weighted 
mean 850$\mu$m signal of $1.20\pm 0.34$~mJy.

In addition, since the optically-narrow AGNs are typically hard and 
the broad-line AGNs soft, most of the submillimeter signal may also be
expected to arise in sources with hard X-ray spectra. Again, this 
effect is marginally seen. For the 18 sources with $\Gamma > 1.4$ 
that satisfy the conditions $z>0.8$, $L_{X}\ge 10^{42}$~ergs~s$^{-1}$, 
and 850$\mu$m flux errors less than
2.7~mJy, the error-weighted mean flux is $0.21\pm 0.30$~mJy, while for
the 21 sources with $\Gamma < 1.4$ that satisfy the same conditions,
the error-weighted mean flux is $0.72\pm 0.26$~mJy.

\subsection{FIR Data}
\label{secfir}

At observed wavelengths of 90$\mu$m and 170$\mu$m, we can
compare the CLASXS X-ray sample of
Yang et al.\ (2004)
with the ISOPHOT catalog of Kawara et al.\ (2004).
Because $f_{\nu}$ is decreasing with increasing $\nu$ at these
wavelengths, sources fade with increasing redshift, and it is
the most nearby sources that will be detected.

We matched the X-ray sources to the $90\mu$m catalog of
Kawara et al.\ (2004) using a $35''$
search radius. Only the two brightest X-ray sources---that is,
the two $f_{2-8~{\rm keV}}>10^{-13}$~ergs~cm$^{-2}$~s$^{-1}$,
optically-narrow AGNs at redshifts $z=0.5159$ and $z=0.7221$---are
detected in the 90$\mu$m catalog.
The expected number of chance projections is small: randomized
samples show that there is an 8\% probability of a chance
projection for an individual source. Thus, it is likely
that both of these sources are correctly matched, despite the
poor resolution of the ISOPHOT data. Neither of these
sources is detected at 170$\mu$m. None of the
12 broad-line AGNs in
the field with $f_{2-8~{\rm keV}}>3\times 10^{-14}$~ergs~cm$^{-2}$~s$^{-1}$
is detected at either 90$\mu$m or 170$\mu$m, but this
result is not very constraining, given the sensitivity limits at
these wavelengths.

Figure~\ref{figbolo90} shows the $\nu f_{\nu}$ bolometric fluxes
at 90$\mu$m for the 23 CLASXS sources in the ISOPHOT field with
$f_{2-8~{\rm keV}}>3\times 10^{-14}$~ergs~cm$^{-2}$~s$^{-1}$
and $L_X\ge 10^{42}$~ergs~s$^{-1}$.
(We restrict our analysis to the higher X-ray flux sources, where
the FIR flux limits are more constraining.)
The 21 CLASXS sources that do not
have a match in the ISOPHOT catalog are shown at the $3\sigma$ upper
limit of 60~mJy for the field.
If the $f_{FIR}$ to $f_X$ ratio were similar to that of the two
brightest sources, then we would not expect to be able to detect
most of the remaining sources, as can be seen by the solid line.

%
%
\begin{inlinefigure}
\psfig{figure=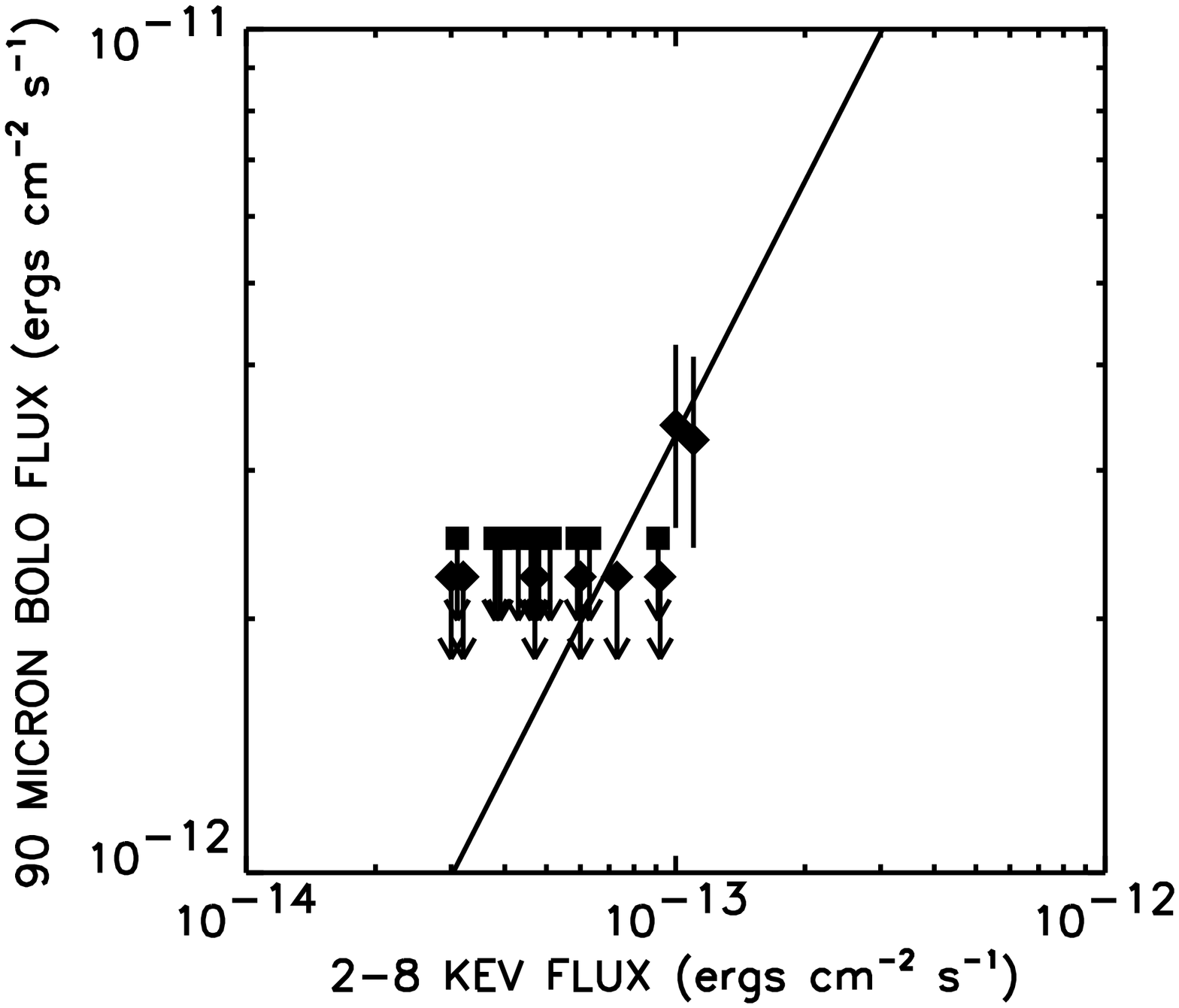,width=3.5in,angle=90}
\figurenum{32}
\caption{
$\nu f_{\nu}$ at $90\mu$m vs. $f_{2-8~{\rm keV}}$ flux
for the 23 CLASXS sources in the ISOPHOT field with
$f_{2-8~{\rm keV}}>3\times 10^{-14}$~ergs~cm$^{-2}$~s$^{-1}$
and $L_X\ge 10^{42}$~ergs~s$^{-1}$. Only the two brightest
X-ray sources have been detected at 90$\mu$m. All of the other
sources are denoted by an upper limit corresponding to the
$3\sigma$ threshold of 60~mJy.
Broad-line AGNs (optically-narrow AGNs) are denoted
by solid squares (solid diamonds). We have slightly displaced
the broad-line AGN upper limits from the optically-narrow AGN upper
limits to distinguish them. Solid line shows the relation
$f_{90\mu{\rm m}} = {\rm constant}\times f_{2-8~{\rm keV}}$
matched to the two detected sources.
\label{figbolo90}
}
\addtolength{\baselineskip}{10pt}
\end{inlinefigure}

\subsection{MIR Data}
\label{secmir}

At observed 15$\mu$m wavelengths, we
can compare the CDF-N X-ray sample of Alexander et al.\ (2003b)
with the ISOCAM catalog of Aussel et al.\ (1999;
see also Alexander et al.\ 2002).
There are six X-ray sources in the well-covered central region
of the ISOCAM image that have
$f_{2-8~{\rm keV}}> 10^{-15}$~ergs~cm$^{-2}$~s$^{-1}$
and $L_X\ge 10^{42}$~ergs~s$^{-1}$, where again we restrict
to the brightest X-ray sources, which are more likely to be detected.
All six are optically-narrow AGNs. Four of the six match to 15$\mu$m
sources in the Aussel et al.\ (1999) catalog using
a $2''$ search radius.  We have generated randomized samples
by placing the same number of sources at arbitrary positions within
this region and running the same matching procedure. Using a large
number of such samples, we find that there is only a 20\% probability
that any identification is a chance projection.

\section{Composite Spectral Energy Distributions by Spectral Type}
\label{secspectype}

For the broad-line AGNs, the strong linear correlation 
(see \S\ref{secfailure}) of the nuclear
rest-frame UV/optical fluxes with the rest-frame $2-8$ keV fluxes
enables us to determine the mean UV and 
optical spectral energy distributions of these objects.
In Figure~\ref{nuc_broad}, we show the correlation between the two
fluxes at rest-frame wavelengths of 3000~\AA, 2000~\AA, and 1200~\AA.
The fits give ratios of $\nu f_{\nu}/f_{\rm 2-8~keV}=2.7$, 4.7, and
4.0, respectively, at these wavelengths, which corresponds to
$f_{\nu}$ scaling roughly as $\nu^{-1}$. This is similar to the
composite quasar spectra of vanden Berk et al.\ (2001)
and Zheng et al.\ (1997), which are
shown in Figure~\ref{figbolo_opt}. Individual
sources scatter by approximately a multiplicative factor of 3 about
these relations, as can be seen in Figure~\ref{nuc_broad}.

%
%
\begin{inlinefigure}
\psfig{figure=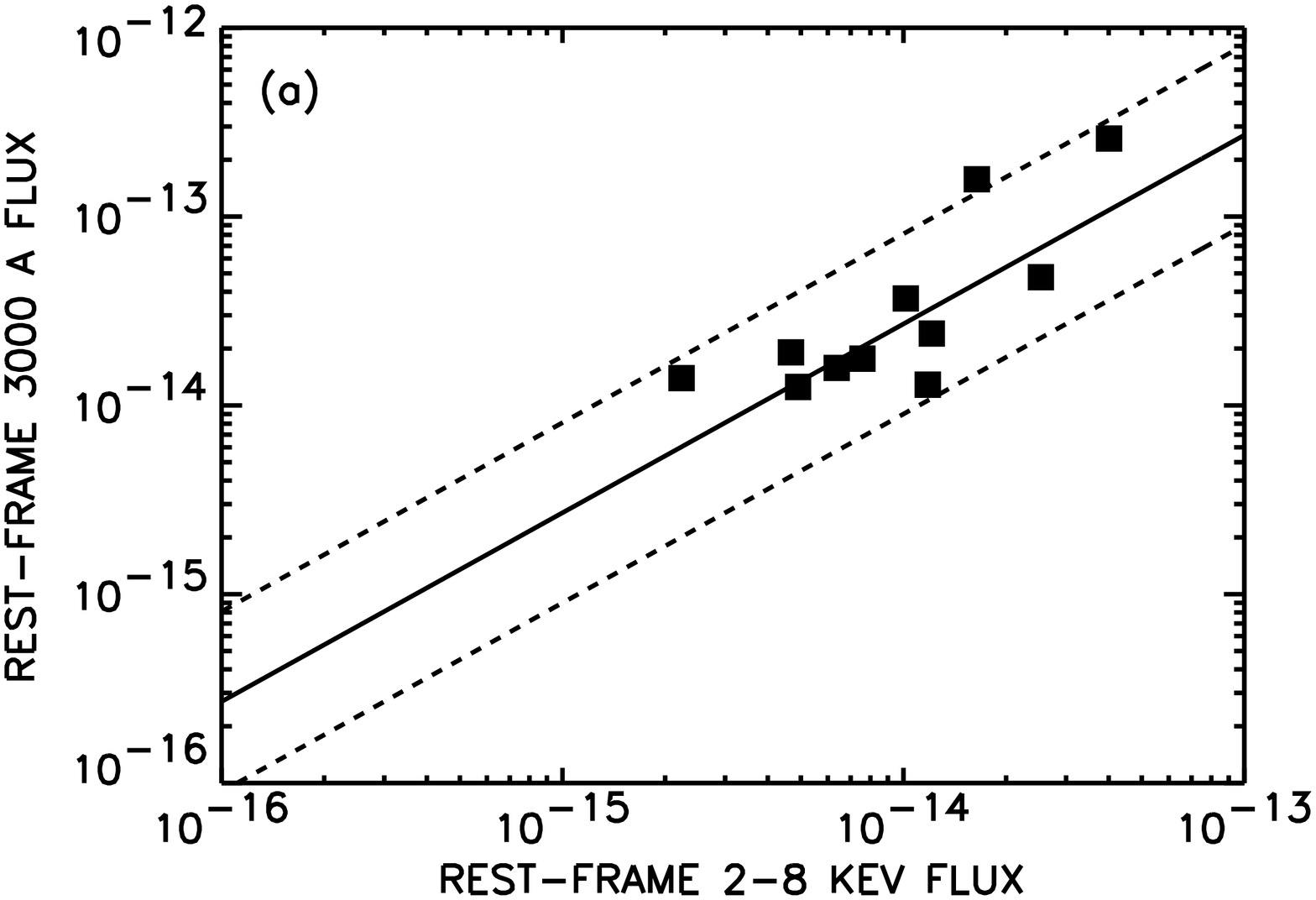,width=3.5in,angle=90}
\psfig{figure=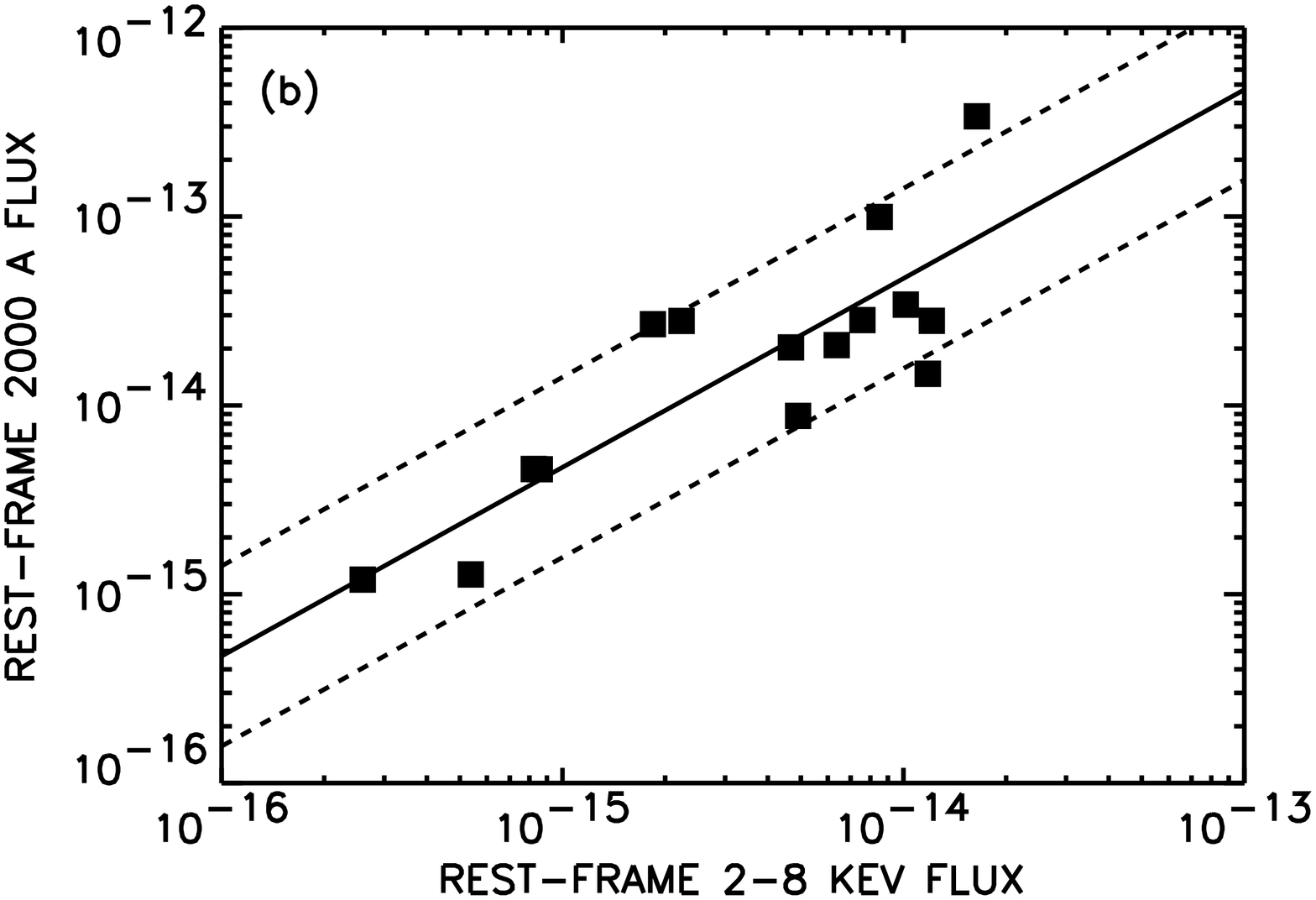,width=3.5in,angle=90}
\psfig{figure=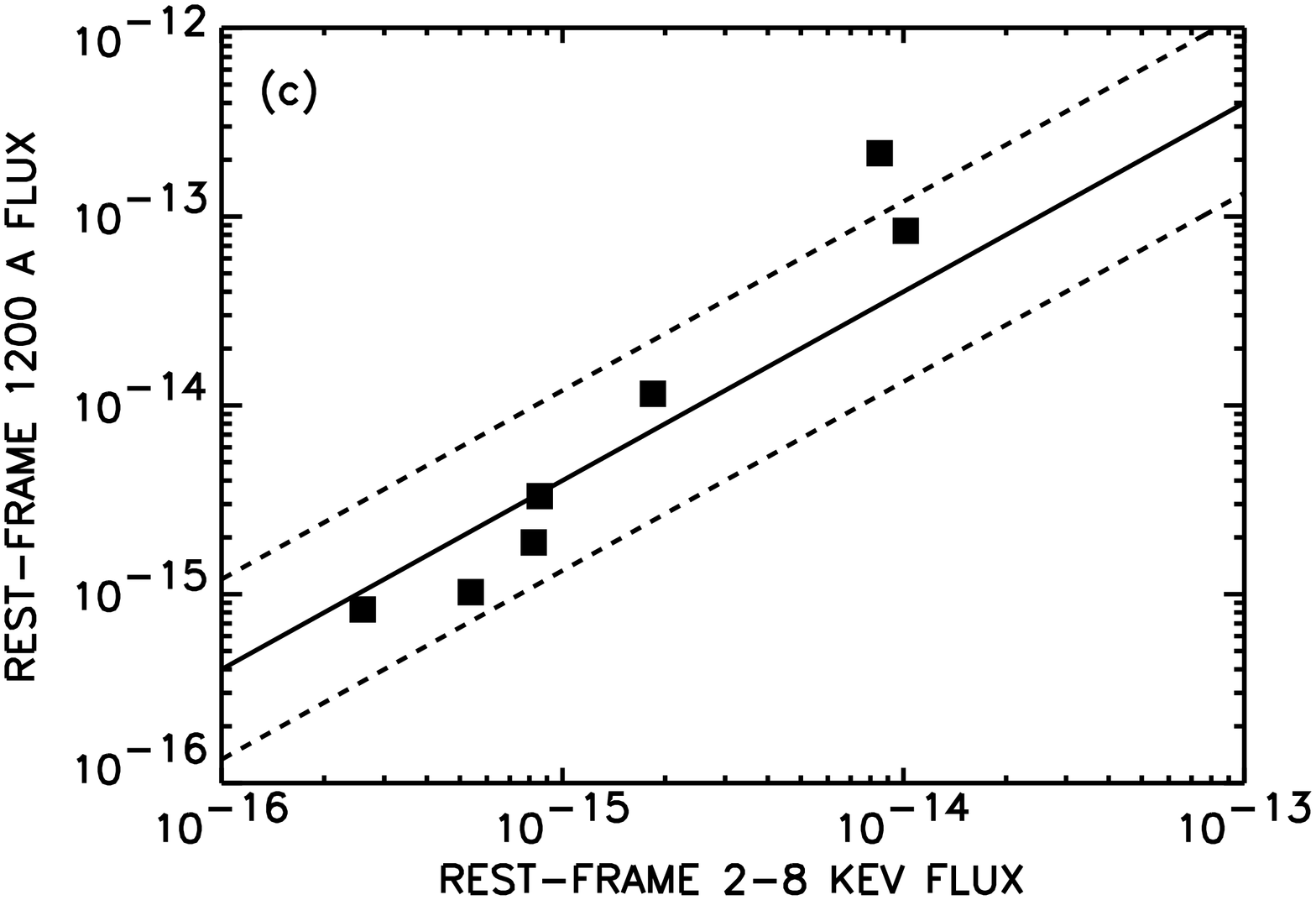,width=3.5in,angle=90}
\figurenum{33}
\caption{
Rest-frame (a) 3000~\AA, (b) 2000~\AA, and (c) 1200~\AA\
fluxes ($\nu f_{\nu}$) vs. rest-frame $2-8$~keV flux.
In each case, we show the best linear fit
{\em (solid line)\/} and the
same fit multiplied and divided by a factor of 3
{\em (dashed lines)\/}. Only broad-line AGNs whose redshifted
wavelengths lie in the $3800-10000$~\AA\ range are
shown in each panel.
\label{nuc_broad}
}
\addtolength{\baselineskip}{10pt}
\end{inlinefigure}

We can also plot for each source the ratio of the nuclear
rest-frame UV/optical flux to the rest-frame $2-8$~keV flux
at the rest-frame frequencies corresponding to the four GOODS bands.
Since the sources lie at a range of redshifts, the individual
measurements spread over a wide range of frequencies. As
long as we assume that there is no redshift dependent evolution 
of the spectral energy distributions, we can determine for the various 
spectral classes the composite spectral energy distributions over the range
$1000-10000$~\AA\ normalized to the X-ray flux. 
In Figure~\ref{figbolo_opt}, we show these composite spectral energy 
distributions for three spectral classes (broad-line AGNs, 
high-excitation sources, and normal galaxies).
These plots show the shape of the spectral energy distributions relative 
to the hard X-ray emission. The sources in the broad-line AGN and 
high-excitation classes all have UV/optical nuclei, while many of 
the sources in the normal galaxy class show no signs of having a
UV/optical nucleus at all. Sources that have no measurable nuclei are 
assigned a nominal magnitude of 29.5.

Consistent with Figure~\ref{nuc_broad}, the broad-line AGNs exhibit only a 
small amount of scatter (roughly a factor of 3) about a mean spectral 
energy distribution that closely matches to composite quasar spectra 
(Zheng et al.\ 1997; vanden Berk et al.\ 2001).
The ratios of the rest-frame UV/optical luminosities to the rest-frame 
$2-8$~keV luminosities match those of the local hard X-ray broad-line AGN 
sample analyzed by Kuraszkiewicz et al.\ (2003).

%
%
\begin{inlinefigure}
\psfig{figure=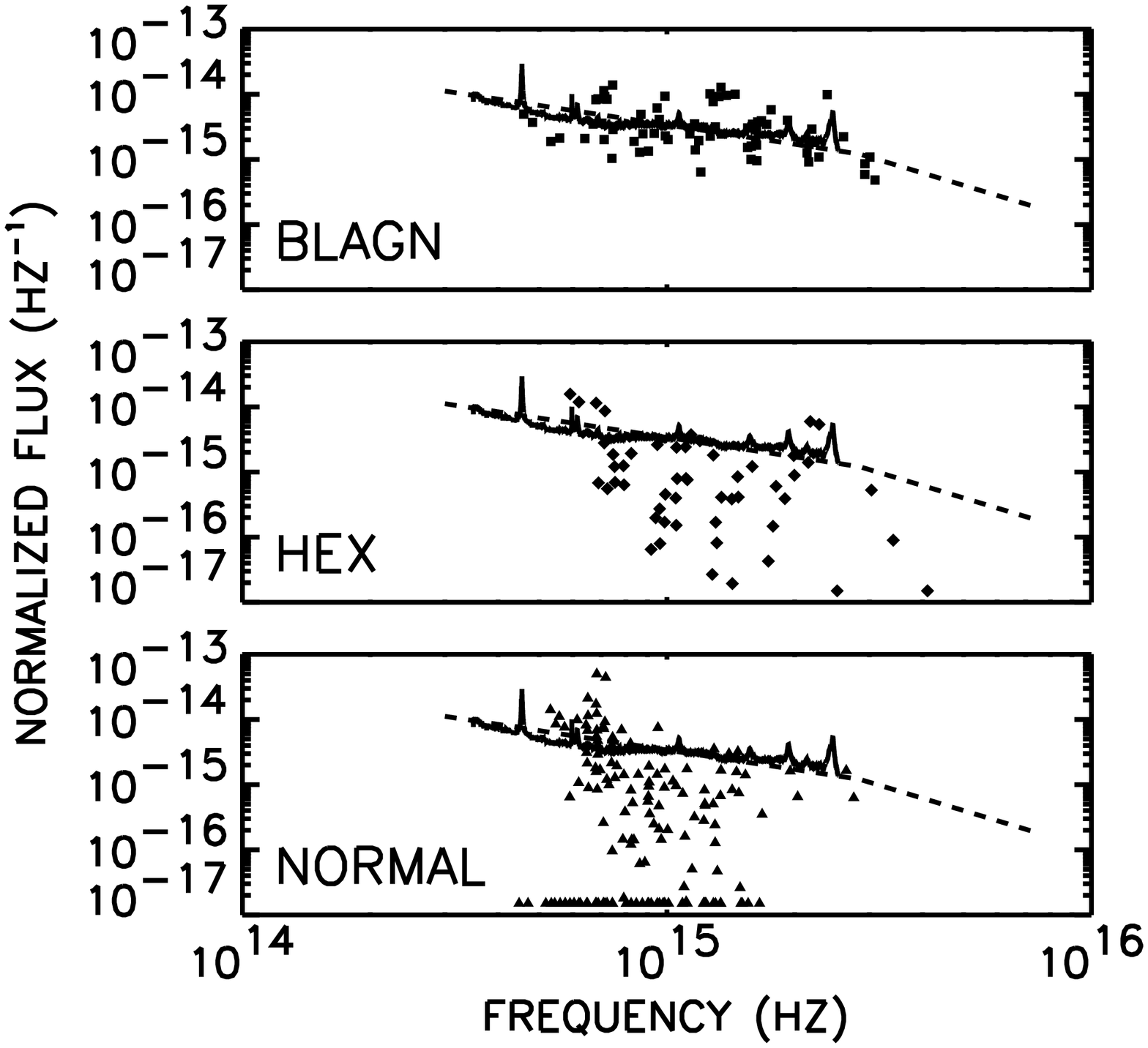,width=4in,angle=90}
\figurenum{34}
\caption{
Composite spectral energy distributions of the nuclear light
over the UV/optical range ($1000-10000$~\AA)
normalized to the $2-8$~keV flux for three spectral classes.
Solid (dashed) curves are the composite quasar spectrum
from vanden Berk et al.\ (2001) (Zheng et al.\ 1997).
\label{figbolo_opt}
}
\addtolength{\baselineskip}{10pt}
\end{inlinefigure}

In contrast, the high-excitation sources scatter to much fainter UV/optical
values than do the broad-line AGNs, and the effect is even more extreme
for the normal galaxies. Mean normalized fluxes 
versus frequency are summarized in Table~\ref{tab3} for each class 
(broad-line AGNs, high-excitation sources, and normal galaxies) and are 
shown in Figure~\ref{mean_bolo}. Numbers are only given if there are more 
than five sources in a frequency interval. The values for the high-excitation 
and normal galaxy classes drop to less than half that of the broad-line AGN
class at $0.8\times 10^{15}$~Hz and continue to fall rapidly at higher 
frequencies.

These results, and the opacity ranges seen in the X-ray
color-color plots of Figure~\ref{figcol},
show that, on average, the optically-narrow AGNs (high-excitation sources
and normal galaxies) are highly suppressed in the frequency range
between the UV and soft X-rays (approximately $10^{15}$~Hz to
$5\times 10^{17}$~Hz) relative to the broad-line AGNs.

%
%
\begin{deluxetable}{ccccccc}
\tablecaption{Mean Normalized Flux by Spectral Class and by Hard X-ray
Luminosity Interval}
\tablehead{Frequency Range & Broad-line & High-excitation & Normal &
$10^{44-45}$~ergs~s$^{-1}$ & $10^{43-44}$ &
$10^{42-43}$ \cr
($10^{14}$~Hz) & 
($10^{-15}$~Hz$^{-1}$) & ($10^{-15}$~Hz$^{-1}$) & ($10^{-15}$~Hz$^{-1}$) &
($10^{-15}$~Hz$^{-1}$) & ($10^{-15}$~Hz$^{-1}$) & ($10^{-15}$~Hz$^{-1}$)
}
\startdata
$5-7$  &   \nodata  &  \nodata  &  5.14 & \nodata & 3.62 & 5.90 \cr
$7-9$  &   5.55  &  2.13  &  2.47 & 4.02 & 1.85 & 3.28 \cr
$9-12$  &  3.48  &  1.12  &  0.664 & 2.03 & 1.64 & 0.737 \cr
$12-20$ & 4.37   &  0.588  &  0.614 & 4.03 & 1.43 & 0.511 \cr
\enddata
\label{tab3}
\end{deluxetable}

%
%
\begin{deluxetable}{lccr}
\tablecaption{Submillimeter Determinations of the Bolometric
Luminosity Ratio}
\tablehead{Redshift  & Number & $<\nu>$ & $<\nu L_\nu/L_X>$  \cr
& & (Hz) & \cr
}
\startdata
\multicolumn{4}{c}{Broad-line} \cr
\cr
0.4-1.0  & 1 & $5.3 \times 10^{11}$   &  $0.04 \pm 0.29$  \cr
1.0-2.0  & 3 & $7.4 \times 10^{11}$   &  $0.59 \pm 0.54$   \cr
2.0-4.0  & 5 & $1.5 \times 10^{12}$   & $-1.20 \pm 0.50$   \cr
\cr
\multicolumn{4}{c}{Optically-narrow} \cr
\cr
0.4-1.0  & 30 & $6.2 \times 10^{11}$   &  $0.17 \pm 0.10$  \cr
1.0-2.0  & 16 & $7.7 \times 10^{11}$   &  $1.38 \pm 0.70$   \cr
2.0-4.0  & 4 & $1.3 \times 10^{12}$   & $16.1 \pm 3.67$   \cr
\enddata
\label{tabbolosmm}
\end{deluxetable}

%
%
\begin{inlinefigure}
\psfig{figure=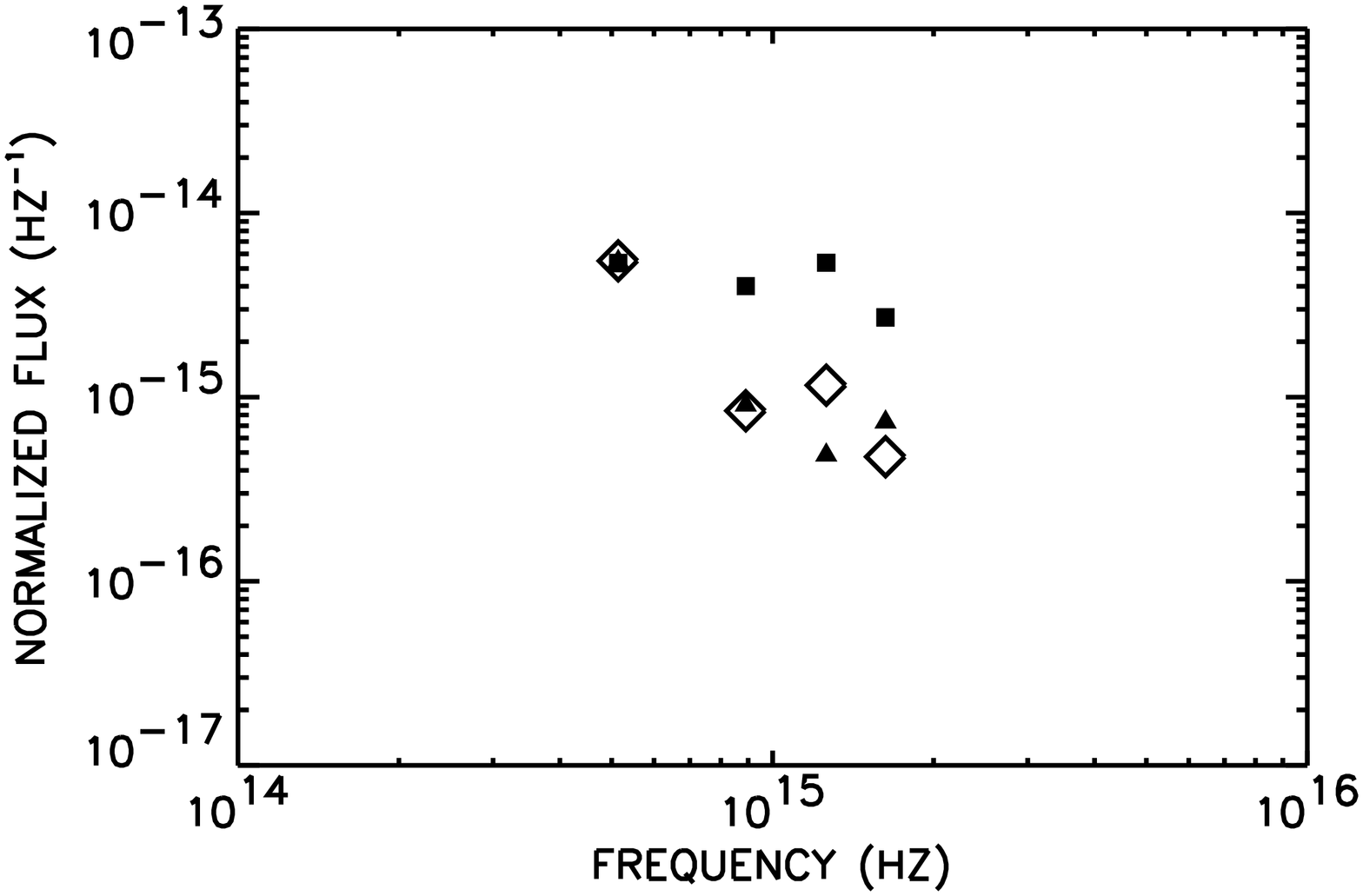,width=3.5in,angle=90}
\figurenum{35}
\caption{
Mean normalized flux vs. frequency for three spectral classes
({\em solid squares\/}---broad-line AGNs;
{\em open diamonds\/}---high-excitation sources;
{\em solid triangles\/}---normal galaxies) over the UV/optical
range ($1000-10000$~\AA).
\label{mean_bolo}
}
\addtolength{\baselineskip}{10pt}
\end{inlinefigure}

Using our long-wavelength data,
we may now construct composite spectral energy distributions 
of both the broad-line AGNs 
and the optically-narrow AGNs. In Figure~\ref{figbolo_all}, we
show the mean bolometric luminosities (relative to
the X-ray luminosities) versus the rest-frame frequencies for the
(a) broad-line AGNs and (b) optically-narrow AGNs.
The dashed curve that appears in both plots shows the median value
for the local hard X-ray--selected sample of
Kuraszkiewicz et al.\ (2003), which
is primarily broad-line AGNs with a small admixture of intermediate-type
Seyferts.

From the SCUBA data (see \S\ref{secsmm}), 
we were able to compute the ratio of the
bolometric luminosity, $\nu L_\nu$, to $L_X$ over most of the
100$\mu$m to 1~mm wavelength range. To do this, we divided all
of the AGNs by redshift bin and by spectral class (broad-line AGNs and
optically-narrow AGNs) and then computed their mean ratios
$\nu L_\nu/L_X$, weighted by the square of the submillimeter
error divided by $L_X$. The results are given in Table~\ref{tabbolosmm}
and plotted in Figure~\ref{figbolo_all}.

The optically-narrow AGNs show a strong rise with increasing frequency,
which is consistent with a strong FIR component in these sources,
while the broad-line AGNs have only upper limits or weak signals, which
is consistent with a flat spectral energy distribution.
The higher redshift (frequency) results are based
on a small number of high $L_X$ sources and so may not
be representative of the lower X-ray luminosity sources.

In the FIR (see \S\ref{secfir}), we used the two X-ray sources in
the CLASXS field that were detected at 90$\mu$m to estimate a ratio
of bolometric luminosity to rest-frame $2-8$~keV luminosity of
$36\pm 6$ {\em (open diamond)\/} for the optically-narrow AGNs
(Figure~\ref{figbolo_all}b).
The second open diamond in the figure shows
the $3\sigma$ upper limit at 170$\mu$m, which is based on
the non-detections of the two X-ray sources in the
Kawara et al.\ (2004) catalog.
For the broad-line AGNs (Figure~\ref{figbolo_all}a), we show the
$3\sigma$ upper limits at both 90$\mu$m and 170$\mu$m, 
which are based on the non-detections of the X-ray bright
broad-line AGNs in the Kawara et al.\ (2004) catalog.

%
%
\begin{inlinefigure}
\psfig{figure=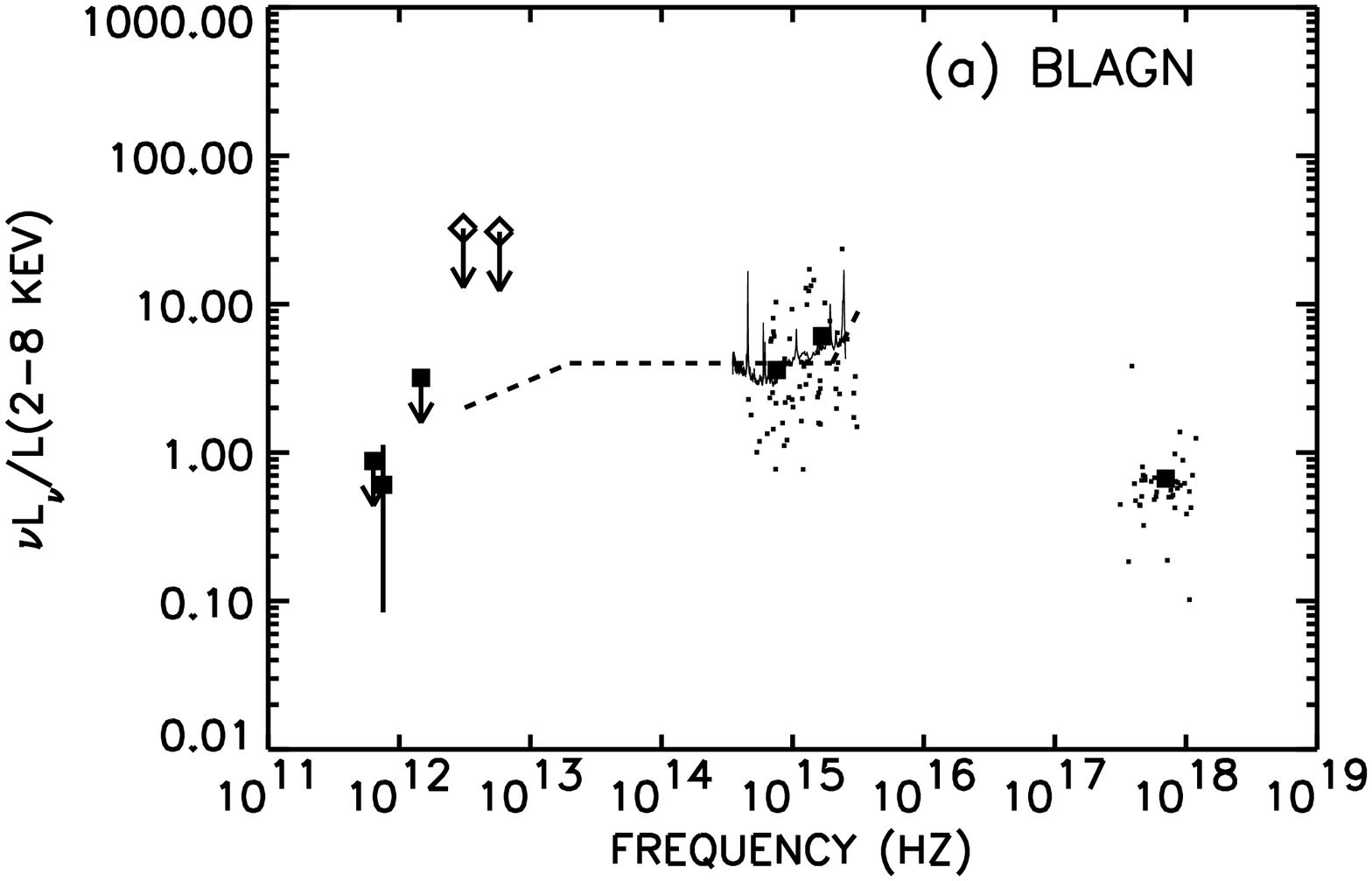,width=3.5in,angle=90}
\psfig{figure=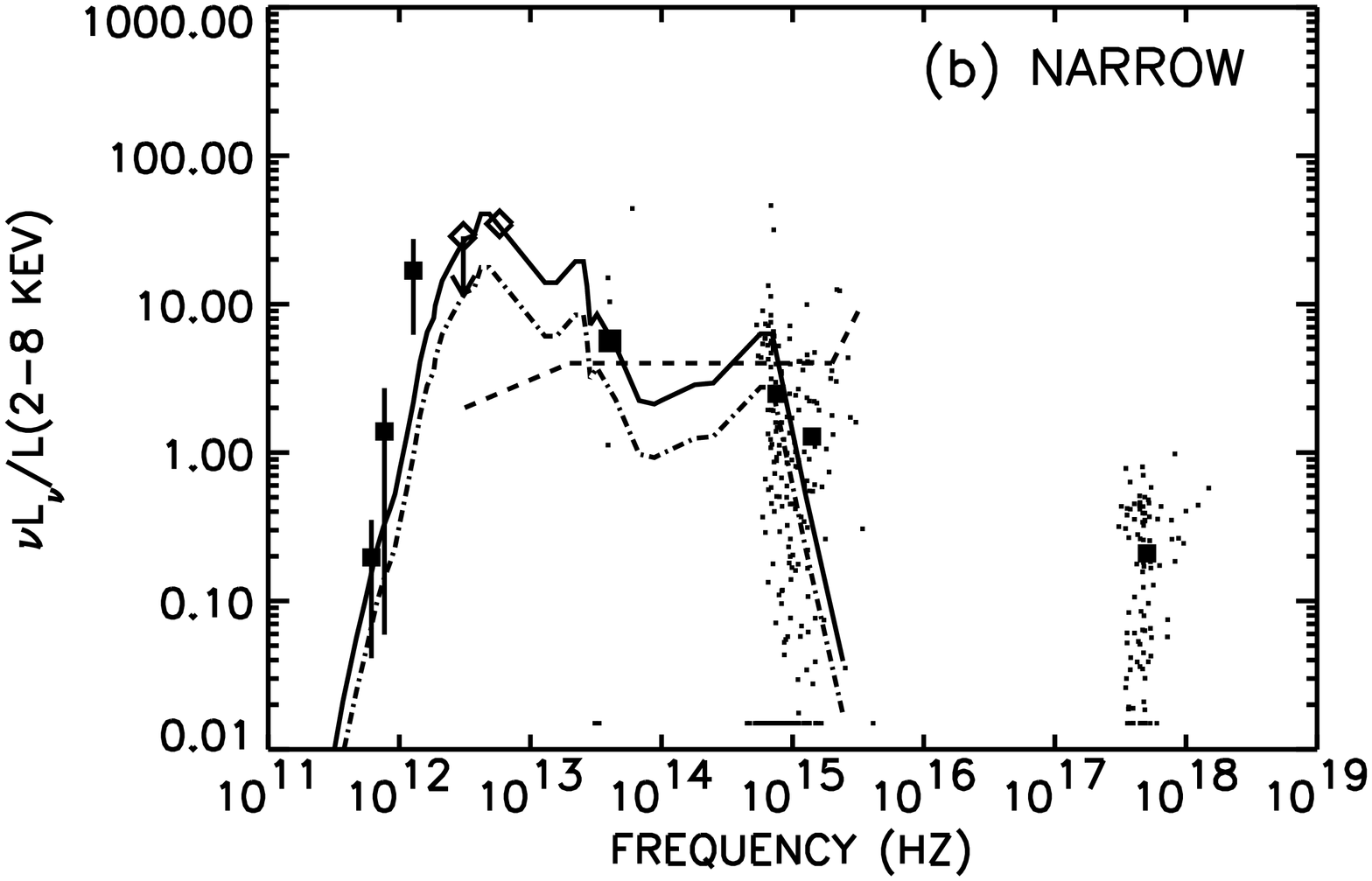,width=3.5in,angle=90}
\figurenum{36} 
\caption{
Composite spectral energy distributions of (a) broad-line AGNs (``BLAGN'')
and (b) optically-narrow AGNs (``NARROW'') in the CDF-N and CLASXS samples.
Squares denote measurements based on the CDF-N 850$\mu$m SCUBA
data of Wang et al.\ (2004), the mid-infrared data of Aussel et al.\ (1999),
the ACS GOODS-North data of Giavalisco et al.\ (2004), and the soft X-ray
data of Alexander et al.\ (2003b). Open diamonds denote the 90$\mu$m
and 170$\mu$m fluxes or limits derived from the
data of Kawara et al.\ (2004) in the CLASXS field.
All of the points are normalized to rest-frame $2-8$~keV luminosity.
Dots show the individual measurements for the mid-infrared, optical, UV, and
soft X-ray samples. Dashed curves show the median value for the local
hard X-ray--selected sample of Kuraszkiewicz et al.\ (2003), which is
primarily broad-line AGNs with a small admixture of intermediate type
Seyferts. In (b), dot-dashed curve denotes the spectral energy
distribution of the optically-thick local galaxy NGC6240 from
Hasinger (1999), normalized to give a bolometric correction of 35.
Solid curve denotes the same spectral energy distribution, normalized
to give a bolometric correction of 85.
\label{figbolo_all}
}
\addtolength{\baselineskip}{10pt}
\end{inlinefigure}

In the MIR (see \S\ref{secmir}), the ratio of the bolometric 
luminosity to the rest-frame $2-8$~keV luminosity is 6
{\em (large solid square)\/} for the six ISOCAM detected 
optically-narrow AGNs (Figure~\ref{figbolo_all}b). 
The six individual points are shown as dots. No MIR point is 
shown in Figure~\ref{figbolo_all}a.

\section{Bolometric Corrections by Spectral Type}
\label{secbolcorrtype}

Various bolometric corrections have been adopted for the 
optically-narrow AGNs in the literature. Barger et al.\ (2001b)
adopted a value of about 35, based on the radio and submillimeter
properties of a {\em Chandra\/} X-ray sample, while
Fabian (2004) suggested a lower value of
about 15, based on the suppression of UV and optical light in the
local Seyfert galaxies Mrk335, NGC3783, and NGC5548.
However, the latter determination neglects the enhancement of the
MIR/FIR light that results from the reradiation of the absorbed
UV/optical light by gas and dust, which we can see
from Figure~\ref{figbolo_all}b is important.

In fact, the composite spectral energy distributions in Figure~\ref{figbolo_all} show that
the broad-line AGNs are consistent with the locally derived results---though
poorly constrained at the intermediate wavelengths---while the
optically-narrow AGNs are clearly much more strongly peaked
in the FIR. To illustrate this,
in Figure~\ref{figbolo_all}b, we have overplotted the highly obscured
local galaxy NGC6240 (Hasinger 2000), which is
often used as a template for absorbed AGNs. This template provides a
fairly good description of the shape of the composite spectrum over
the frequency range from the submillimeter to the UV. By normalizing
the shape of the template to the data {\em (solid curve)\/},
we would infer a bolometric correction of about 85 for the optically-narrow sources,
with most of the light emerging around 100$\mu$m in the rest-frame.
This inferred bolometric correction is larger than that determined for the
broad-line AGNs. As we can see from the dot-dashed curve in
Figure~\ref{figbolo_all}b, a bolometric correction of 35 would be too low to
fit the optically-narrow AGN spectral energy distribution.

Part of this effect is due to internal absorption in
the $2-8$~keV band, which is higher for the optically-narrow sources
and negligible for the broad-line AGNs. At $z=0.5$, we can use the
$4-8$~keV and $2-4$~keV counts to estimate a mean correction
of about 1.3 for the $2-8$~keV flux in the optically-narrow
sources (see Figure~\ref{figcol}). However, even with
this correction, the bolometric correction inferred for the 
optically-narrow AGNs is higher than that for the broad-line AGNs. 
One additional concern is that the 
FIR emission from these sources may be partially due to star formation,
though the correlation of $L_X$ and $L_{FIR}$ would seem to suggest
that $L_{FIR}$ is related to the nuclear activity (or that star
formation is correlated with the nuclear activity).

Given the limited MIR/FIR/submillimeter data sets,
there are clearly considerable uncertainties in our determinations.
However, the data do suggest that it is unlikely that the bolometric 
corrections for the optically-narrow AGNs are significantly lower than
the bolometric corrections for the broad-line AGNs, as was proposed by
Fabian (2004).

\section{Composite Spectral Energy Distributions by Hard X-ray Luminosity}
\label{sechxlum}

If the luminosity dependent unified model holds, such that the 
covering fraction varies with X-ray 
luminosity, then the low X-ray luminosity sources with their high
covering fractions should have spectral energy distributions 
that are, on average, suppressed
in the UV to the soft X-rays with the light reradiated in the
FIR. In contrast, the high X-ray luminosity sources with their low
covering fractions should have little reradiation. 

In Figure~\ref{figbolo_lx}, we show the UV/optical spectral 
energy distributions according
to X-ray luminosity rather than according to spectral type. 
We again summarize the mean normalized fluxes versus frequency in
Table~\ref{tab3}, this time for each luminosity interval. Numbers are
only given if there are more than five sources in a given frequency 
interval. At high X-ray luminosities, most of the
sources lie close to the unobscured quasar spectral energy distribution, 
and the mean UV/optical
to X-ray flux ratios are only slightly lower than the broad-line AGN values
(Table~\ref{tab3}). In contrast, for the low and
intermediate X-ray luminosity sources, there is considerably more
spread in the individual UV/optical to X-ray flux ratios, with the
mean ratios lying well below the broad-line AGN values at the higher 
frequencies.  Thus, as expected, the lowest X-ray 
luminosity sources are the most suppressed.
This behavior is similar to that seen in samples divided
by optical luminosity, where the extinctions are higher
at lower luminosities (e.g., Gaskell et al.\ 2004). The
same reference also shows that there may be extinction
without spectral curvature, a point which emphasizes the
necessity of looking at the full spectral energy distributions
of the sources in deciding the bolometric corrections.

This is just another statement of the Steffen effect, though one
which is completely independent of the optical spectral typing:
low X-ray luminosity sources have much fainter UV/optical nuclei
relative to their $2-8$~keV fluxes than do high X-ray luminosity
sources.

%
%
\begin{inlinefigure}
\psfig{figure=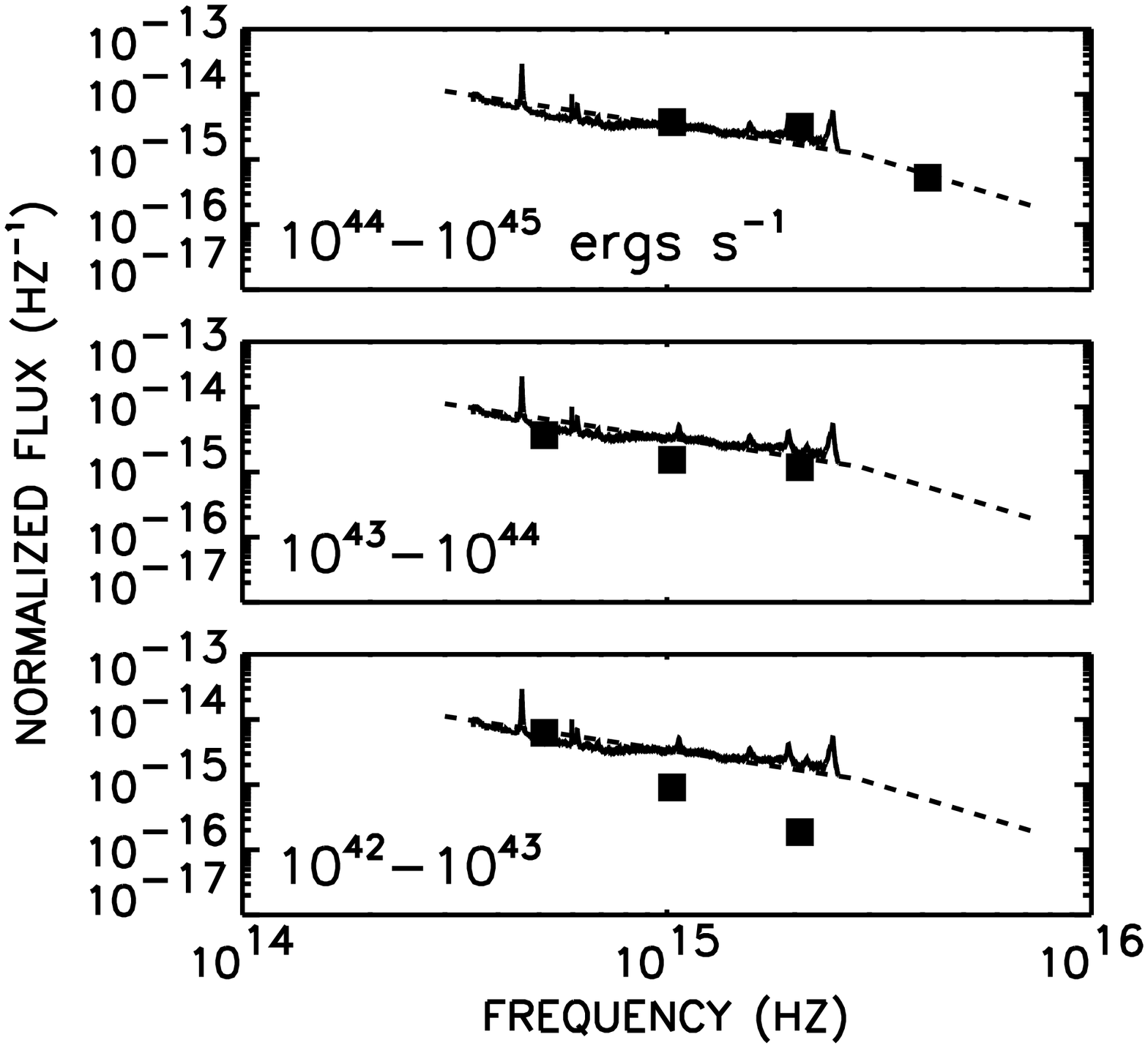,width=4in,angle=90}
\figurenum{37}
\caption{
Mean composite spectral energy distributions of the nuclear light
over the UV/optical range ($1000-10000$~\AA) for three hard X-ray
luminosity intervals.  Solid (dashed) curves are the composite quasar
spectrum from vanden Berk et al.\ (2001) (Zheng et al.\ 1997).
\label{figbolo_lx}
}
\addtolength{\baselineskip}{10pt}
\end{inlinefigure}

\section{Bolometric Corrections by Hard X-ray Luminosity}
\label{secbolcorrlum}

In the luminosity dependent unified model, the covering fraction 
of the sources in a given luminosity range is equal to the fraction 
of non--broad-line AGNs in that luminosity range. If we assume that
the bolometric corrections are not dependent on X-ray luminosity, then 
we can predict how much luminosity should be reradiated into the FIR
at a given X-ray luminosity: the FIR luminosity should just be the 
bolometric luminosity times the covering fraction. This simple
model predicts that the brightest 
FIR sources should be the intermediate X-ray luminosity sources.
Higher X-ray luminosity sources have small covering fractions
and do not reradiate much light into the FIR, while lower
X-ray luminosity sources have low bolometric luminosities
to start off with.

We illustrate this in Figure~\ref{lum_model}, where we show
the error-weighted mean submillimeter fluxes (converted to
FIR luminosities, see below) of the ensemble
of X-ray sources in the redshift range $z=0.8-1.6$ versus their
mean hard X-ray luminosities {\em (diamonds)\/}. We convert the 
submillimeter fluxes to FIR luminosities assuming a FIR spectrum 
with the shape of NGC6240 (Hasinger 2000). 
The solid and dashed curves show the predicted values for 
the luminosity dependent unified model, 
assuming a bolometric correction of 35. The solid curve was 
calculated using the spectroscopically identified sources only, 
and the dashed curve was calculated by placing all of the 
spectroscopically unidentified sources in the spectroscopically observed
sample into the redshift interval as non--broad-line AGNs, which gives a 
very extreme upper limit. 
The dotted curve (shown only for the spectroscopically identified
sources) shows the same calculation for a bolometric correction of 85 
(see \S\ref{secbolcorrtype}), which provides a better fit 
to the data at the low-luminosity end.
At the high-luminosity end, we would expect the bolometric correction 
of 35 to be valid, because we know that value is appropriate for 
broad-line AGNs.

%
%
\begin{inlinefigure}
\psfig{figure=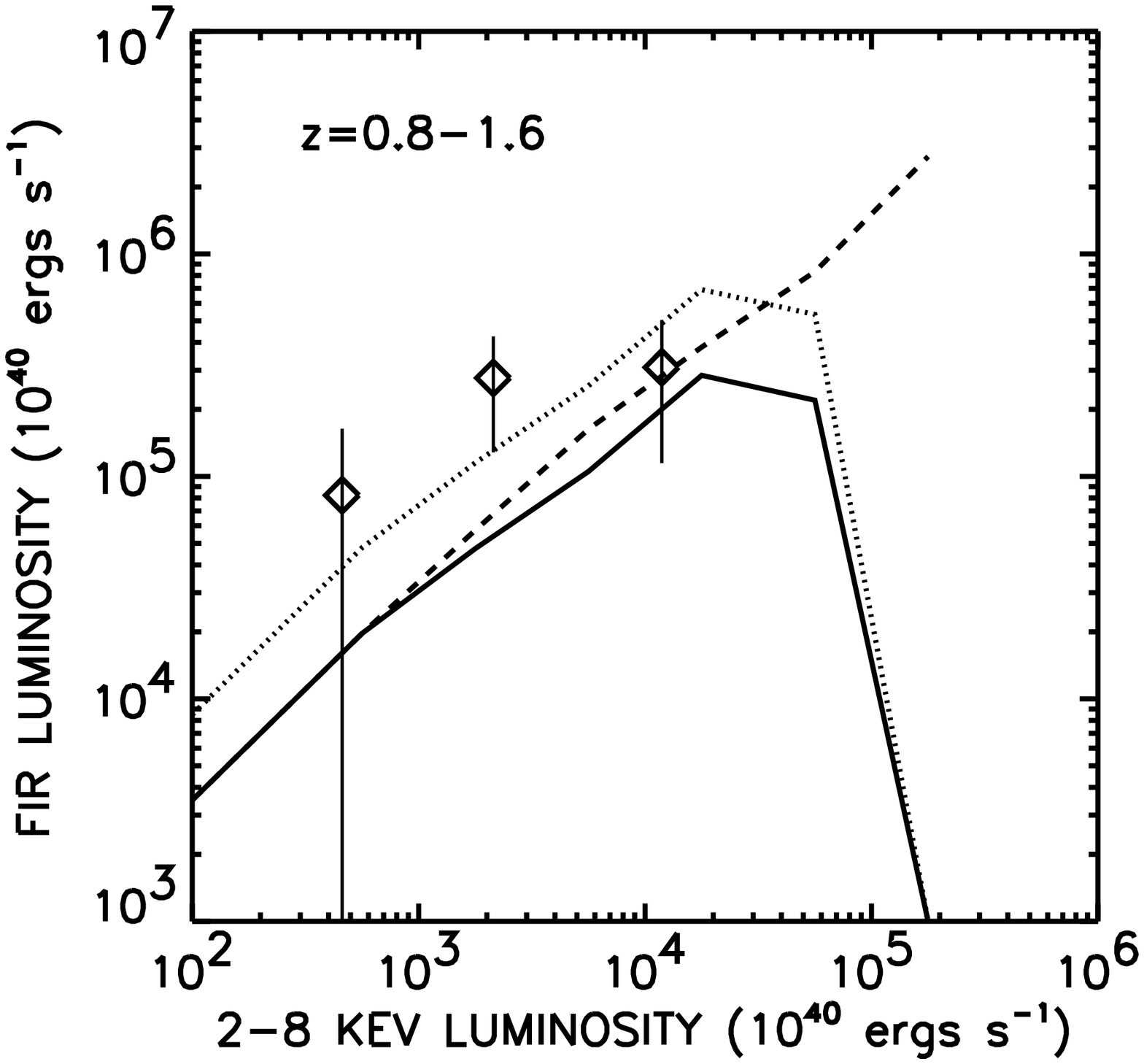,width=4in,angle=90}
\figurenum{38}
\caption{
Predicted reradiated luminosity in the far-infrared over the redshift
interval $z=0.8-1.6$ for a bolometric correction of 35 {\em (solid curve)\/}
and a bolometric correction of 85 {\em (dotted curve)\/}, if the
luminosity dependent unified model holds.
Dashed curve shows the upper bound for a bolometric correction of 35
(see text for details).
Diamonds show the error-weighted mean 850$\mu$m fluxes in this
redshift range converted to far-infrared luminosities, assuming a
NGC6240 spectral energy distribution. Uncertainties are $1\sigma$.
\label{lum_model}
}
\addtolength{\baselineskip}{10pt}
\end{inlinefigure}

Thus, the data appear to be consistent with the luminosity dependent
unified model, but two caveats 
should be kept in mind. First, the errors on the submillimeter data 
are significant, and the consistency of the results with the model
is not highly significant. Also, bearing in mind the 
Page et al.\ (2001, 2004) results for the brighter 
quasars, better data are clearly required. Second, some part of the 
FIR luminosity could be produced by star formation,
which may in turn be correlated with the nuclear activity.

\section{Accretion History of the Universe}
\label{secacc}

As was first pointed out by So\l tan (1982), we 
may use the evolution of the parameterization of the rest-frame
$2-8$~keV energy density production rate per comoving volume,
${\dot\lambda_X}$, to infer the present-day supermassive black hole mass density accreted 
by AGNs, $\dot{\rho}_{\bullet,acc}$. The mass inflow rate onto a supermassive black hole,
$\dot{M}_\bullet$, is related to the bolometric luminosity of
the AGN, $L_{BOL}$, by $\dot{M}_\bullet = L_{BOL}(1-\epsilon)/\epsilon c^{2}$,
where $\epsilon$ is the accretion efficiency (typically taken to be
about 0.1). Assuming $\epsilon=0.1$, we can use this to go from 
${\dot\lambda_X}$ to the accretion rate density, $\dot{\rho}_{\bullet,acc}$, 
if we know the corrections to bolometric luminosity
(see, e.g., Barger et al.\ 2001a). 

In the present
section, we set $L_{BOL} = C_X L_X$, where $C_X$ is the
properly averaged bolometric correction in the $2-8$~keV band 
for a given population.
Recognizing the uncertainties, we use our determinations of $C_X$.
We do not make any correction for internal absorption in the $2-8$~keV 
band, since this is present in both the hard X-ray luminosity functions 
and the bolometric corrections. This 
could result in some redshift dependent correction; in particular,
we may be slightly underestimating the energy density production
rate at $z=0$ relative to that at $z=1$, since we are selecting in a
higher energy band at the higher redshifts. However, we do not
expect this effect to be large, given that the overall correction
for absorption is not large.

Our parameterization of the production rate for all spectral types is
\begin{equation}
{{\dot\lambda_X}}= A~{\left(1+z\over 2\right)}^{\alpha} \,,
\label{bargereq}
\end{equation}
where $A=4.6\times 10^{39}$~ergs~s$^{-1}$~Mpc$^{-3}$,
and $\alpha=3.2$ at $z<1$ and $\alpha = -1$ at higher
redshifts. The integral of Equation~\ref{bargereq}
is only weakly sensitive to the poorly determined
$\alpha$ at the higher redshifts, since it is dominated
by the production rate at $z=1$ (see \S\ref{secenergy}). Thus,
we have chosen our value of $\alpha$ at $z>1$ so that the 
parameterization is between the measured spectroscopic values 
and the incompleteness corrected values of Figure~\ref{figevol}.

At low redshifts, roughly 40\% of the
hard X-ray energy density production rate is due to broad-line AGNs and
the remainder to optically-narrow AGNs.
Assuming this constant ratio to determine what fraction 
of each bolometric correction to use, and integrating through time, 
we find the accreted supermassive black hole mass density for all spectral types to be
\begin{equation}
\rho_{\bullet,acc}(z=0)=4.0\times 10^{5}~{\rm M}_\odot~{\rm Mpc}^{-3} \,,
\label{eqall1}
\end{equation}
if we use $C_X=85$ for the optically-narrow AGNs, and
\begin{equation}
\rho_{\bullet,acc}(z=0)=2.1\times 10^{5}~{\rm M}_\odot~{\rm Mpc}^{-3} \,,
\label{eqall2}
\end{equation}
if we use $C_X=35$ for the optically-narrow AGNs.
(We always use $C_X=35$ for the broad-line AGNs.)
Making the energy density production rate
history flat ($\alpha=0$) at higher redshifts (i.e., using the 
incompleteness correction of Figure~\ref{figevol})
would increase these numbers by at most 40\%.
These numbers do not depend on the Hubble constant and are
insensitive to the assumed cosmological geometry
(e.g., Chokshi \& Turner 1992).

The accreted supermassive black hole mass density from the broad-line AGNs may be more exactly 
calculated, since the bolometric correction is better known, and the history of the 
energy production should be less susceptible to the incompleteness 
corrections. The ratio of the hard X-ray energy density production rate 
due to broad-line AGNs relative to that due to optically-narrow AGNs 
is clearly higher at higher redshifts (Figure~\ref{figevolblagn}). 
Thus, we parameterize the rest-frame $2-8$~keV comoving energy density 
production rate of the broad-line AGNs alone as
\begin{equation}
{{\dot\lambda_X}}= B~{\left(1+z\over 2\right)}^{\beta} \,,
\label{bargereq2}
\end{equation}
where $B=1.9\times 10^{39}$~ergs~Mpc$^{-3}$~s$^{-1}$,
and $\beta =3.0$ at $z<1$ and $\beta = 0$ at higher
redshifts. This fit is shown in Figure~\ref{figevolblagn}.
Integrating Equation~\ref{bargereq2}, we obtain
\begin{equation}
\rho_{\bullet, acc}^{BLAGN} (z=0) = 1.2\times 10^5
(C_X/35)[0.1(1-\epsilon)/\epsilon]~{\rm M}_\odot~{\rm Mpc}^{-3} \,,
\label{bargereq3}
\end{equation}
which would say that about one-half to one-quarter of the supermassive black hole
mass density was fabricated in broad-line AGNs and the remainder in 
optically-narrow AGNs.

With the energy density production rate of Equation~\ref{bargereq}, 
half of the accreted supermassive black hole mass density has formed by $z=1.07$  
($6\times10^{9}$~yr) and the remaining amount after this redshift. 
The broad-line AGNs (which are the high X-ray luminosity population, and 
hence correspond to the more massive black holes) form slightly 
earlier, with half the mass density in place by $z=1.45$.
For a constant bolometric correction, regardless of type
or luminosity, roughly half of the supermassive black hole mass 
density is formed in X-ray sources with $L_X=10^{42}$~ergs~s$^{-1}$ 
to $L_X=5\times10^{43}$~ergs~s$^{-1}$ and half from more luminous
sources. Using Equation~\ref{eq4}, this would correspond to a black 
hole mass of about $8\times10^{7}~{\rm M}_\odot$. If the bolometeric
correction is larger for the lower luminosity sources, then this would
weight the result to a lower break point.

The broad-line AGN supermassive black hole mass density calculation 
should also provide the most precise comparison to previous optical 
determinations, since these 
have been computed only for the broad-line AGN population.
Yu \& Tremaine (2002) determined the accreted supermassive black hole
mass density during optically-bright quasar phases to be
\begin{equation}
\rho_{\bullet, acc}^{QSO} (z=0) = 2.1\times 10^5
(C_B/11.8)[0.1(1-\epsilon)/\epsilon]~{\rm M}_\odot~{\rm Mpc}^{-3} \,,
\end{equation}
where $C_B$ is the bolometric correction in the $B$ band, defined
by $L_{BOL}=C_B \nu_B L_{\nu_B}$ and found to be about 11.8 by
Elvis et al.\ (1994).
This value of $\rho_{\bullet, acc}^{QSO} (z=0)$ is similar to that 
derived by Chokshi \& Turner (1992).

%
%
\begin{inlinefigure}
\psfig{figure=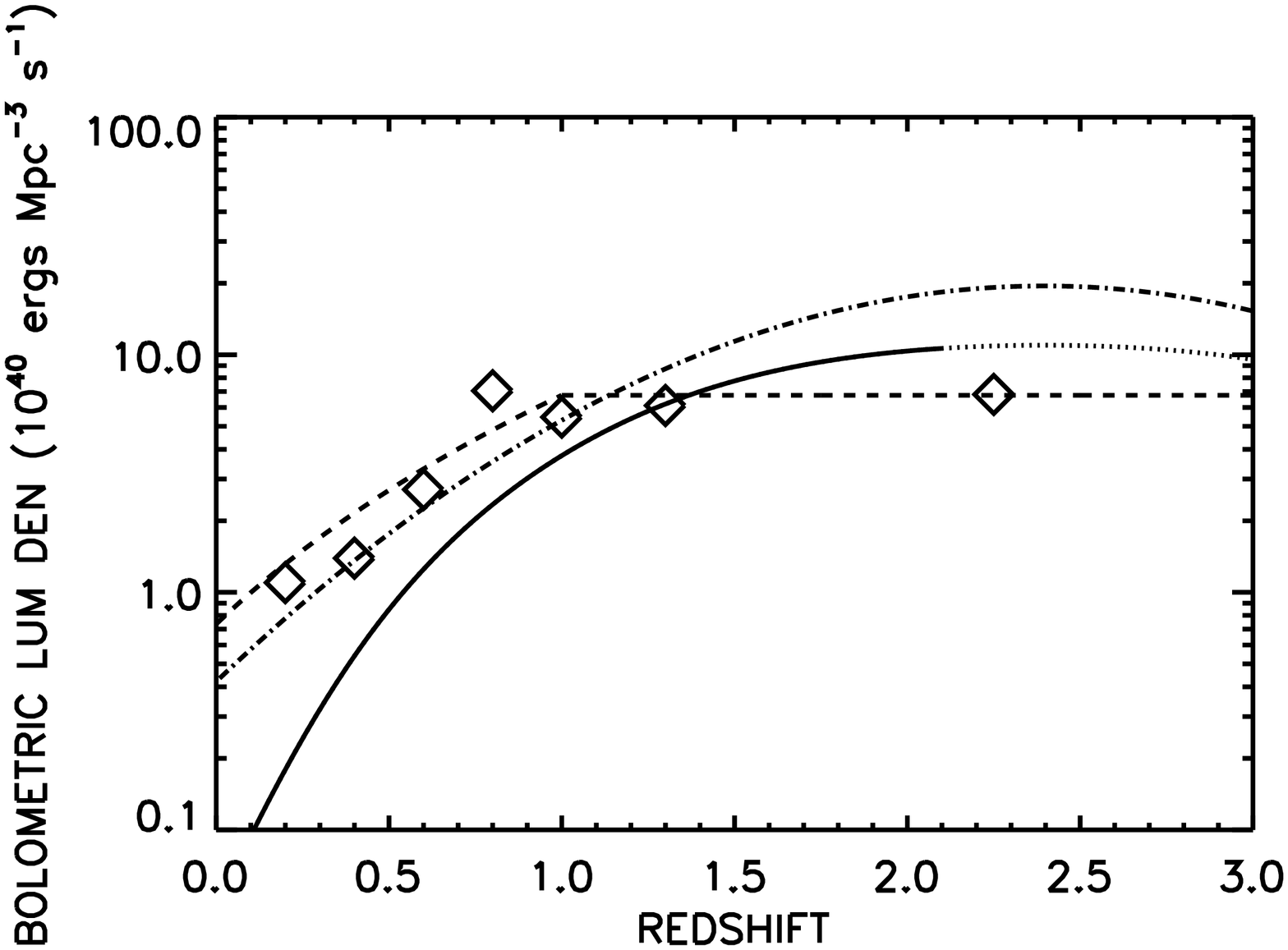,width=8.5cm,angle=90}
\figurenum{39}
\caption{
Bolometric light production rate per unit comoving volume vs. redshift.
Solid curve shows the production rate determined from the optical
QSO luminosity function of Croom et al.\ (2004), and dotted curve
shows the redshift extrapolation of their fit.
Dot-dashed curve shows the production
rate determined after extrapolating (to low and high optical
luminosities) the functional form that Croom et al.\ determined in
the observed $M_{B}=-23$ to $-26$ range.
Diamonds show the production rate calculated from our hard X-ray
luminosity functions,
and the dashed line shows our parameterization of this.
Our hard X-ray luminosity function determination of the light
production is higher than the observed optical QSO luminosity
function determination at low redshifts, where much of the light
is in lower luminosity sources. The extrapolated optical QSO
luminosity function determination is more consistent with the hard
X-ray luminosity function determination at low redshifts, but it
is higher at high redshifts.
\label{lum_den_opt}
}
\addtolength{\baselineskip}{10pt}
\end{inlinefigure}

We can see that the Yu \& Tremaine (2002) calculation is about 1.75 
times higher than our value. The reason for this difference lies in 
the extrapolation by Yu \& Tremaine of the optical QSO luminosity 
function determinations 
outside the measured absolute magnitude range. This extrapolation is 
quite sensitive to the poorly determined shape of the faint-end optical 
QSO luminosity function. Yu \& Tremaine 
used the optical QSO luminosity function of Boyle et al.\ (2000), which 
found a faint-end slope of $-1.58$, whereas a more recent determination 
by Croom et al.\ (2004) finds a shallower slope of
$-1.09$. Thus, mass density estimates based on the 
Croom et al.\ paper will be lower than those obtained by Yu \& Tremaine.
However, even the Croom et al. extrapolation does not show the
downturn seen in the hard X-ray luminosity functions at low luminosities.

In Figure~\ref{lum_den_opt}, we use the new Croom et al.\ (2004) 
QSO luminosity function determination to show the effect of the 
extrapolation outside the measured absolute magnitude range. The 
solid curve shows the
bolometric energy density production determined directly from the 
Croom et al.\ data and the dotted curve the redshift 
extrapolation of this determination. The corresponding results determined 
from our X-ray--selected population are shown by the diamonds and dashed 
curve. We can see that the restriction of the measured optical 
QSO luminosity function to absolute magnitudes $M_B=-23$ to $-26$ 
means that just using the Croom et al.\ fit underestimates the 
broad-line AGN light production at $z<1$,
while above this redshift, where the light is dominated by higher 
luminosity sources, the two measurements agree.
Extrapolating the Croom et al.\ data outside the fitted 
absolute magnitude range {\em (dot-dashed curve)\/}
improves the agreement at the 
low-redshift end, where much of the light is in low-luminosity
sources. However, at the high-redshift end, this extrapolation 
results in an overestimation, because the faint-end extrapolation
is flatter than the hard X-ray luminosity function determinations, 
which turn down at the low-luminosity end (see Figure~\ref{blagn_opt}).

In terms of the effect that the extrapolation has on the accreted 
supermassive black hole 
mass density from optical QSOs, if we only include the sources in the 
$M_{B}=-23$ to $-26$ range and use the Croom et al.\ (2004) fit, 
we obtain 
\begin{equation}
\rho_{\bullet, acc}^{QSO} (z=0) = 0.8\times 10^5
(C_B/11.8)[0.1(1-\epsilon)/\epsilon]~{\rm M}_\odot~{\rm Mpc}^{-3} \,,
\end{equation}
while if we extrapolate to brighter and fainter luminosities
using the Croom et al.\ fit, we instead find
\begin{equation}
\rho_{\bullet, acc}^{QSO} (z=0) = 1.4\times 10^5
(C_B/11.8)[0.1(1-\epsilon)/\epsilon]~{\rm M}_\odot~{\rm Mpc}^{-3} \,.
\end{equation}
These values bracket our broad-line AGN hard X-ray luminosity 
function determination.

In conclusion, whether one uses the recent optical QSO luminosity 
function determinations of Croom et al.\ (2004) or our broad-line 
AGN hard X-ray luminosity function determinations,
assuming $\epsilon=0.1$, both give about half the accreted 
supermassive black hole mass 
density found by Yu \& Tremaine (2002). This means that there is still
room for obscured accretion from the optically-narrow AGNs when we
compare with the local supermassive black hole mass density.
In fact, for $\epsilon\approx 0.1-0.2$, we find reasonable agreement 
between our accreted supermassive black hole mass density for all 
spectral types (Equation~\ref{eqall1} or \ref{eqall2}) and the local 
supermassive black hole mass density determined by Yu \& Tremaine (2002):
\begin{equation}
\rho_\bullet (z=0)=(2.5\pm0.4)\times
10^5\Bigl(\frac{h}{0.65}\Bigr)^2~{\rm M}_\odot~{\rm Mpc}^{-3} \,.
\end{equation}
However, we note that there is very little room for further obscured 
sources or for any low efficiency accretion periods.

\section{Summary}
\label{secsummary}

In this paper, we presented a thorough analysis of the nature and 
evolution of hard X-ray--selected AGNs using both deep and wide-area 
{\em Chandra\/} surveys with nearly complete optical spectroscopic 
follow-up observations. Our results are as follows.

$\bullet$ We determined median redshifts and $1\sigma$ 
median redshift ranges with X-ray flux for the spectroscopically
and spectroscopically plus photometrically identified X-ray
samples. We found that the median redshifts are fairly constant 
with X-ray flux at $z\sim 1$ and that the lower X-ray flux sources, 
where the surveys are more spectroscopically incomplete, are
dominated by sources with hard X-ray luminosities below
$10^{44}$~ergs~s$^{-1}$ and redshifts near one. 

$\bullet$ We spectrally classified the optical counterparts to the 
X-ray sources and measured the FWHM line widths of the spectra. We 
found that the X-ray--selected broad-line AGNs recover essentially all 
of the optically-selected type 1 AGNs and that
the only distinction that can be made from the X-ray 
colors is between the broad-line AGNs and the sources 
that do not show broad lines (${\rm FWHM} < 2000$~km~s$^{-1}$;
the ``optically-narrow'' AGNs). Most of the broad-line AGNs show 
essentially no visible absorption in X-rays, while the optically-narrow 
AGNs show a wide range of absorbing column densities. However, these
absorbing column densities show little dependence on the optical 
spectral type or on the line widths of the optically-narrow AGNs.

$\bullet$ We constructed up-to-date low and high-redshift hard X-ray 
luminosity functions for all spectral types and for broad-line AGNs 
alone using our highly spectroscopically complete observed samples. 
We did maximum likelihood fits over the redshift range $z=0-1.2$.
We found that all of the hard X-ray luminosity functions are consistent 
with pure luminosity evolution, 
with $L_\ast=(1+z)^{3.2\pm 0.8}$ for all spectral types
and $L_\ast=(1+z)^{3.0\pm 1.0}$ for broad-line AGNs.
Thus, all AGNs drop in luminosity by almost an order of magnitude
over this redshift range. 

$\bullet$ We directly compared our broad-line AGN hard X-ray 
luminosity functions with the optical QSO luminosity functions 
and found that at the bright end, the luminosity functions agreed 
extremely well at all redshifts. However, we found that the optical 
QSO luminosity functions do not probe faint enough to see the 
downturn in the broad-line AGN hard X-ray luminosity functions and 
that they may even be missing some sources at the very lowest 
luminosities to which they probe.

$\bullet$ We determined the evolution with
redshift of the hard X-ray energy density production rate and
found that at $z<1$, it bears a striking resemblance to the 
overall redshift evolution of the star formation rate density.
We also determined the evolution according to spectral type,
applying incompleteness corrections to the production rates of the
optically-narrow AGNs. We found that at $z<1.5$, most of the hard
X-ray energy density production is due to optically-narrow AGNs,
while at higher redshifts, the broad-line AGNs dominate the production.

$\bullet$ We estimated black hole masses for the broad-line AGNs using our 
measured MgII 2800~\AA\ line widths and nuclear optical luminosities 
of sources in the {\em HST\/} ACS GOODS-North region of the CDF-N. For 
the resulting small sample,
we found a surprisingly tight correlation between hard X-ray luminosity
and black hole mass at $z=1$, with the black holes accreting
at a fairly constant rate of 16\% of the Eddington luminosity.

$\bullet$ From the X-ray sources in the ACS GOODS-North region of the 
CDF-N, we found that the most luminous X-ray sources lie in the most
optically-luminous host galaxies, as would be expected from the 
bulge luminosity--black hole mass relation. We also found that 
the host galaxy optical luminosities do not change
much with redshift for a given interval in hard X-ray luminosity,
but the most optically-luminous host galaxies that existed at
high redshifts have completely disappered at low redshifts.
We therefore concluded that the observed drop in the hard X-ray
luminosities is due to AGN downsizing rather than to an evolution
in the accretion rates onto the supermassive black holes.
That is, the highest mass supermassive black holes in the most 
optically-luminous
host galaxies are switching off between $z=1$ and the present
time, leaving only the lower mass supermassive black holes in the less 
optically-luminous galaxies active.

$\bullet$ A primary aim of the paper was to explore the origin 
of the Steffen effect, which is the observation that broad-line AGNs 
dominate the number densities at the higher X-ray luminosities, while 
non--broad-line AGNs dominate at the lower X-ray luminosities.
We ruled out galaxy dilution as being a partial explanation of
the Steffen effect by seeing how the nuclear UV/optical properties
of the X-ray sources in the CDF-N (measured from the ACS GOODS-North data)
relate to the spectral characteristics of the sources.
We found that the UV/optical nuclei of the optically-narrow
AGNs are much weaker than expected if the optically-narrow AGNs
were similar to the broad-line AGNs.
Since the simple unified model cannot explain the Steffen effect, 
we postulated that the simplest 
interpretation of the Steffen effect is a luminosity dependent 
unified model. Another possible interpretation is that the broad-line AGNs
and the optically-narrow AGNs are intrinsically different. 
Regardless of which interpretation is correct, a large 
fraction of the low-redshift, hard X-ray energy density production 
is in optically-narrow AGNs; thus, bolometric corrections for these 
sources are needed if we are to map the accretion history with redshift.

$\bullet$ To infer the bolometric corrections, we
constructed composite spectral energy distributions (including
long-wavelength data from the mid-infrared to the submillimeter)
as a function of both spectral type and X-ray luminosity (to
cover both interpretations of the Steffen effect).
In contrast to the broad-line AGNs, we found that the optically-narrow AGNs
are, on average, highly suppressed in the frequency range between
the UV and soft X-rays and have a strong far-infrared component.
We inferred a bolometric correction of about 85 for these sources.
Likewise, we found that the low X-ray luminosity sources have much
fainter UV/optical nuclei relative to their $2-8$~keV fluxes than
do high X-ray luminosity sources. The far-infrared luminosities
of these low X-ray luminosity sources are also consistent with a
bolometric correction of 85. 

$\bullet$ Using our bolometric corrections and our parameterizations
of the evolution of the rest-frame hard X-ray energy density
production rates, we inferred the accreted supermassive black hole mass densities
for all spectral types and for broad-line AGNs alone. We found that
only about one-half to one-quarter of the supermassive black hole mass density
was fabricated in broad-line AGNs.

$\bullet$ We compared our accreted supermassive black hole mass 
density for broad-line AGNs
alone with the optical QSO determination made by Yu \& Tremaine (2002)
using the Boyle et al.\ (2000) QSO luminosity function, assuming
$\epsilon=0.1$. Our value is 1.75 
times smaller than their value, but this difference is very sensitive
to their extrapolation of the optical QSO luminosity function outside 
the optically observed luminosity range. The new Croom et al.\ (2004) 
QSO luminosity function 
has a shallower faint-end slope than Boyle et al.\ (2000).
We recalculated the accreted supermassive black hole mass densities for 
optical QSOs using the Croom 
et al.\ fit, first by restricting to the measured absolute magnitude 
range, and second by extrapolating the fit to higher and lower 
luminosities, and we found that the two values bracketed our hard 
X-ray luminosity function 
determination. Thus, both the new optical QSO luminosity function 
and the hard X-ray luminosity function 
determinations of the supermassive black hole mass densities are a 
factor of almost two
lower than the Yu \& Tremaine (2002) determination. This leaves
room for obscured accretion by the optically-narrow AGNs in a
comparison with the local supermassive black hole mass density.

$\bullet$ For $\epsilon\approx 0.1-0.2$, we found reasonable
agreement between our hard X-ray luminosity function supermassive 
black hole mass density for all spectral types and the local 
supermassive black hole mass density; however, there
is little room for further obscured sources or for any low
efficiency accretion periods.

\acknowledgments
We thank the referee, Robert Antonucci, for helpful comments that
improved the manuscript. We thank Qingjuan Yu and Scott Tremaine
for helpful conversations concerning their paper.
We gratefully acknowledge support provided by NASA through
grants HST-GO-09425.30-A (A.~J.~B.) and
HST-GO-09425.03-A (L.~L.~C.) from the Space Telescope Science
Institute, which is operated by the Association of Universities for
Research in Astronomy, Incorporated, under NASA contract NAS5-26555.
We also gratefully acknowledge support from NSF grants
AST-0084847 and AST-0239425 (A.~J.~B.) and AST-0084816 (L.~L.~C.), 
NASA CXC grants GO2-3191A (A.~J.~B.) and DF1-2001X and GO2-3187B 
(L.~L.~C.), the University of Wisconsin Research 
Committee with funds granted by the Wisconsin Alumni Research 
Foundation, the Alfred P. Sloan Foundation, and the David and
Lucile Packard Foundation (A.~J.~B.), the IDS program of
R.~F.~M., and NASA's National Space Grant College and Fellowship 
Program and the Wisconsin Space Grant Consortium (A.~T.~S.).


\end{document}